\documentclass[graybox]{svmult}

\usepackage{mathptmx}       
\usepackage{helvet}         
\usepackage{courier}        
\usepackage{type1cm}        
\usepackage{makeidx}         
\usepackage{graphicx}        
\usepackage{multicol}        
\usepackage[bottom]{footmisc}
\usepackage{hyperref}        
\usepackage{soul}            
\hypersetup{colorlinks=true,urlcolor=blue}
\usepackage[square,numbers]{natbib}
\makeindex             
                       
\usepackage{textgreek}

\begin{document}
\title*{Cluster outskirts and their connection to the cosmic web}
\author{Stephen Walker \thanks{corresponding author} and Erwin Lau}
\institute{Stephen Walker \at Department of Physics and Astronomy, The University of Alabama in Huntsville, Huntsville, AL 35899, USA, \email{stephen.walker@uah.edu}
\and Erwin Lau \at Smithsonian Astrophysical Observatory, 60 Garden St, Cambridge, MA 02138, USA , \email{erwin.lau@cfa.harvard.edu}}
%
%
\maketitle
\abstract{We review the latest developments in our X-ray observational and theoretical understanding of the outskirts of galaxy clusters, and their connection to the cosmic web. The faint cluster outskirts are challenging regions to observe in X-rays, requiring highly sensitive telescopes with low and stable background levels. We present our latest understanding of the thermodynamic profiles of clusters in the outskirts, and the biases that gas clumping and non-thermal pressure support can introduce. Features in the outskirts due to merging activity are discussed, along with the chemical enrichment of the outskirts ICM. We describe future prospects for X-ray observations to explore further out in the cluster outskirts and probe their connections to the cosmic web.}

\section{Keywords} 
Galaxy clusters, Intracluster medium, Cosmic web

\section{Introduction}
As the largest gravitationally bound structures in the universe, galaxy clusters continue to grow and accrete matter in their outskirts. {In the outskirts, infalling cold gas is heated through an accretion shock, eventually virializing
to high temperatures (10$^{7-8}$K) to form the intracluster medium (ICM), which contains the vast majority of the baryonic matter in clusters. The majority of the gas lies in the outskirts beyond the cluster's virial radius (the boundary within which gas in the cluster is virialized) and in the intergalactic medium (IGM) within filaments that connect clusters to the cosmic web.} {This gas emits X-ray's due to thermal bremsstrahlung, and the brightness of the X-ray emission is proportional to the square of the density of the gas. While the cores of relaxed clusters are typically very X-ray bright due to the relatively high gas density of $\sim$0.1cm$^{-3}$, in the outskirts the density declines to below 10$^{-4}$ cm$^{-3}$.} 

This low gas density means that these regions are extremely faint in X-rays, making measurements challenging, and pushing X-ray telescopes to their limit of sensitivity. Cosmological simulations of galaxy cluster formation predict the outskirts to be a hive of activity with gas continuing to accrete, and ongoing mergers with small sub-clusters and clumps of gas. A plethora of structure formation physics is believed to be operating in the outskirts. These physical processes are fundamentally different from the physics in the cores of clusters that has been the focus of X-ray cluster science over the past several decades. 

The outskirts of galaxy clusters is a new territory for addressing the following outstanding questions at the crossroads of cosmology and astrophysics: How do galaxy clusters grow? How do they connect to the cosmic web? Here we provide a review of the current state of the art theoretical and observational view of cluster outskirts, as explored by X-ray telescopes. 


{This review is structured as follows. In section~\ref{sec:definition} we describe the technical definition of what is meant by the `Cluster Outskirts'. In section~\ref{sec:Observations} we describe X-ray observations of the cluster outskirts. Section~\ref{sec:theory} describes our current state of the art theoretical understanding of the cluster outskirts.}

{Finally, we conclude with a discussion of future prospects for observations of the cluster outskirts in X-rays in section~\ref{sec:futureobservations}.}

{Unless noted otherwise, we adopt the standard $\Lambda$CDM model of cosmology, using the parameters from Planck 2018 \citep{planck18_cosmo}, with $H_0=67.36\,\mathrm{km\,s^{-1}\,Mpc^{-1}}$, $\Omega_{M,0} = 1-\Omega_\Lambda = 0.3153$, throughout the chapter. Here $\Omega_{M,0} \equiv \rho_{m}(z=0)/\rho_c(z=0)$ and $\Omega_\Lambda$ are the energy density fractions of matter, and the Cosmological Constant $\Lambda$, respectively. }

\section{Definition of Cluster Outskirts}\label{sec:definition}

\begin{figure}
    \centering
    \includegraphics[scale=0.4]{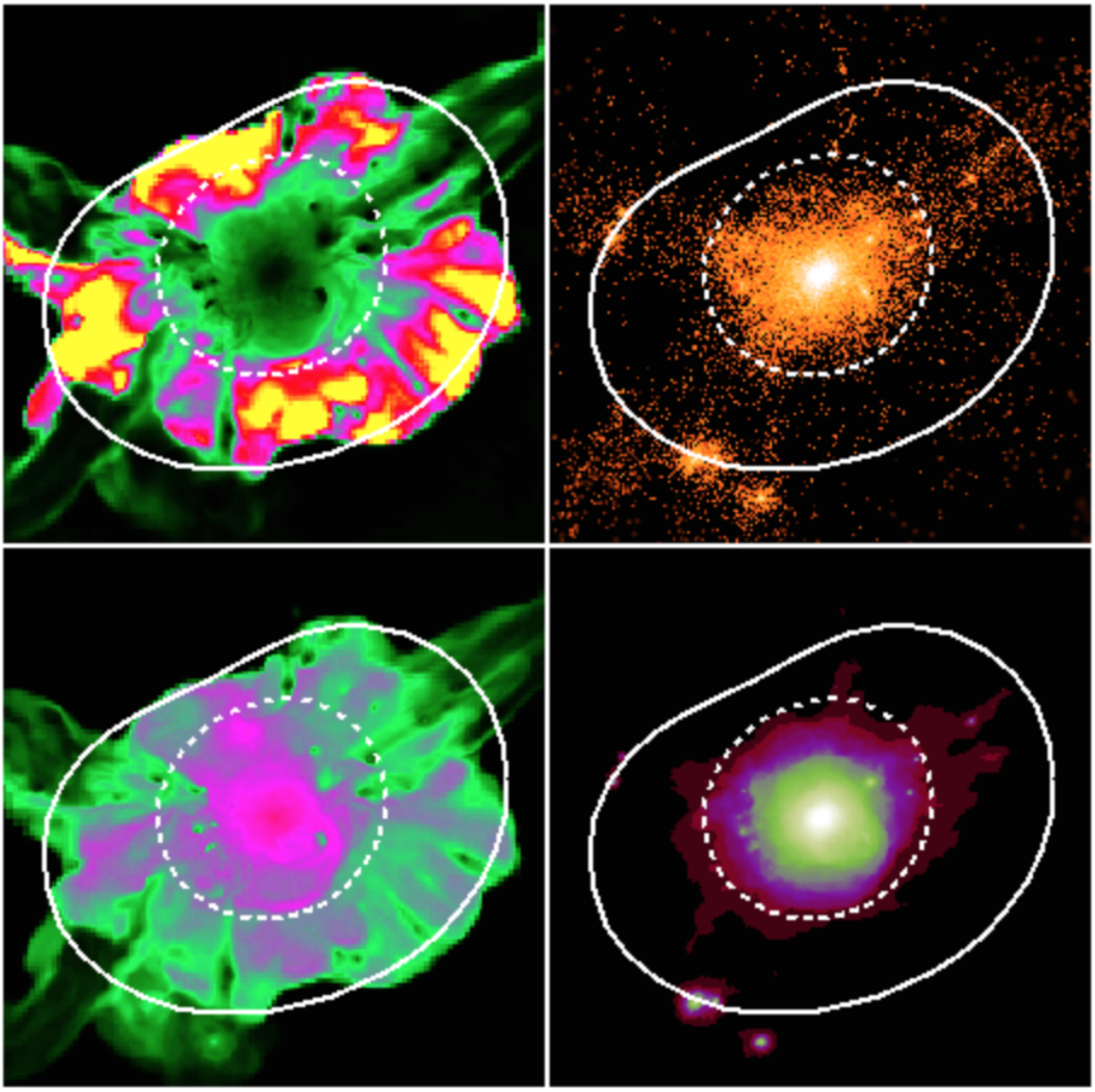}
    \caption{Maps in gas entropy (top left), dark matter (top right), gas temperature (bottom left) and gas pressure (bottom right) for a cluster halo with mass $M_{200c} = 7.4 \times 10^{14}\,M_\odot$, taken from a cosmological simulation. Each map is $22$~Mpc wide with a projection depth of $5.6$~Mpc. The dashed line traces the dark matter splashback shell, while the solid line shows the accretion shock shell. Figure taken from \cite{aung20}, reprinted with permission.}
    \label{fig:shock_splashback_map}
\end{figure}

The outskirt regions of galaxy clusters contain rich information on physics of the growth and assembly of galaxy clusters. Under the \textLambda CDM model of structure formation, more massive systems, such as galaxy clusters, grow by mergers from less massive systems (galaxies and galaxy groups), and from accretion from the cosmic web filaments. Under this picture of cluster formation, accreted matter piles up in the outer regions of the cluster in an inside-out fashion: more recently accreted matter is deposited farther away from cluster center. Cluster outskirts therefore are where matter actively accretes and virializes with the gravitational potential wells of the clusters. By probing the properties of matter in the cluster outskirts, we are probing the growth of galaxy clusters in action. 

To properly define the outskirts of clusters, we first have to define the outer boundary of a galaxy cluster. Conventionally, the radius and the mass of a given dark matter halo are defined in terms of overdensity $\Delta_{\rm ref}$ with respect to some reference background density $\rho_{\rm ref}$: 
\begin{equation}\label{eq:radius_defn}
    M_{\Delta,{\rm ref}} = \frac{4\pi}{3}\Delta_{\rm ref} \rho_{\rm ref} R_{\Delta,{\rm ref}}^3, 
\end{equation}
{where the reference background density can be either the critical density:
\begin{equation}
\rho_{c}(z) \equiv \frac{3H(z)^2}{8\pi G}
\end{equation}
with $\rho_{c}(z=0)=1.259 \times 10^{11}\,M_\odot\mathrm{Mpc}^{-3}$, or the mean density of the universe: 
\begin{equation}
\rho_{m}(z) = \rho_c(z) \Omega_M(z) = \rho_c(z)\frac{\Omega_{M,0}}{\Omega_{M,0} + \Omega_\Lambda (1+z)^{-3}},    
\end{equation}
with $\rho_{m}(z=0)=3.971 \times 10^{10}\,M_\odot\mathrm{Mpc}^{-3}$.  The subscripts $c$ or $m$ of any quantity denote that it is defined with respect to the critical, or mean density of the Universe. } 

A common choice of the radius of the halo boundary is $R_{200c}$, which is the radius encompassing matter with an average density 200 times the critical density. 
{ The 200 value is commonly chosen because it is close to the overdensity value ($\Delta_c=178$) for the virial radius obtained in the spherical collapse model in a matter-dominated universe ($\Omega_M = 1$). 
For other cosmologies (with negligible radiation density), a widely used approximation for the virial overdensity is $\Delta_c = 18\pi^2 + 82x -39x^2$, where $x\equiv \Omega_m(z)-1$ \cite{bryan_norman98}. The virial overdensity at $z=0$ is approximately $100$ for the standard \textLambda CDM model. } 

Another radius that is often used as reference  is $R_{500c}$, the radius enclosing a total mass with an average of 500 times the critical density of the Universe. 
{ The density and temperature profiles of hot gas in galaxy clusters are routinely measured out to $R_{500c}$ with the current generation of X-ray telescopes \cite{vikhlinin06}. }

Note that these halo radius definitions based on overdensities are not  ``physical'', in that they do not correspond to any actual physical boundaries. Recent theoretical developments \cite[e.g.][]{diemer14, adhikari14}, however, have suggested the existence of physically-motivated boundary of dark mater halos, called the ``splashback'' radius, or $R_{sp}$ (see Fig. \ref{fig:shock_splashback_map}). The splashback radius refers to the apocenters (farthest point of the particle orbit with respect to the halo potential minimum) of infalling dark matter through the pericenter (closest point of the particle orbit with respect to the minimum of the gravitational potential of the halo). It is called ``splashback'' radius because it is the radius where infalling dark matter that is splashed back from the halo center to the outskirts accumulates. { The splashback radius is found to be dependent on the mass accretion rate of the halo \cite{more15}, defined as 
\begin{equation}\label{eq:mar}
    \Gamma \equiv \frac{\Delta \ln M(a)} { \Delta \ln a},
\end{equation}
where $M(a)$ is the mass of the cluster at scale factor $a = 1/(1+z)$. 
Cosmological simulations show that $R_{sp}$ is closely tracked by $R_{200m}$ \cite{more15}, the radius where the enclosed average density is 200 times the {\em mean} density of the Universe. Thus, $R_{200m}$ has been used as a proxy for the $R_{sp}$ and $\Gamma$.  }

{ Another relevant radius in cluster outskirts is the accretion shock radius $R_{sh}$. It is the radius where the infalling gas from the surrounding environment gets shock heated for the first time. The accretion shock radius can be identified with a strong jump in temperature, pressure, and entropy, corresponding to Mach number $\mathcal{M} \sim 10-100$. Here $\mathcal{M}=v/c_{\rm s}$ where $v$ is the local gas velocity, and $c_{\rm s}=1480(T/10^{8} {\rm K})^{1/2}\,{\rm km\,s^{-1}}$ is the speed of sound. Cosmological simulations show that the accretion shock radius $R_{sh}$ is approximately $1.6$ times the splashback radius, independent of halo mass \citep{aung20}. }

For reference, at redshift of zero the ratio of $R_{500c} : R_{200c} : R_{200m} : R_{sp} : R_{sh}$ is approximately $1:1.4:3:3:4.8$.


\section{Observations}
\label{sec:Observations}

\begin{figure*}[t]
 \hbox{ 
  \includegraphics[width=0.33\textwidth]{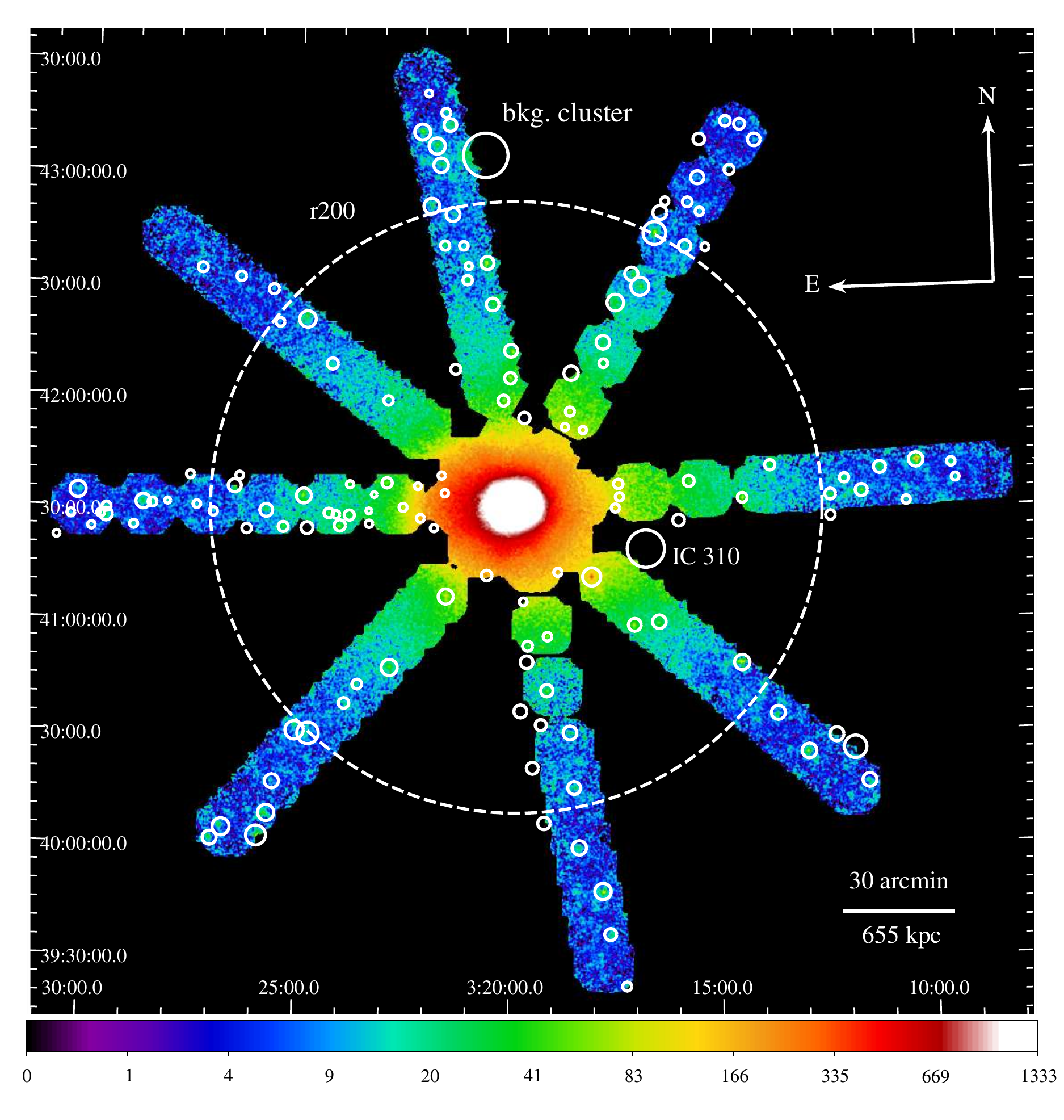}
 \includegraphics[width=0.28\textwidth]{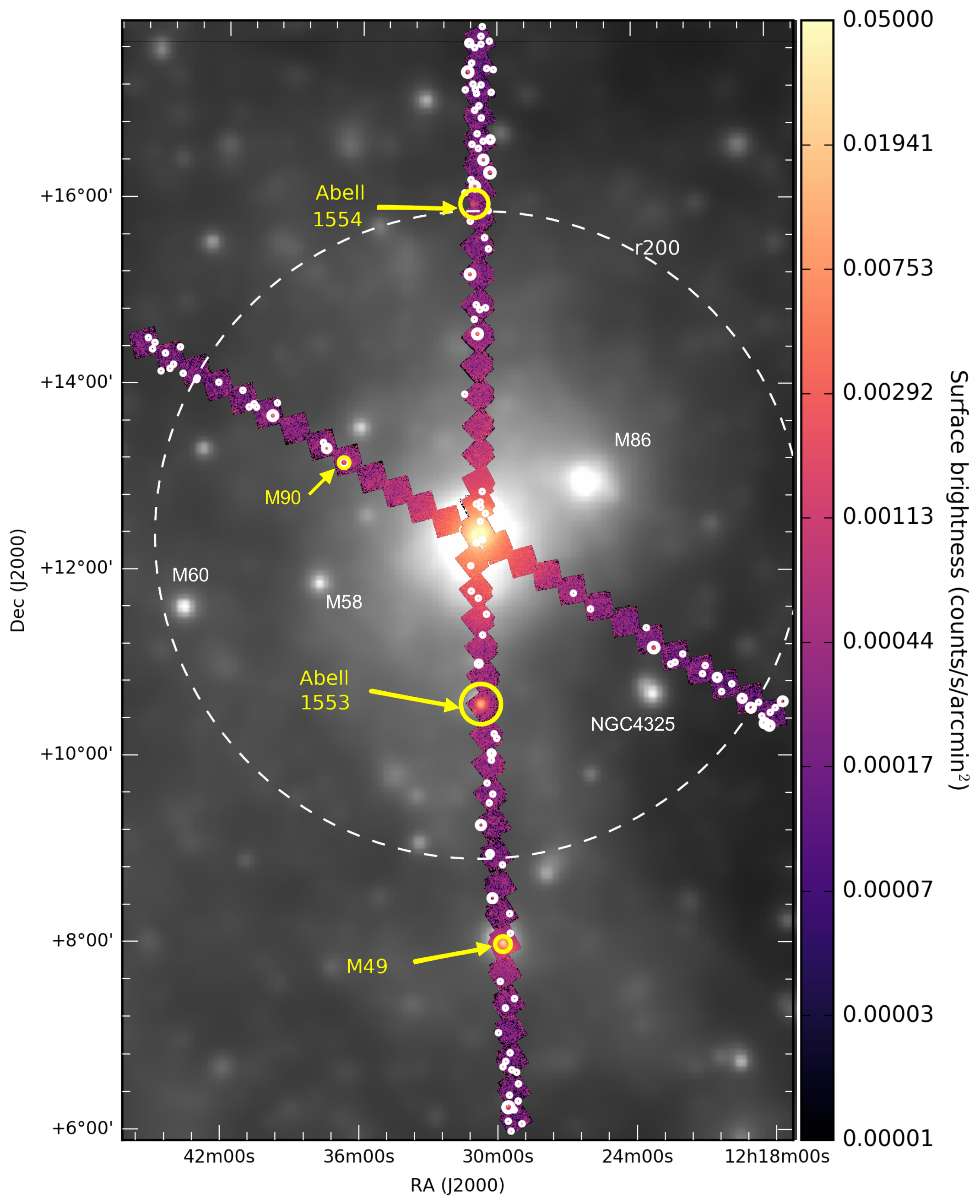}
 \includegraphics[width=0.33\textwidth]{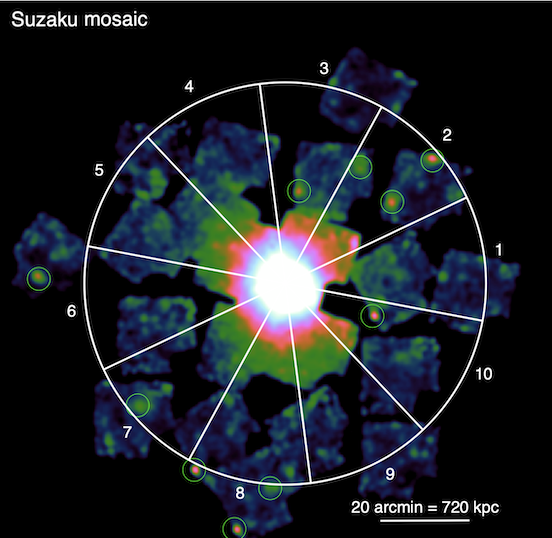}
 
 }
\caption{Results from large Suzaku Projects on the outskirts of the Perseus cluster \cite{urban14} (left), the Virgo cluster \cite{Simionescu17} (center) and Abell 2199 \cite{MirakhorA2199} (right), reprinted with permission. Suzaku's low instrumental background allows the ICM temperature, density and entropy in the outskirts to be measured using spectral fitting. }
\label{fig:Suzaku_large_projects}
\end{figure*}

In order to see the low surface brightness emission in the outskirts of galaxy clusters, an X-ray observatory needs to have a low and stable instrumental background which can be accurately measured and subtracted, together with a large collecting area to gather enough photons. The two main observatories which have driven the modern era of X-ray astronomy, NASA's Chandra X-ray observatory and ESA's XMM-Newton observatory (both 1999-present), are both in long period elliptical orbits, allowing them to make long, uninterrupted observations. However these orbits take Chandra and XMM-Newton outside the magnetopause of the Earth, giving them much higher instrumental background levels than X-ray observatories located at low Earth orbit, such as ROSAT (1990-1999) and {JAXA}'s Suzaku observatory (2005-2015). As a result of this, Chandra and XMM-Newton are only able to reliably measure gas temperatures out to around 66 percent of the virial radius in galaxy clusters (\cite{Vikhlinin2005}, \cite{Leccardi2008}).

ROSAT pre-dated the use of CCD’s to detect X-rays, and instead used gas proportional counters. This resulted in much poorer spectral resolution than can be achieved with CCD’s, while the short focal length of the telescopes resulted in a narrow passband (0.1-2.4 keV), {such that ROSAT was not designed for temperature measurements of the ICM.} 

Suzaku was the first X-ray observatory to allow X-ray spectroscopy out to the virial radius ($R_{200c}$) of clusters, allowing direct temperature measurements to be made. Observing strategies with Suzaku broadly followed one of two approaches, necessitated by Suzaku's large point spread function (PSF) (a half power diameter, HPD of 2 arcmins) and modest field of view (18$\times$18 arcmins), which is just slightly larger than Chandra's. One approach was to study the nearby brightest clusters, which allowed the highest resolution profiles of temperature and entropy to be made. However because of the large angular extent of these clusters, it was not possible to completely cover all of their azimuth, and so strategies were limited to observing strips from the cluster core to the outskirts (e.g. Perseus, \cite{simionescu11,urban14}, Centaurus \cite{walker13} and Virgo \cite{Simionescu17}, see Fig. \ref{fig:Suzaku_large_projects}). The other approach was to look at intermediate redshift clusters which could be completely covered by Suzaku with modest mosaics of around 4 observations (e.g. Abell 1689 \cite{kawaharada10}, Abell 2029 \cite{walker12}). This allowed full azimuthal coverage to be achieved, at the expense of the spatial resolution of the observed thermodynamic profiles. In only one case, for Abell 2199 \cite{MirakhorA2199}, was a nearby cluster covered with very high azimuthal coverage.

Here we describe X-ray observations of the cluster outskirts, starting with a discussion of methods for measuring thermodynamic parameters (section~\ref{sec:methods_measuring_thermo}), before summarizing the latest measurements in section~\ref{sec:observed_thermo}. Potential observational biases are discussed in section~\ref{sec:observed_biases}. The phenomena of cold fronts and shocks in the outskirts are discussed in sections~\ref{sec:coldfronts} and \ref{sec:shocks} respectively. The observed abundance of metals in the cluster outskirts is covered in section \ref{sec:metals}, and the connections with the cosmic web in section~\ref{sec:cosmicweb}.

\subsection{Methods for measuring thermodynamic properties}
\label{sec:methods_measuring_thermo}
Two different approaches have been used for measuring gas temperatures in the low surface brightness outskirts of clusters. The first is to use the X-ray spectra from a mission with a low and stable instrumental background, and good spectral resolution over a wide pass band (0.5-7.0 keV). So far only Suzaku and eROSITA meet all of these criteria, as the instrumental backgrounds of Chandra and XMM-Newton are significantly higher, and ROSAT's proportional counters had poor spectral resolution and a small passband. The main challenges with this method are to accurately subtract the non X-ray background from the observations (produced {by} particle interactions with the X-ray instrumentation). One then needs to carefully model all of the foreground components from the Milky Way and the background components from the cosmic X-ray background (unresolved point sources from distant AGN). This spectral fitting method allows the temperature ($kT$), electron density ($n_e$) and metal abundance ($Z$) of the ICM to be obtained. {The pressure, $P$, can then be obtained using $P=n_ekT$, while the entropy, $K$, can be obtained using $K =kT/n_{e}^{2/3}$. The entropy definition here is related to the conventional thermodynamical entropy $S$ by $S \equiv \ln K + \mathrm{constant}$. For further details of the thermodynamic properties of galaxy clusters, please see the chapter `Thermodynamical profiles of clusters and groups, and their evolution'.  }

\begin{figure*}
 \hbox{ 
  \includegraphics[width=1.0\textwidth]{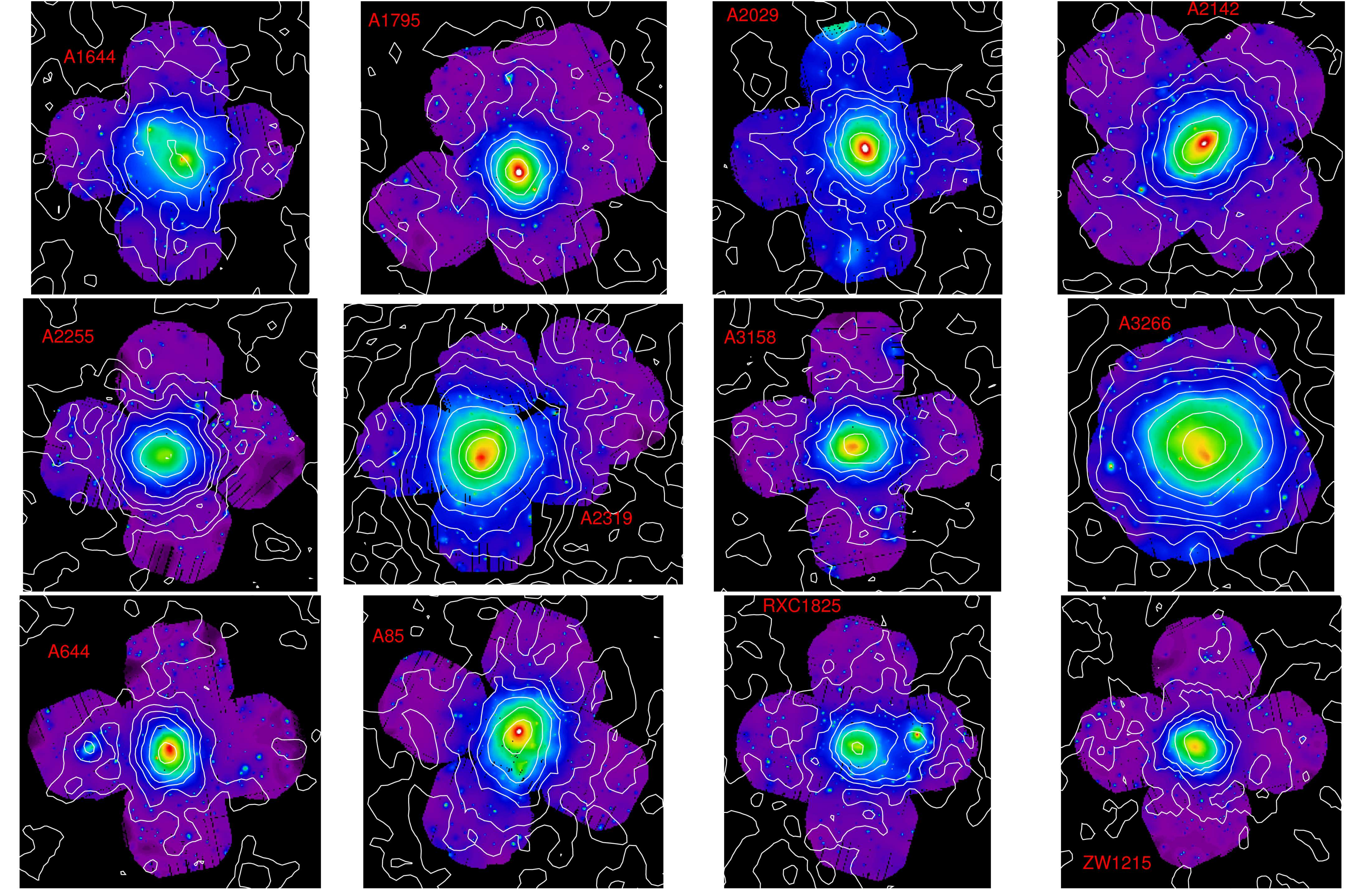}
 }
\caption{XMM-Newton images of the cluster outskirts from the X-COP sample of clusters, with Planck SZ contours overplotted in white, from \cite{Ghirardini2019}, reprinted with permission. By combining the X-ray surface brightness measured with XMM-Newton with the pressure profiles measured by Planck in the SZ, it is possible to obtain the temperature and entropy profiles of galaxy clusters in the outskirts. }
\label{fig:X-COP_images}
\end{figure*}

A second method gets around the need for {X-ray spectroscopy}, but requires using both X-ray observations and measurements of the Sunyaev-Zeldovich (SZ) effect from telescopes operating in the microwave band. {The Sunyaev Zeldovich effect is the inverse Compton scattering of photons from the Cosmic Microwave Background (CMB) by the hot electrons in the intracluster medium. This increases the energy of the CMB photons, distorting the CMB spectrum. The SZ effect is proportional to the line of sight integral of the electron pressure, and so these measurements allow the pressure profiles of galaxy clusters to be measured.}

In this method, the X-ray surface brightness profile in the soft band {(between 0.7-1.2 keV)} is needed. For hot ICM (kT $>$ 2keV) the X-ray surface brightness in the soft band is relatively independent of temperature and metal abundance, so one can simply convert the X-ray surface brightness into an emission measure, and then into a density under the assumption of spherical symmetry. This therefore provides a density profile, but not a temperature or metal abundance profile. These density profiles are then combined with pressure profiles obtained from SZ measurements, which allows the temperature {($kT=P/n_e$)} and entropy {($K=P/n_e^{5/3}$)} profile to be obtained. Because this method does not require direct spectral fitting, and just needs a measurement of the X-ray surface brightness in the outskirts, it requires less accurate modelling of the background. This means that it is possible to use X-ray surface brightness measurements from XMM-Newton (see Fig. \ref{fig:X-COP_images}) to achieve this.

\subsection{Observed thermodynamic profiles in the outskirts}
\label{sec:observed_thermo}
\begin{figure*}
 \hbox{ 
  \includegraphics[width=0.5\textwidth]{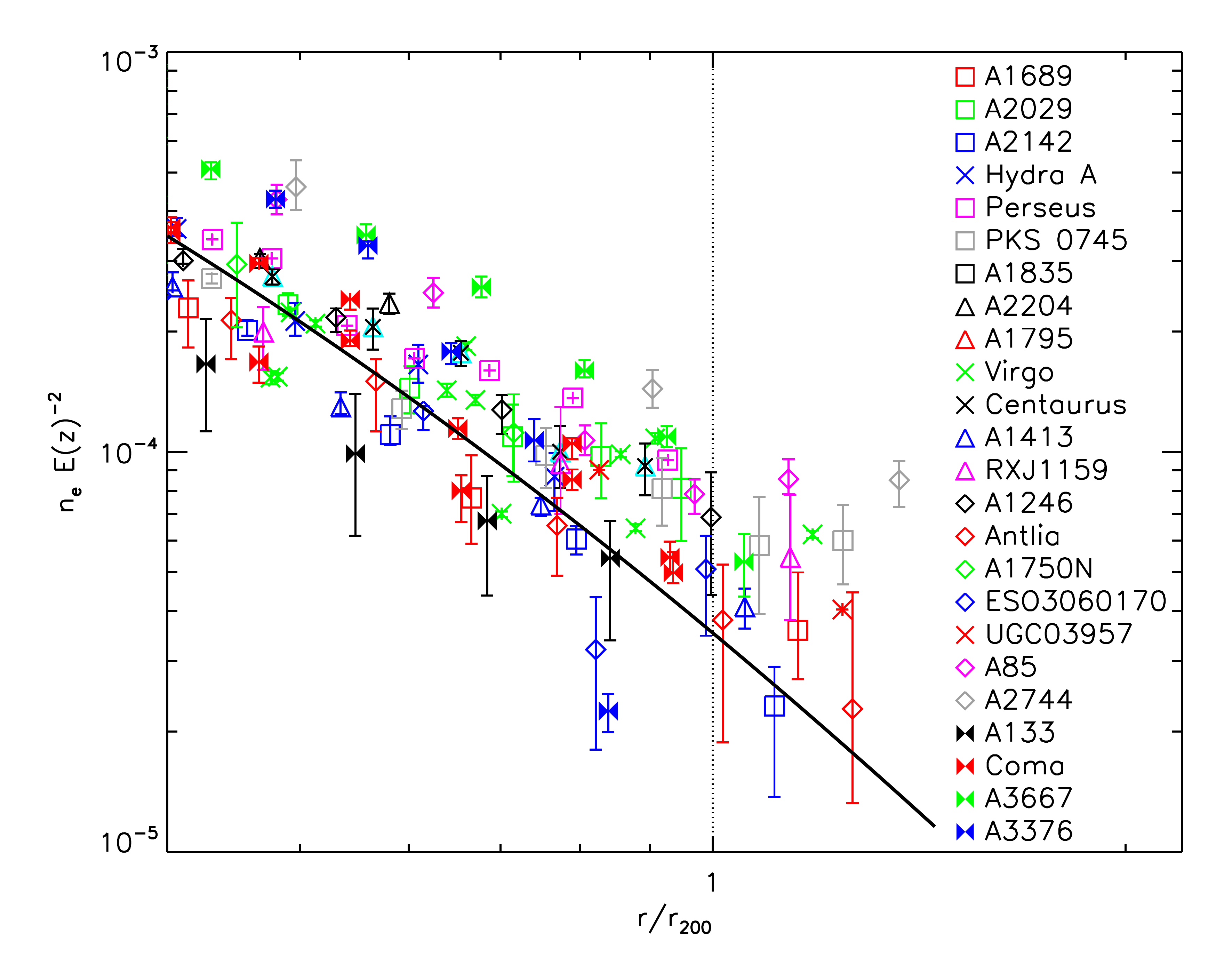}
 \includegraphics[width=0.5\textwidth]{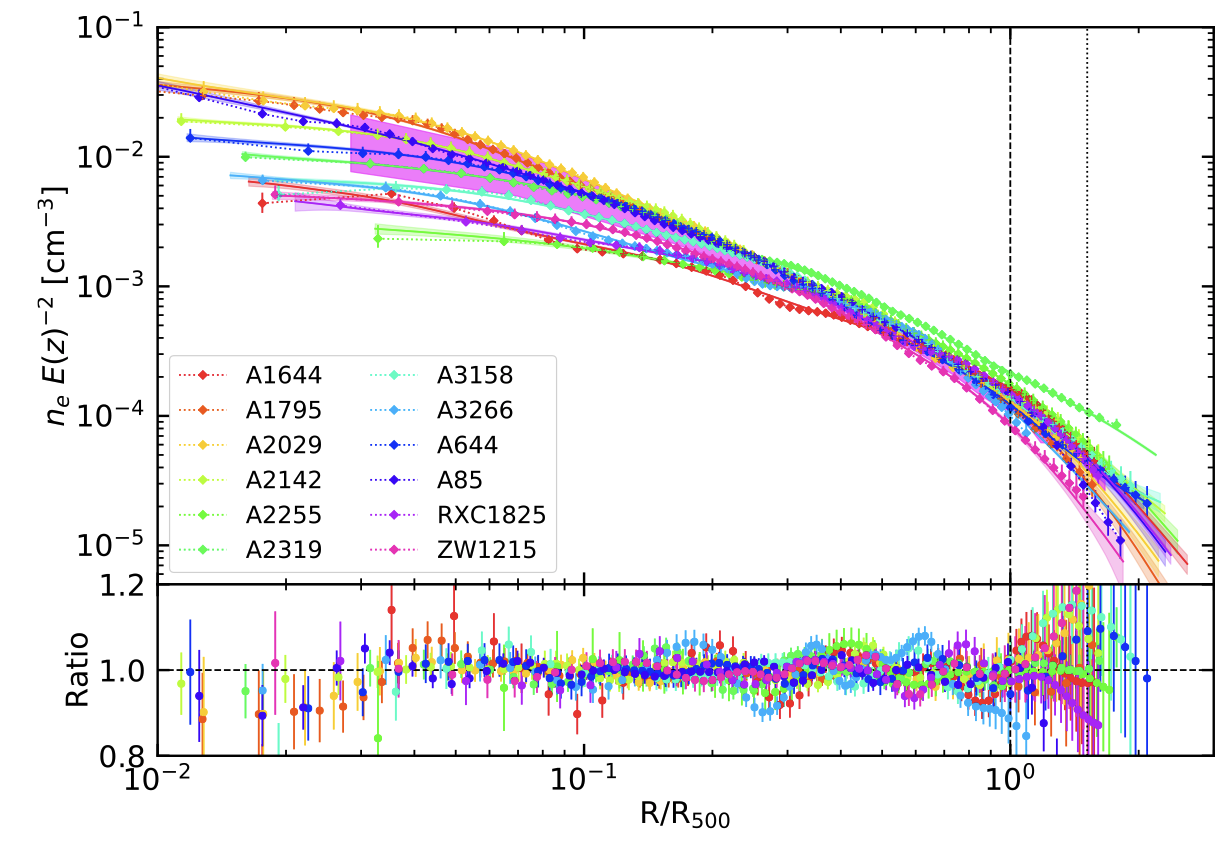}
 }
 \hbox{ 
  \includegraphics[width=0.5\textwidth]{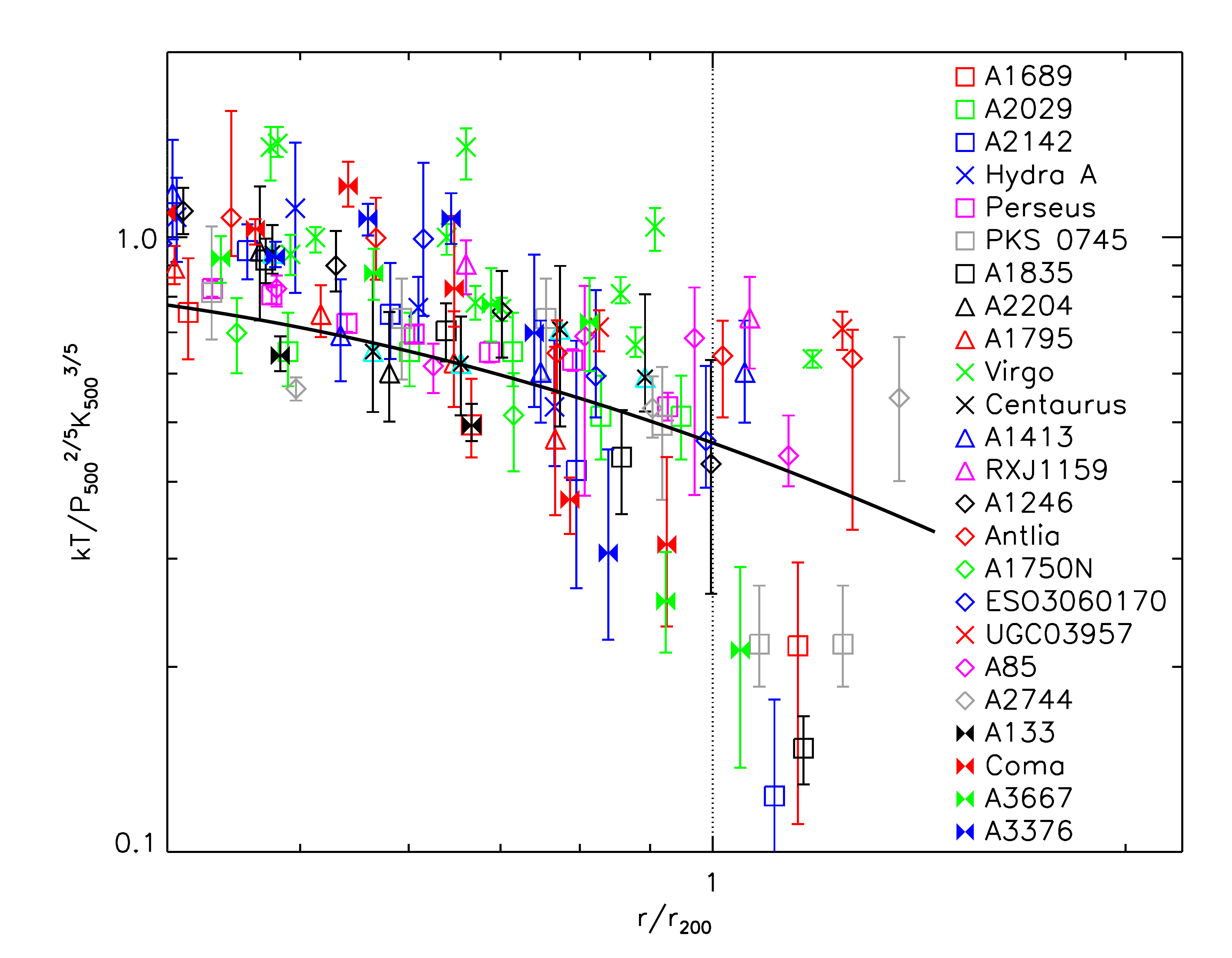}
  \includegraphics[width=0.5\textwidth]{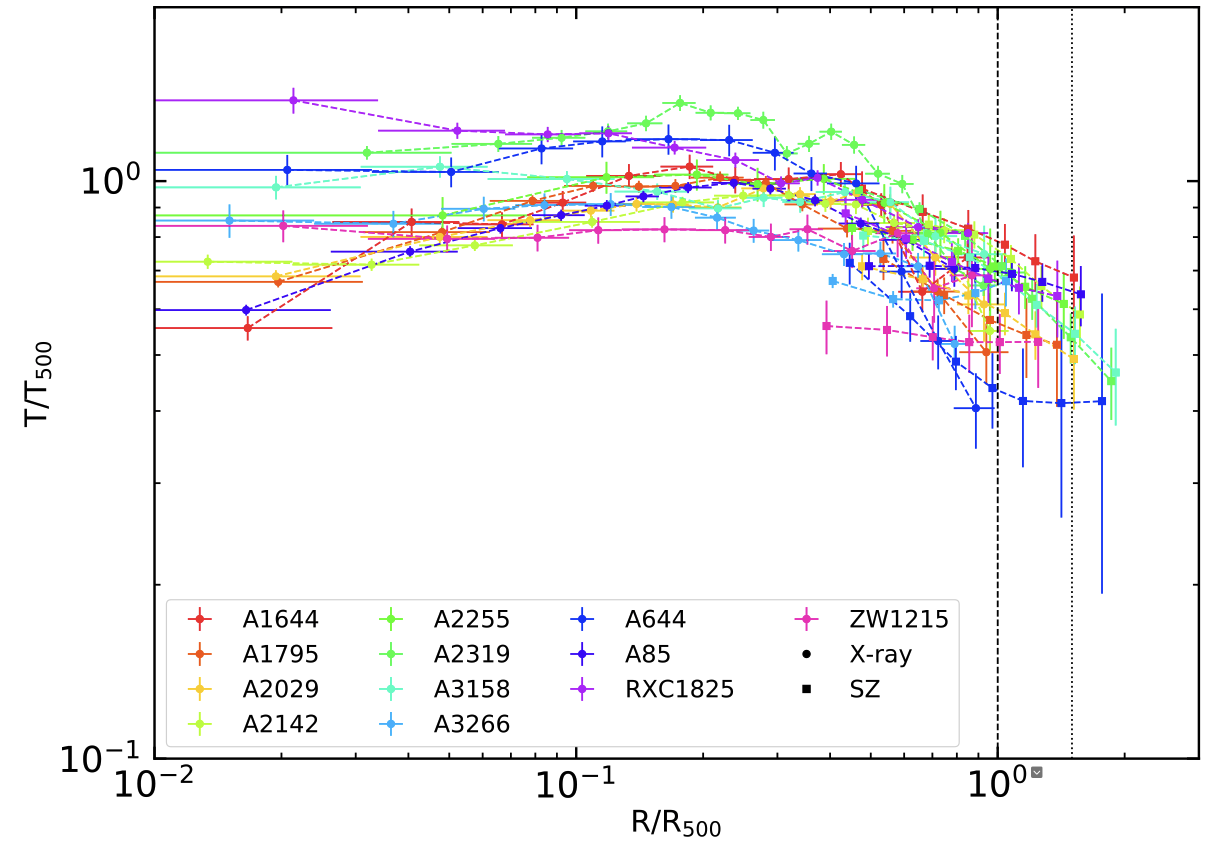}
 }
 \hbox{ 
  \includegraphics[width=0.5\textwidth]{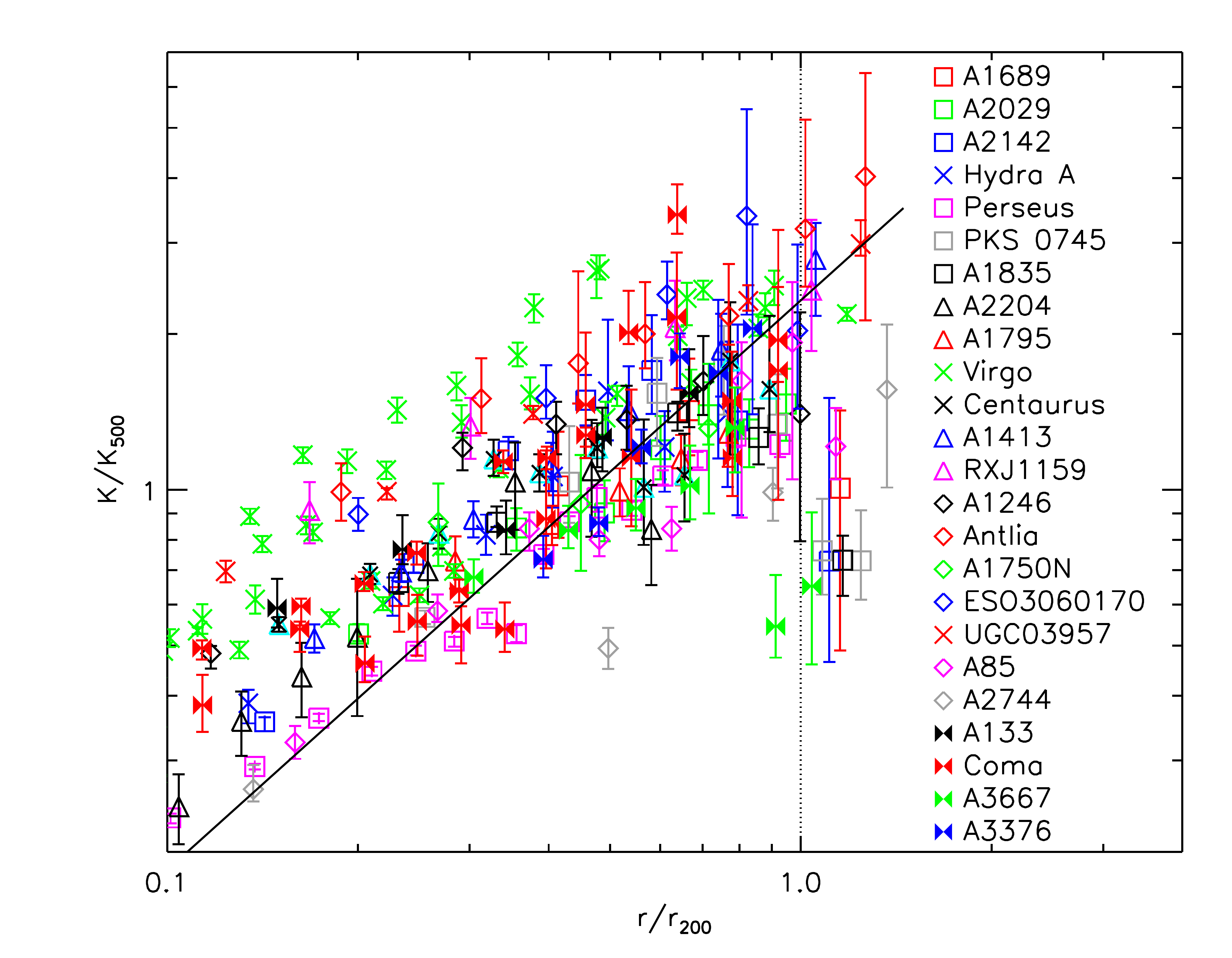}
  \includegraphics[width=0.5\textwidth]{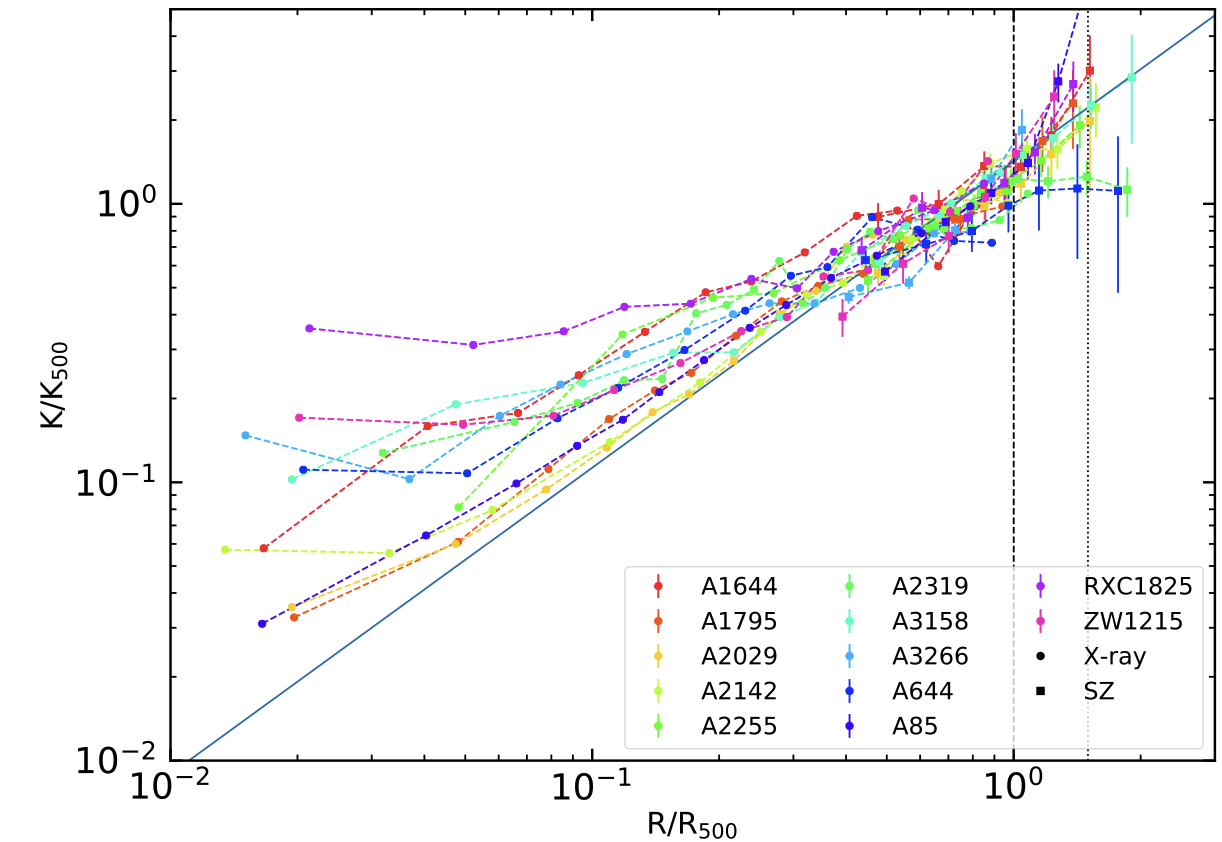}
 }
  \hbox{ 
  \includegraphics[width=0.5\textwidth]{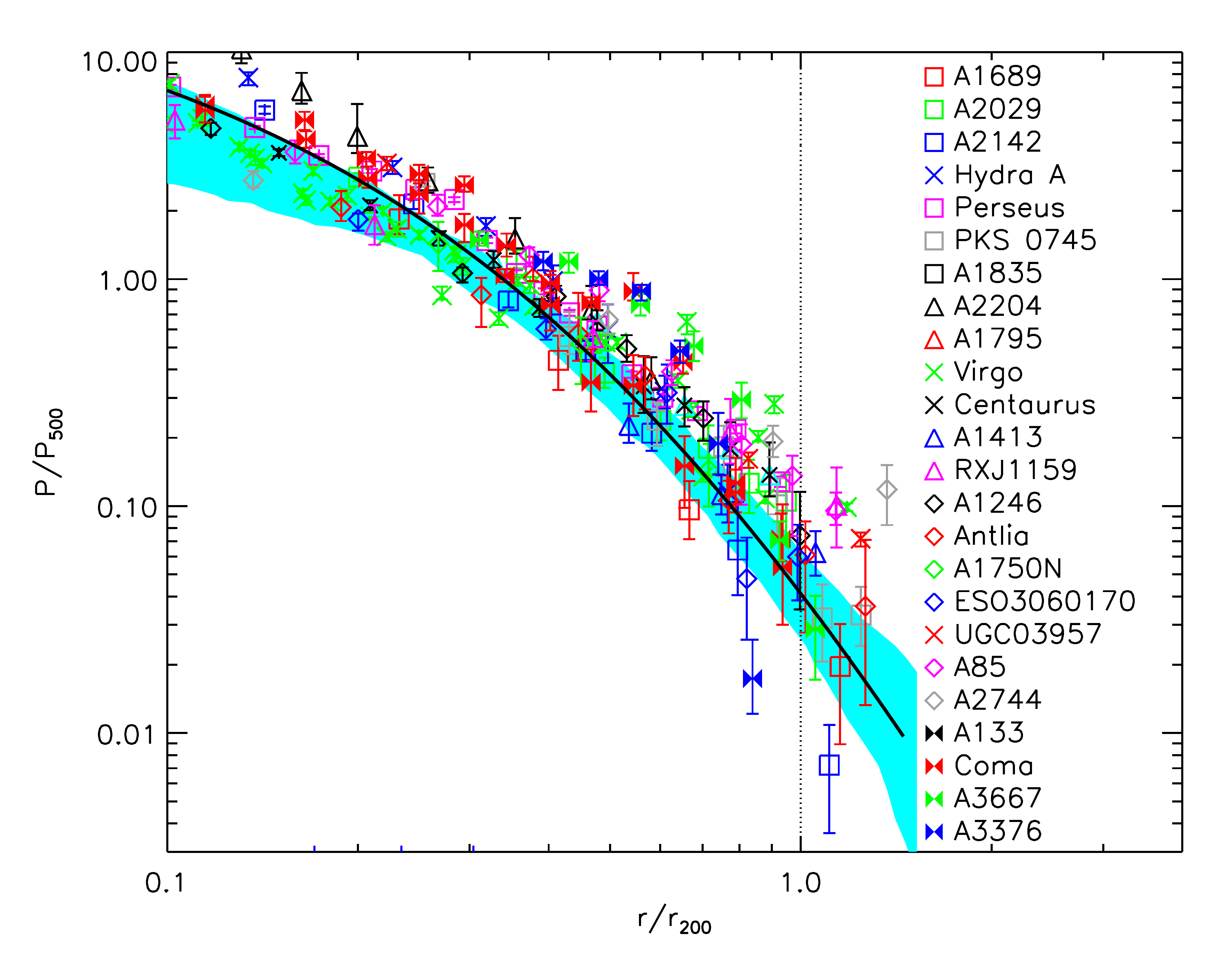}
    \includegraphics[width=0.5\textwidth]{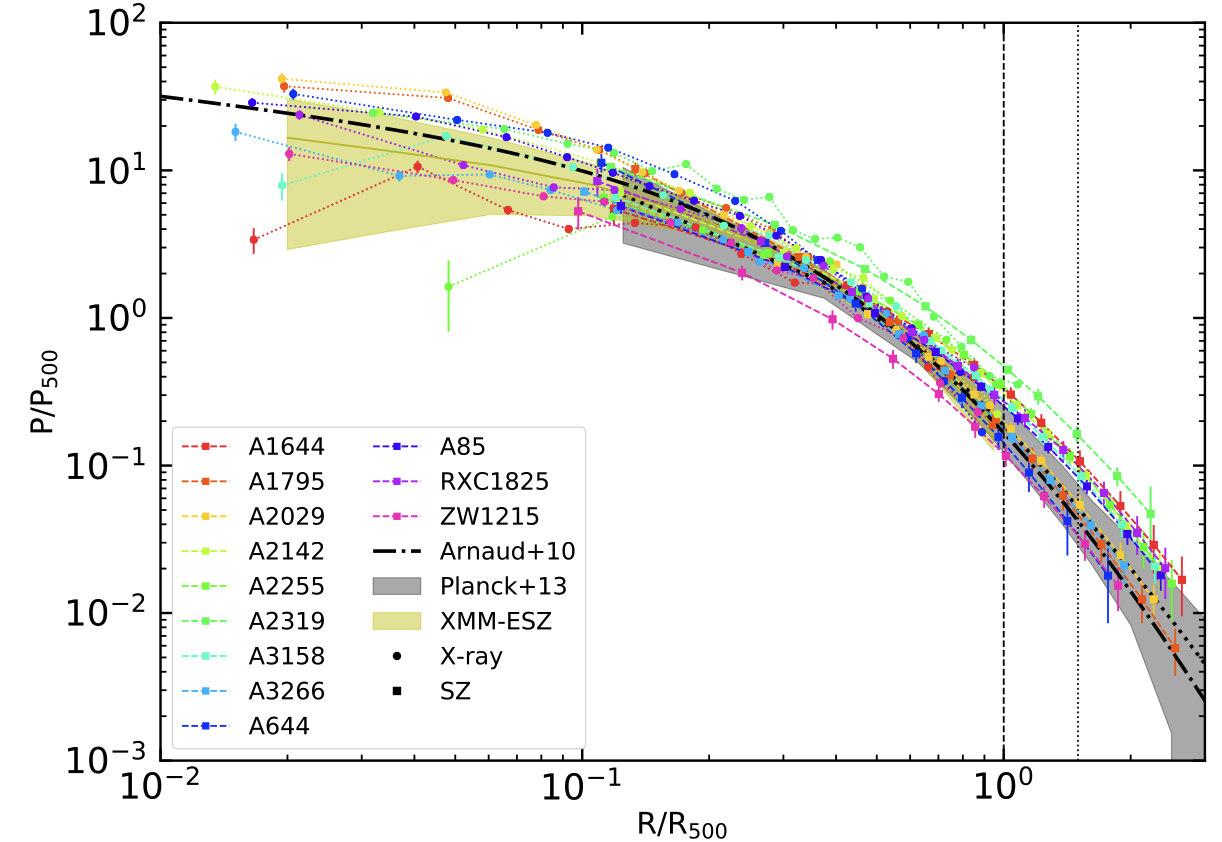}

 }

\caption{Left panels: profiles of thermodynamic properties of galaxy clusters in their outskirts obtained from {X-ray} spectral fitting to Suzaku data, from \cite{Walker2019}, reprinted with permission. From top to bottom we show the density, temperature, entropy and pressure profiles. {The vertical dashed line shows the location of $R_{200c}$. The solid black lines show the the expected profiles for these parameters using the universal pressure profile of \cite{Arnaud10} and the baseline entropy profile of \cite{voit05}.} Right panels show the same profiles, but this time obtained by {combining density profiles derived from XMM-Newton surface brightness profiles with Planck SZ pressure profiles}, \cite{Ghirardini2019}, reprinted with permission. {The vertical dashed and dotted lines show the locations of $R_{500c}$ and $R_{200c}$ respectively.} }
\label{fig:Thermodynamic_profiles}
\end{figure*}

In Fig. \ref{fig:Thermodynamic_profiles} we show a compendium of all of the profiles of temperature, density, entropy and pressure obtained using X-ray spectral fitting to Suzaku data (left column, adapted from \cite{Walker2019}). In the right column of Fig. \ref{fig:Thermodynamic_profiles}, we show the profiles obtained by combining X-ray surface brightness profiles (from XMM-Newton) with SZ derived pressure profiles (from Planck), from the X-COP (XMM-Newton Cluster Outskirts Project) sample presented in \cite{Ghirardini2019}. {The former set of profiles is reported in units of $R_{200c}$, while the latter in units of $R_{500c}$.}

Early observations with Suzaku found that the observed entropy in the outskirts of clusters was lower than expected (e.g. \cite{simionescu11,walker12}). Rather than rising as a power law with an index of $r^{1.1}$ (see Fig. \ref{fig:entropy_profile_sims}) as expected from numerical simulations taking into account only gravitational physics (\cite{voit05}), the first clusters observed {out to large radii} with Suzaku had flat entropy profiles in the outskirts. Additionally, some measurements of the gas mass fraction in {these external regions} were higher than the mean cosmic baryon fraction, suggesting that the density of the gas was being overestimated in the outskirts. 

As can be seen in the Suzaku profiles in the left hand column of Fig. \ref{fig:Thermodynamic_profiles}, the density measurements (coloured data points in the top left panel) tend to lie above the expected density level from simulations (the solid black line). This is consistent with the idea that the gas density is overestimated due to the effect of gas clumping (see section \ref{sec:observed_biases}). The Suzaku temperatures (second panel in the left column) agree well with theoretical expectations out to $R_{200c}$, but outside $R_{200c}$ some measurements are systematically below the expected temperature. This higher density and lower temperature are qualitatively consistent with what would be expected if the ICM in the outskirts contains unresolved, cold and dense gas clumps. The effect of the high density and the low temperature in some systems outside $R_{200c}$ is to cause the entropy profile (third panel in left column) to lie below the expected baseline entropy profile in the outskirts. When considering the pressure profile (bottom panel in the left hand column), the effect of the higher than expected density and the lower than expected temperature tend to cancel each other out, leading to pressure profiles that agree reasonably well with expectations.

In the Suzaku data, the entropy bias mostly only appears for massive clusters ($M>2\times10^{14}$ M$_{\odot}$), while low mass clusters and groups have been found to mostly agree with the baseline entropy profile \cite{Walker2019}. Possibly this is the result of gas clumping being more prevalent in the outskirts of more massive clusters (see section \ref{sec:observed_biases}).

The right hand column of Fig. \ref{fig:Thermodynamic_profiles} shows the same profiles for temperature, density, entropy and pressure, but obtained by {combining density profiles derived from X-ray surface brightness measurements from XMM-Newton with Planck SZ pressure profiles} \cite{Ghirardini2019}. {This sample consists of massive clusters in the range $M_{200c}=6.1-15.1 \times 10^{14} M_{\odot}$, and does not include galaxy groups or low mass clusters.} The higher spatial resolution of the XMM-Newton data allows the effect of clumping to be corrected for (see the next section on clumping corrections for details), which lowers the density, raises the temperature and causes the entropy profiles to better agree with the baseline entropy profile. Some entropy profiles remain below the baseline profile in the outskirts, most notably Abell 2319 \cite{Ghirardini2018}. In this case the leading candidate for the low entropy is the presence of significant non-thermal pressure support in the outskirts. This is discussed in more detail in the next section.

\subsection{Biases due to gas clumping and non-thermal pressure support}
\label{sec:observed_biases}
\begin{figure*}
 \hbox{ \hspace{1.0cm}
  \includegraphics[width=0.8\textwidth]{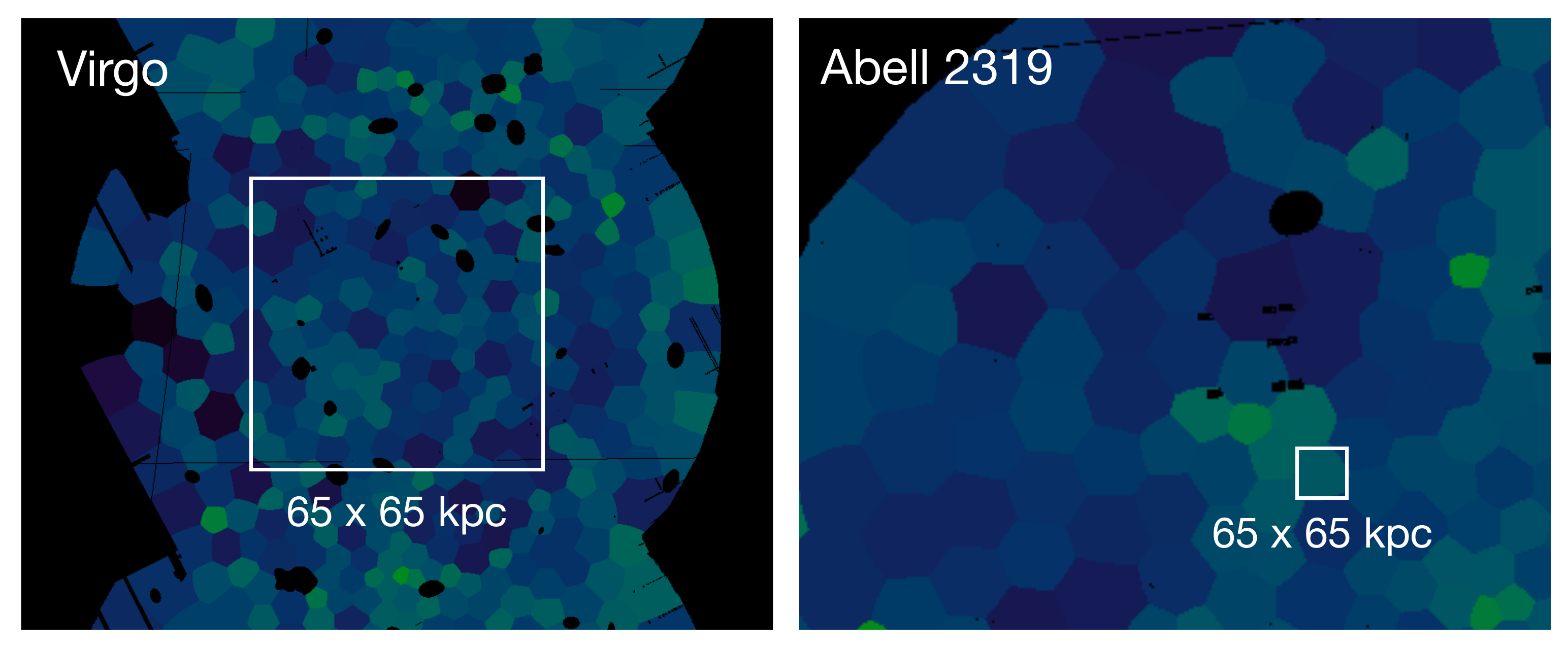}
  
 }

 \hbox{ 
  \includegraphics[width=0.45\textwidth]{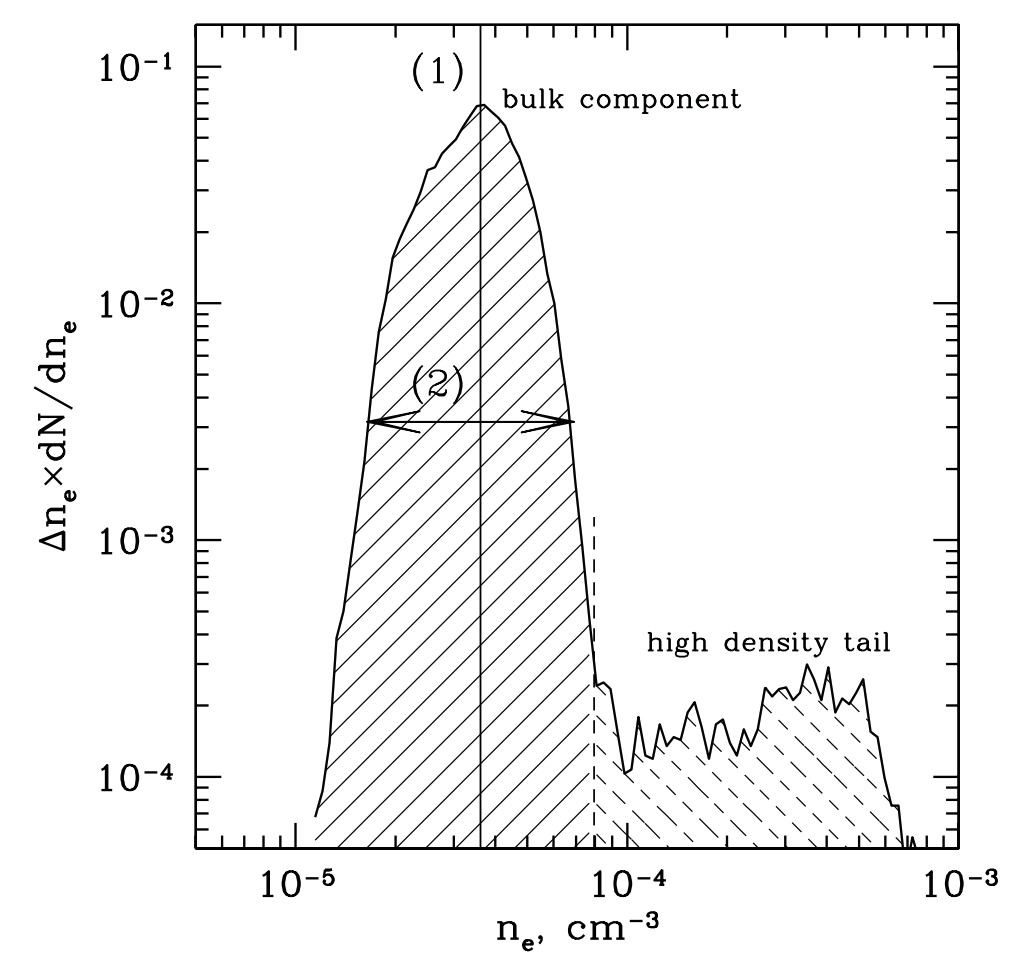}
  \hspace{0.0cm}
    \includegraphics[width=0.55\textwidth]{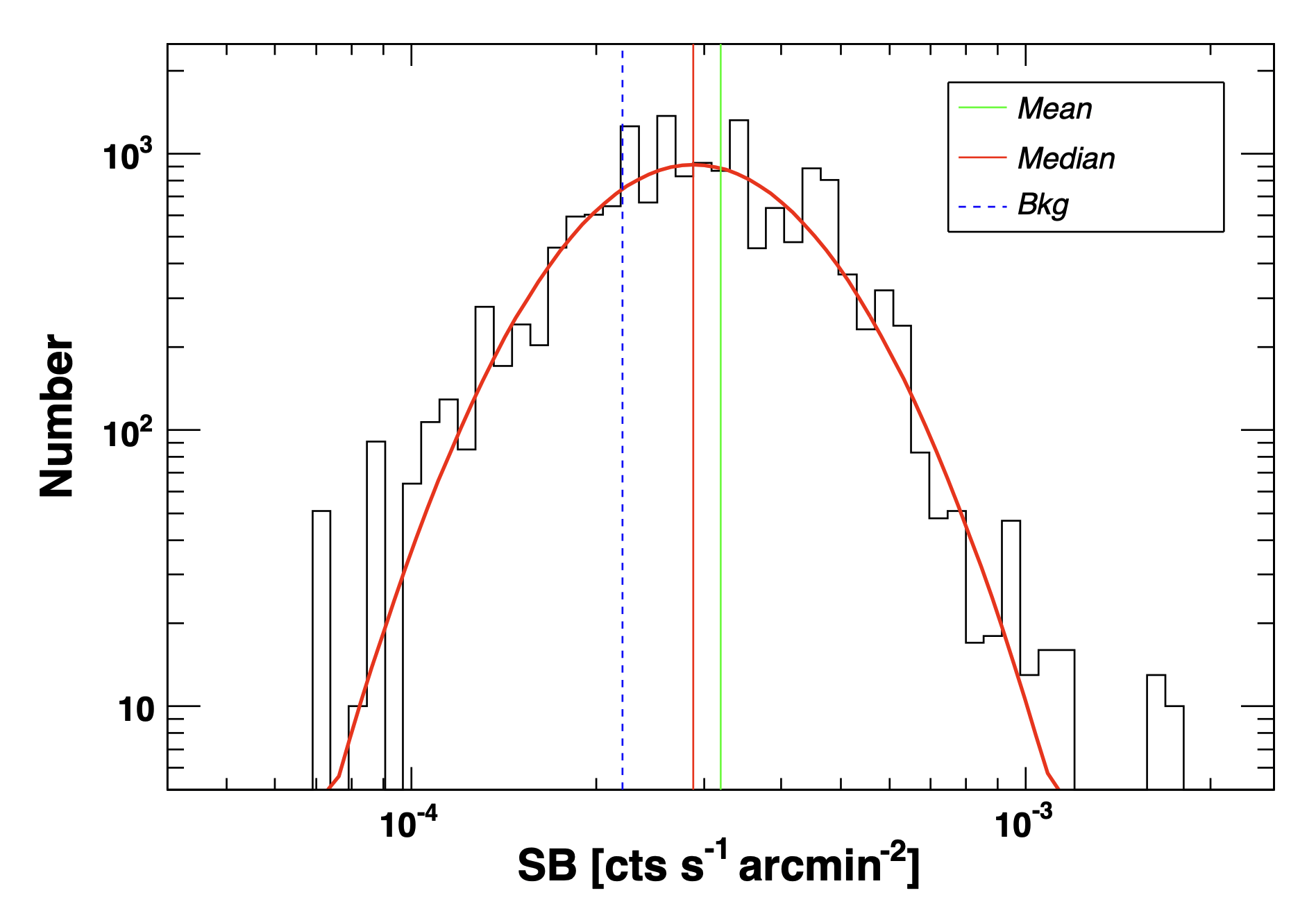}
 }

\caption{Top panels: Voronoi tessellations of the X-ray surface brightness seen using XMM-Newton observations of a nearby cluster (Virgo) and an intermediate redshift cluster (A2319) \cite{MirakhorVirgo2021}, reprinted with permission. {Each Voronoi region contains 20 X-ray photons.} Bottom left: simulations from \cite{zhuravleva13}, reprinted with permission, showing that the distribution of gas densities in the ICM can be divided into a bulk component and a high density tail resulting from gas clumping. The median density is shown by the vertical solid black line and is little affected by the gas clumping, while the mean density (dashed vertical line) is biased high. Bottom right: Observed distribution of surface brightness for an XMM-Newton observation of the outskirts of A2142 from \cite{Eckert15}, reprinted with permission, showing a log normal bulk component and a high brightness tail. }
\label{clumping_method}
\end{figure*}

\begin{figure*}
 \hbox{ 
\includegraphics[width=0.5\textwidth]{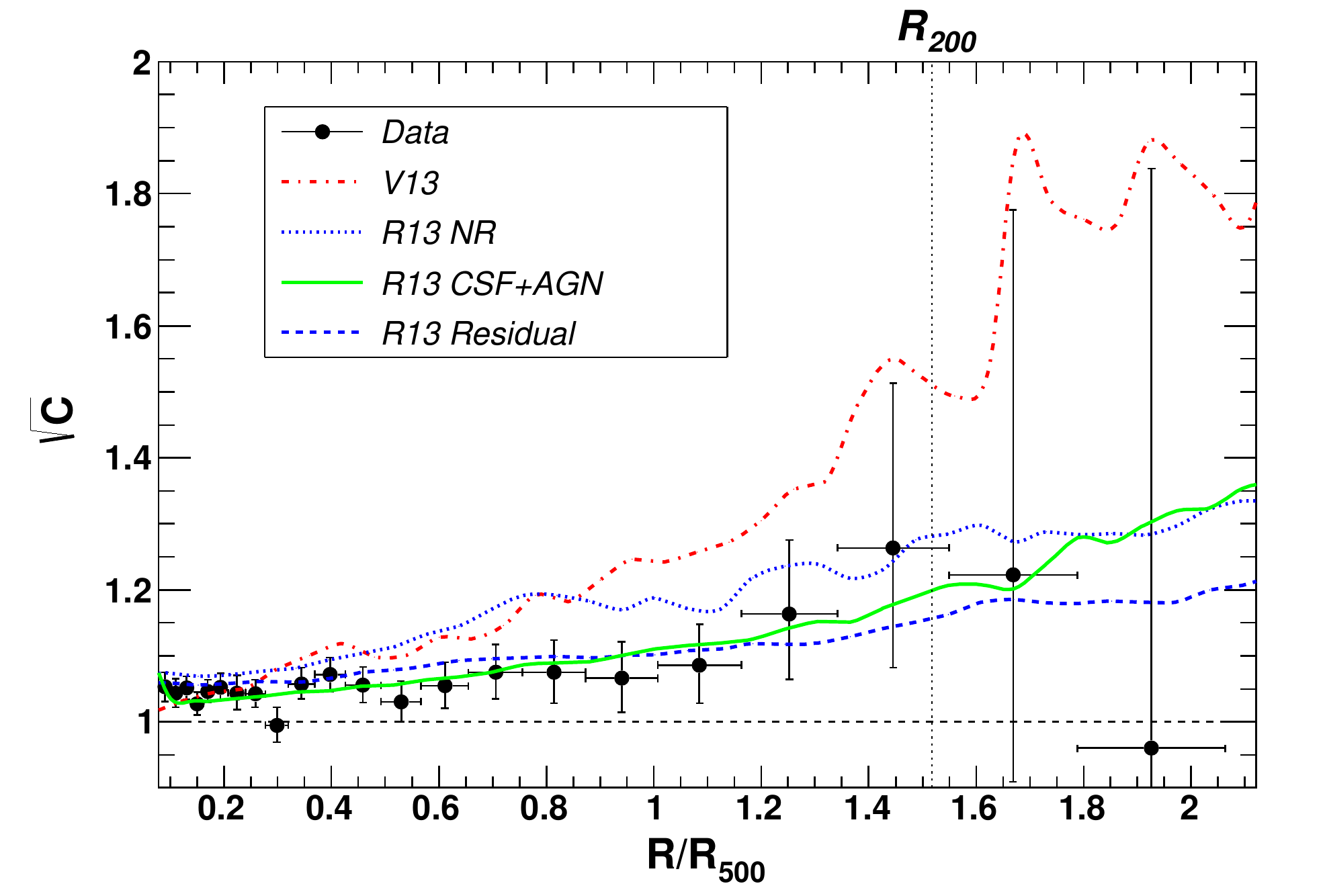}
 \includegraphics[width=0.49\textwidth]{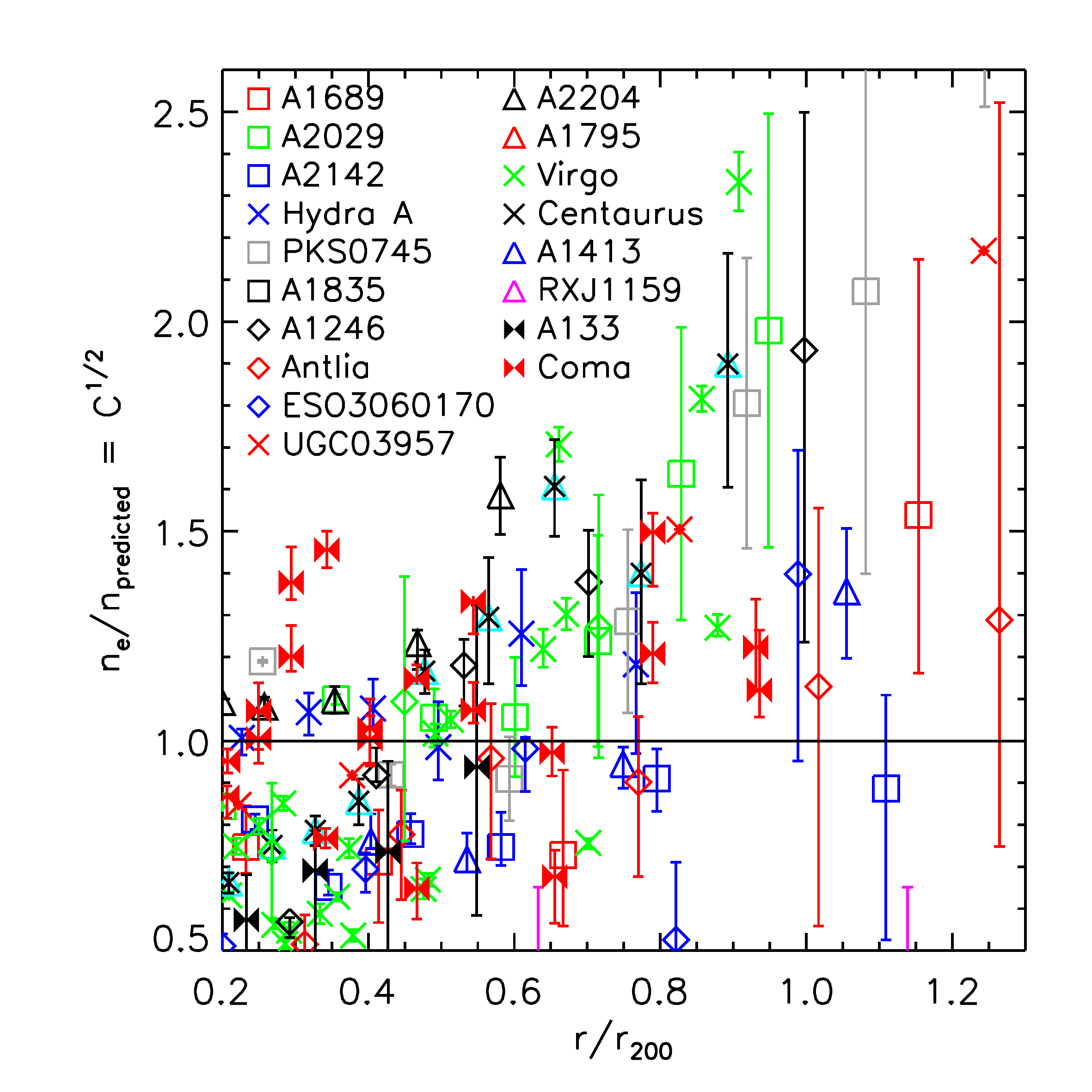}

  }
\caption{Left: Clumping factors obtained in \cite{Eckert15}, reprinted with permission, by comparing the median and mean X-ray surface brightness profiles of ROSAT clusters (black points). The curves show the predictions of several sets of numerical simulations with different hydrodynamic solvers and baryonic physics: non-radiative adaptive- mesh-refinement (AMR) \citep[red,][]{vazza13}, non-radiative smoothed-particle-hydrodynamics (SPH, blue),  SPH including cooling, star formation and AGN feedback (green), and ``residual'' clumping after masking substructures \citep[dashed blue; all from][]{roncarelli13}. Right: Modified figure from \cite{walker13} showing the clumping overestimate $\sqrt{\rm C}$ needed to bring the observed Suzaku density profiles into agreement with the density profile expected from theory if the pressure and entropy agree with their baseline profiles (merging clusters have been excluded from this plot).    }
\label{fig:clumping_profiles}
\end{figure*}

Flatter than expected entropy profiles, and higher than expected gas mass fractions in the outskirts can be explained if the cluster outskirts contain gas clumps, as expected from simulations (see Sec.~\ref{sec:clumping_theory}), which are not resolved out and removed from the observations. As the X-ray emission from bremsstrahlung is proportional to the gas density squared, if there are high density gas clumps that are not removed from the X-ray observations, these will increase the X-ray surface brightness of the ICM in the outskirts. Therefore, if clumping is not taken into account, and the ICM is assumed to be a uniform gas, the increase in X-ray surface brightness would lead us to overestimate the gas density of the bulk ICM. 

Another factor is the role of non-thermal pressure support in the cluster outskirts (see Sec.~\ref{sec:non-thermal_theory}). Through X-ray measurements of gas density and {temperature}, we can only directly measure the thermal pressure of the gas. If there is a significant non-thermal pressure support component in the cluster outskirts, the thermal pressure measurements would then underestimate the total pressure, causing the hydrostatic equilibrium masses calculated from the thermal pressure to underestimate the true mass of the cluster. This would therefore also cause the gas mass fraction measurements to be overestimated. In addition, the increased kinetic energy fraction in the outskirts can suppress the ICM temperature, causing the gas entropy to be lower.

These two different phenomena, gas clumping and non-thermal pressure support, can therefore both lead to similar biases. In order to unravel the contribution from each, we need a method of separating their contributions from one another. Ideally one would simply want to resolve out all of the gas clumps in X-ray images and remove them, however the X-ray surface brightness of individual gas clumps is typically so low that they are unresolved, even with high spatial resolution X-ray telescopes like Chandra. One way forward is to bin the X-ray images into a Voronoi tessellation (see Fig. \ref{clumping_method}, top panels) and examine the distribution of the X-ray surface brightness of the tessellated bins in a particular region of the cluster (such as an annulus), as shown in the bottom right of Fig. \ref{clumping_method} from \cite{Eckert15}. The effect of gas clumping is to cause some of the tessellated bins to have a higher X-ray surface brightness than they should if the ICM were perfectly uniform, leading to a bright tail in the surface brightness distribution, which can be separated from the bulk ICM, whose distribution of surface brightness has been found from simulations to be log-normal (see bottom left of Fig. \ref{clumping_method}, \cite{zhuravleva13}).

\begin{figure}
\begin{center}
    \includegraphics[width=0.5\textwidth]{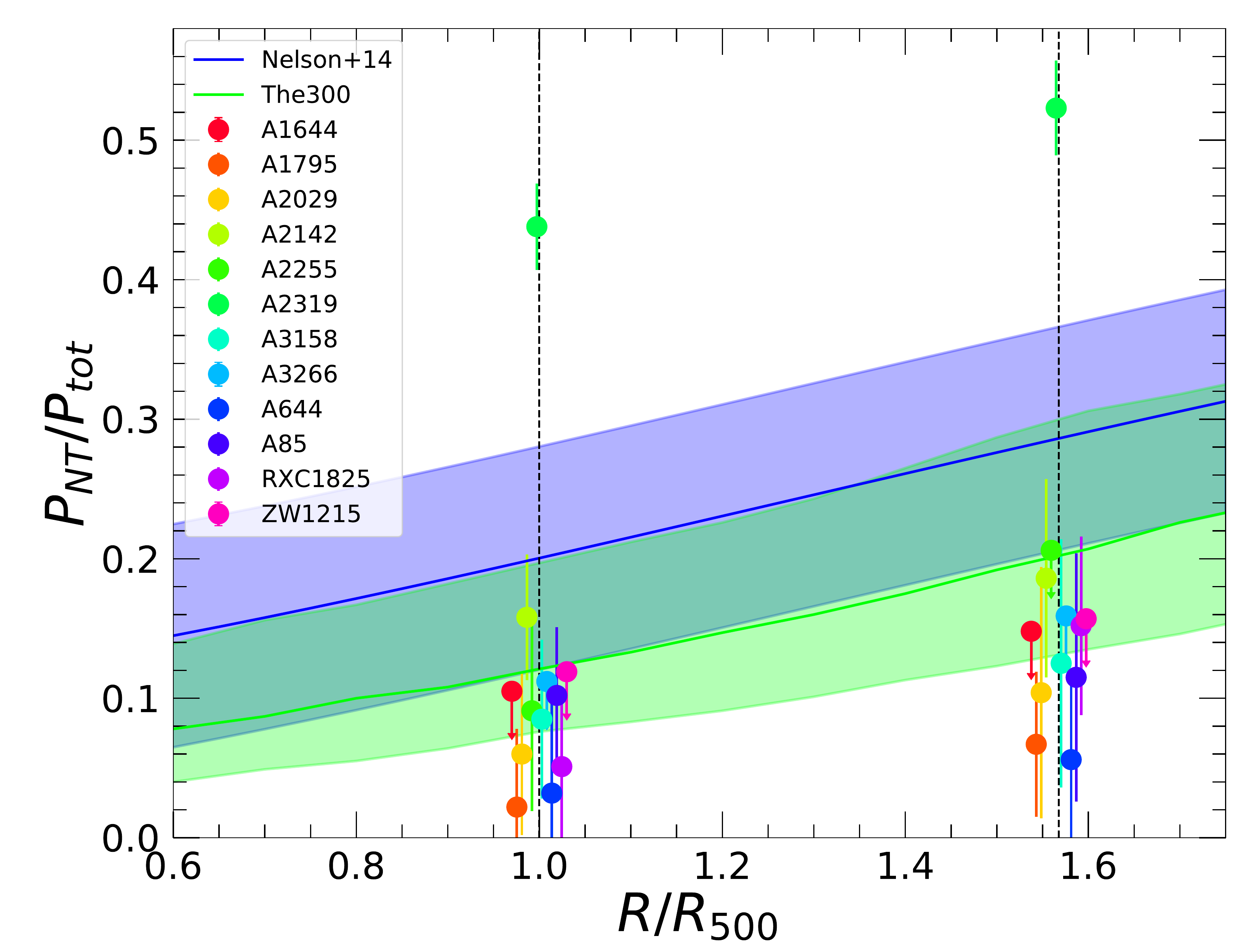}
\end{center}

\caption{Constraints on non-thermal pressure support in cluster outskirts in the X-COP sample \cite[reproduced from][]{Eckert18}, reprinted with permission, showing consistency with theoretical expectations (the shaded regions). }
\label{fig:xcop_pnt}
\end{figure}

The effect of the high density tail from gas clumping can be quantified by comparing the mean and median surface brightness of the tessellated bins contained within an annulus of the cluster. While the median surface brightness will be mostly unaffected by the high density tail, the mean surface brightness will be biased high (see the bottom right hand panel of Fig. \ref{clumping_method}). A clumping correction factor can therefore be measured, and the median surface brightness can be used to obtain a density measurement that is representative of the actual bulk ICM, without being biased high by gas clumping. For the mathematical definition of the clumping factor, $C$, see section \ref{sec:clumping_theory}.

The clumping factor profile obtained in \cite{Eckert15} using this Voronoi tessellation method is shown in the left hand panel of Fig. \ref{fig:clumping_profiles}, which is found to be in agreement with the level expected from simulations. Whilst clumping cannot be directly imaged in the Suzaku data due to its worse spatial resolution, one can estimate the level of clumping using the Suzaku densities, by calculating the level of clumping needed to bring the Suzaku densities into agreement with the expected gas density profile for clusters. The right hand panel of Fig. \ref{fig:clumping_profiles} shows the resulting estimated clumping values, which are consistent with the Voronoi tessellation method and with simulations. {The level of clumping rises outwards with radius, with values in the range $\sqrt{C} =1-2$ at $R_{200c}$.}

Once clumping is corrected for, in some cases an entropy deficit remains \cite{Ghirardini2018,MirakhorVirgo2021}. This remaining entropy deficit is most likely the result of non thermal pressure support. One can therefore estimate the level of non thermal pressure support needed to bring the entropy profile into agreement with the expected profile. Fig. \ref{fig:xcop_pnt} from \cite{Eckert18} shows that for the clusters in the X-COP sample, the estimated non-thermal pressure support is in reasonable agreement with the level expected from simulations. {For the majority of the clusters in the XCOP sample, the non thermal pressure fraction is around 0.1 at $R_{500c}$ and $R_{200c}$, though in Abell 2319 it is significantly higher at around 0.5.}

Another possible bias is the effect of electron-ion nonequilibrium, as suggested by \cite{Hoshino10}, which can also lead to the gas temperatures (and entropies) in the outskirts being underestimated. However with current X-ray telescopes it is not possible to directly measure this effect. For further details on the expected electron-ion nonequilibrium biases on temperature measurements, see section \ref{sec:noneq_theory}.
 
\subsection{Cold fronts in the cluster outskirts}
\label{sec:coldfronts}

Cold fronts around the central cores of relaxed cool core clusters
have been well studied with Chandra (see \cite{markevitch07} for a review), where the high surface brightness allows
them to be easily resolved. The exceptional sharpness of these cold fronts,
smaller than the Coulomb mean free path, indicates that transport processes in
the ICM and hydrodynamic instabilities are suppressed, possibly due to magnetic
draping. Cold fronts are believed to
be formed due to the sloshing of the cold cluster core (similar to the sloshing of wine in a wine glass) as it responds
to the gravitational disturbance created by an infalling subcluster's
dark matter halo during an off-axis minor merger, (e.g.
\cite{Ascasibar2006}). Simulations predict that the geometric
features of older cold fronts should propagate outwards into
the lower pressure regions of the cluster as they age (\cite{Roediger2011} and \cite{ZuHone2011}).

One remarkable recent development in the study of cold fronts is the discovery
of large scale cold fronts reaching out to very large radii, far outside the
cooling radius. In the Perseus cluster, a cold
front was found at the colossal distance of 700 kpc (nearly half the virial radius) from
the cluster core \cite{simionescu12}, using XMM-Newton data, which appears to be a continuation of the sloshing seen in the Perseus's core \cite{Walker2017}. Similar large scale cold fronts reaching out to $\sim$0.5$R_{200c}$ have since
been found in just a handful of other clusters; Abell 2142
\cite{Rossetti13},  RXJ2014.8-2430 \cite{Walker14}, the Virgo cluster \cite{Simionescu17} and Abell 1763 \cite{Douglass2018}. In all of these
systems, there is a large scale spiral pattern of concentric cold fronts on
opposite sides of the cluster at increasing radii,
suggesting that the structure is one continuous outwardly moving sloshing
motion.

\begin{figure}[t]
\begin{center}

\includegraphics[width=0.45\textwidth]{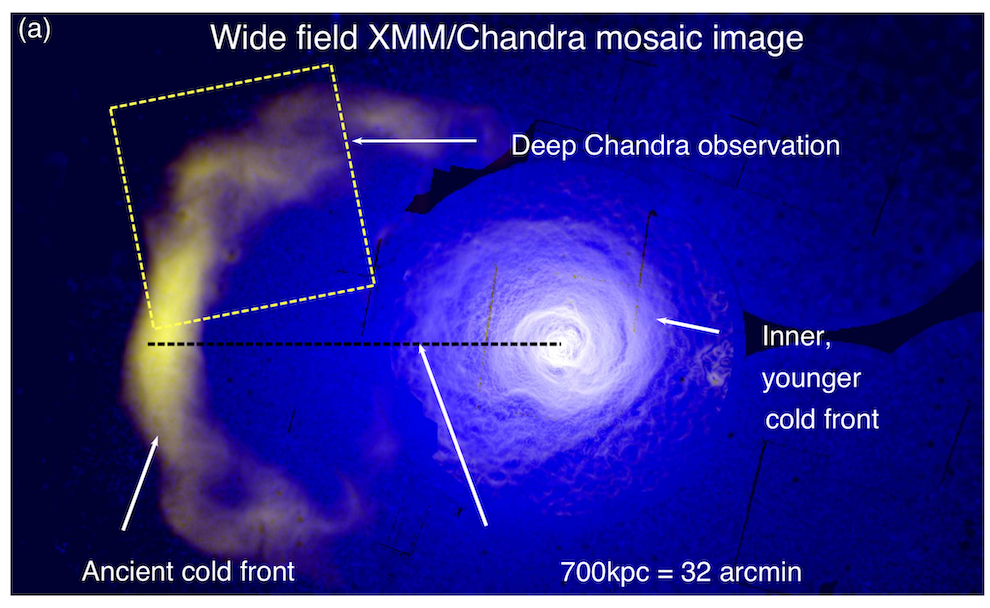} 
\includegraphics[width=0.45\textwidth]{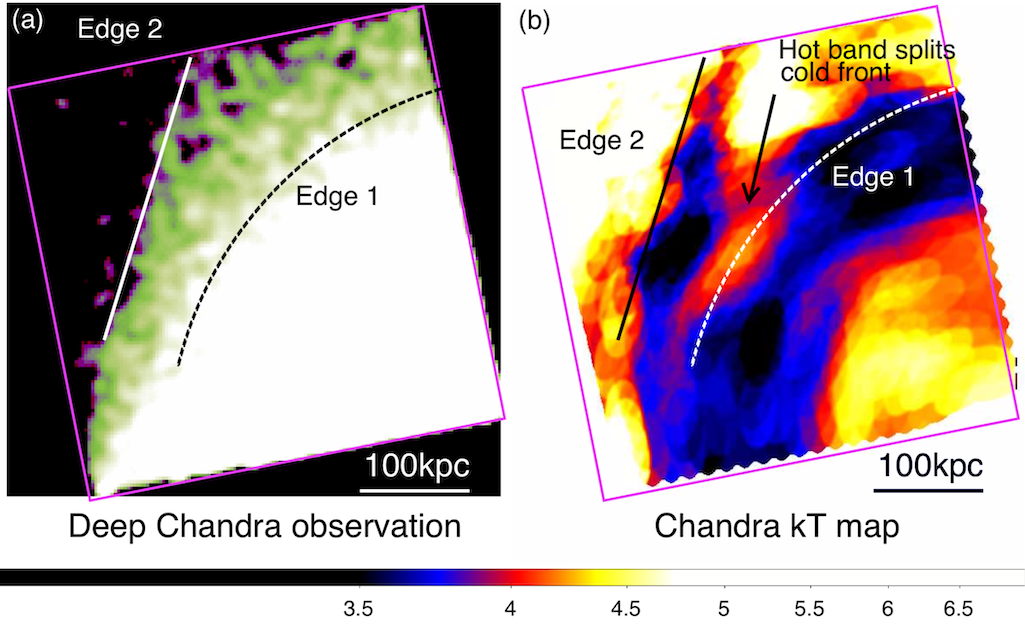}

\end{center}
\caption{Figures from \cite{Walker2018}, reprinted with permission, which studied the eastern, 730~kpc radius cold front in the Perseus cluster. The left hand panel shows the XMM-Newton mosaic of Perseus, enhanced by an edge detection filter. A deep Chandra observation (whose location is shown by yellow dashed box in the left panel) is shown
in the center and right hand panels. The X-ray image (center) shows 2 sharp edges separated by around 100 kpc, while the temperature map (right) shows that the cold front is split, 
separated by a hot band.
  }
  \vspace{-0.2cm}
\label{Perseus}
\end{figure}

These large scale cold fronts are much older than the cold fronts commonly found
in cluster cores, as they have risen outwards and grown with time. They are over
an order of magnitude farther out from the core ({reaching radii of nearly 1 Mpc, compared to the typical radii of around 100 kpc for cold fronts in the cluster core}). Simulations \cite{Roediger2011, zuhone11} find that cold fronts rise outwards with constant velocity, so we would expect these giant cold fronts to be an order of magnitude older than those found in cluster cores. Based on numerical simulations of cold front evolution, \cite{Walker2018}
have calculated the age of the large Perseus cold front to be 5 Gyr, and have also found evidence for the cold front splitting into multiple edges (see Fig. \ref{Perseus}). These results indicate that gas sloshing can continue out into the outskirts of galaxy clusters for vast periods of time, and suggest that magnetic draping can continue to support the cold fronts against instabilities out into the outskirts.

{For further details on cold fronts and sloshing, we direct the reader to the chapter `The Merger Dynamics of the X-ray Emitting Plasma in Clusters of Galaxies'.}

\subsection{Merger shocks in the cluster outskirts}
\label{sec:shocks}

 X-ray telescopes are at present not sensitive enough to observe the accretion shock shell in the outskirts of galaxy clusters (which is expected to be at around 6$R_{500c}$). However, shocks produced by major mergers of galaxy clusters, lying near the virial radius, are sufficiently bright in X-rays to be observed by instruments such as Suzaku. These shocks occur when a cluster moves faster than the speed of sound in the intracluster medium during a collision, and appear in X-ray images as surface brightness edges. The shock acts to heat up the gas and raise its density. These temperature and density jumps in the outskirts have been explored using X-ray spectroscopy. The location of the temperature and density jumps in the ICM have been found to coincide well with radio relic features seen in radio images. The temperature and density jumps found from the X-ray data can be used to measure the Mach number {(the ratio of the speed of the object to the speed of sound of the ICM)} of the shock. {The speed of sound, $c_{\rm s}$ is related to the gas temperature by $c_{\rm s}=1480(T/10^{8} {\rm K})^{1/2}$ km/s, and thus it is typically around 1000 km/s in the outskirts of massive clusters.} While the radio relics in the cluster outskirts are relatively easy to see, it is only recently that X-ray telescopes with high sensitivity to low surface brightness X-ray emission, like Suzaku, have been able to measure the thermal structure of the ICM at the positions of these radio relics (e.g. \cite{Akamatsu2013a, Ogrean2014}). 

\begin{figure*}[t]
 \hbox{ 
  \includegraphics[width=\textwidth]{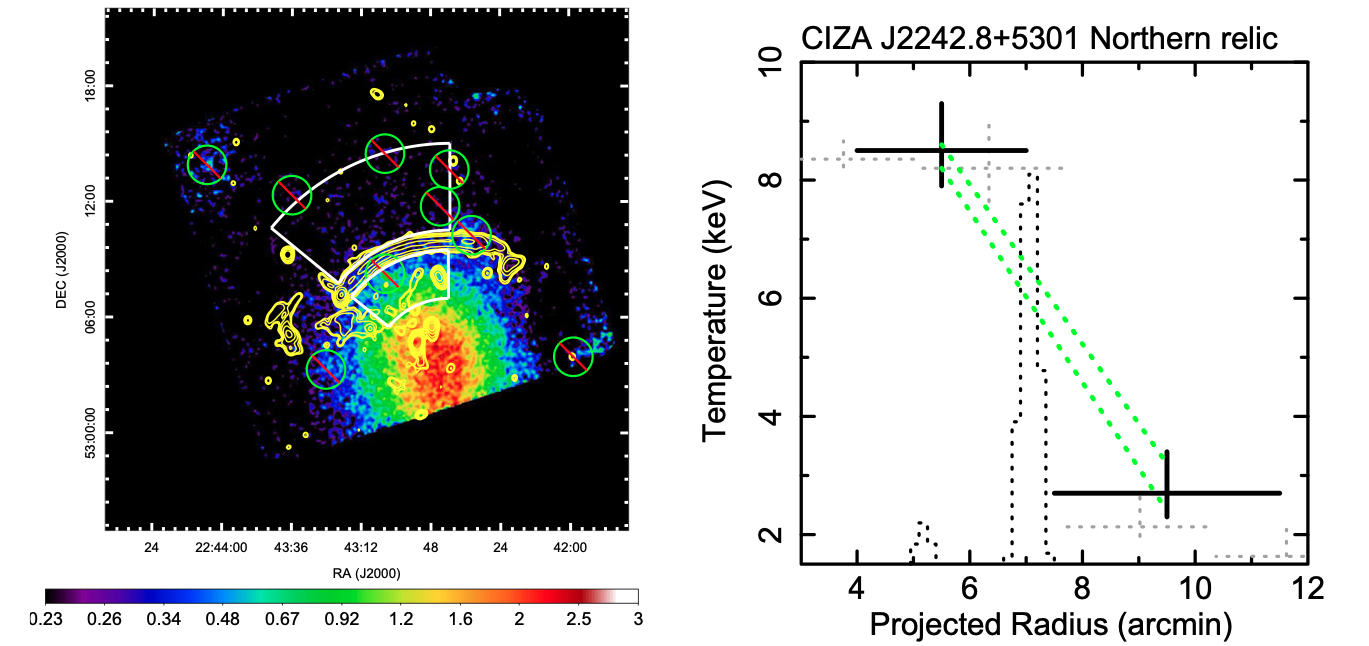}
 }

\caption{Suzaku observations of the radio relic cluster CIZA J2242.8+5301 (a.k.a. the `Sausage' cluster), from \cite{Akamatsu2015}, reprinted with permission. In the left panel, the sausage shaped radio relic at the shock location is shown as the yellow contours, which are plotted on top of the Suzaku X-ray image. The right panel shows the temperature on either side of the radio relic, obtained by X-ray spectral fitting to the Suzaku data. We see that the temperature (solid black crosses) increases by around a factor of 3 across the shock, on either side of the radio relic (dashed black lines).   }
\label{sausage}
\end{figure*}

In Fig. \ref{sausage} we show the Suzaku observation from \cite{Akamatsu2015} of the radio relic `Sausage' cluster (CIZA J2242.8+5301, \cite{vanWeeren2010}), a merging cluster with a sausage shaped radio relic in its northern outskirts (shown by the yellow contours, \cite{vanWeeren2010}). In the right hand panel of Fig. \ref{sausage}, we see that the temperature measured using the X-ray data jumps by around a factor of 3 across the shock location.

One interesting observation is that in a number of cases, the Mach number measured from the X-ray data is significantly lower than the Mach number for the same shock measured using just the radio data alone \citep{vanWeeren2019}. For example in the Sausage cluster, the Mach number obtained from the radio data \citep{vanWeeren2010} is $M=4.6^{+1.3}_{-0.9}$, but when using Suzaku X-ray data, \cite{Akamatsu13} found a much lower Mach number of $M=2.7^{+0.7}_{-0.4}$. 

A number of solutions to this difference in Mach number have been suggested. One possibility is that the X-ray measured Mach numbers are underestimated due to the effects of the viewing angle (i.e. colder X-ray gas may be projected in front of the shock heated region, leading to the temperature increase being underestimated).  This effect may be compounded if the shock surface has a complicated shape. Further possible solutions to this discrepancy are explored in \citep{Akamatsu2017}. Finally, possible biases in the radio measurement may cause the radio derived Mach number to be overestimated, {as discussed in \cite{Stroe2014}}. This joint X-ray and radio approach of viewing shocks provides a powerful probe of the physics underlying the way the ICM is heated during mergers.

{For further details on cluster mergers, we direct the reader to the chapter `The Merger Dynamics of the X-ray Emitting Plasma in Clusters of Galaxies'.}

\subsection{Metals in the cluster outskirts}
\label{sec:metals}

\begin{figure*}[t]
 \hbox{ 
  \includegraphics[width=0.5\textwidth]{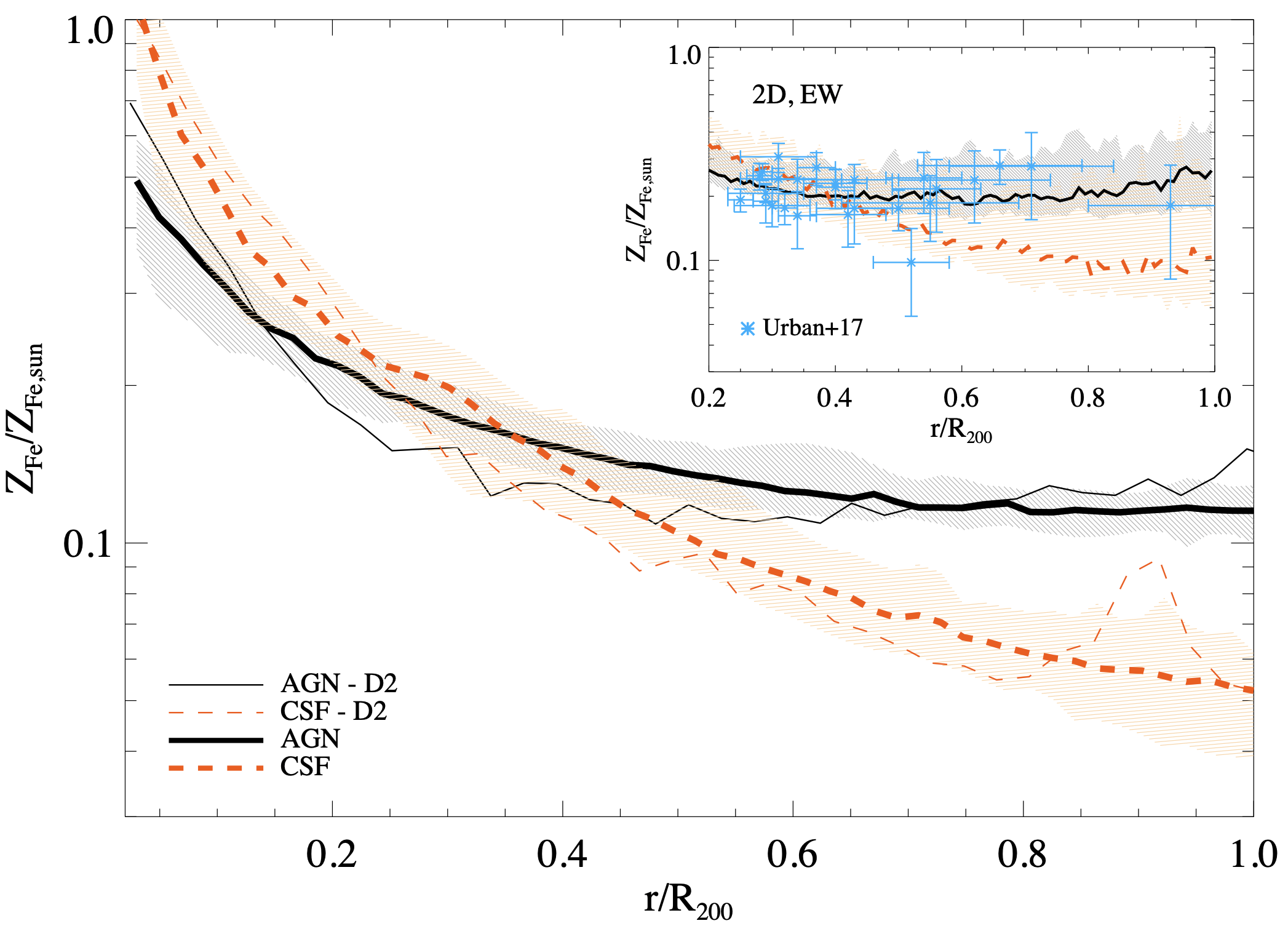}
  \includegraphics[width=0.5\textwidth]{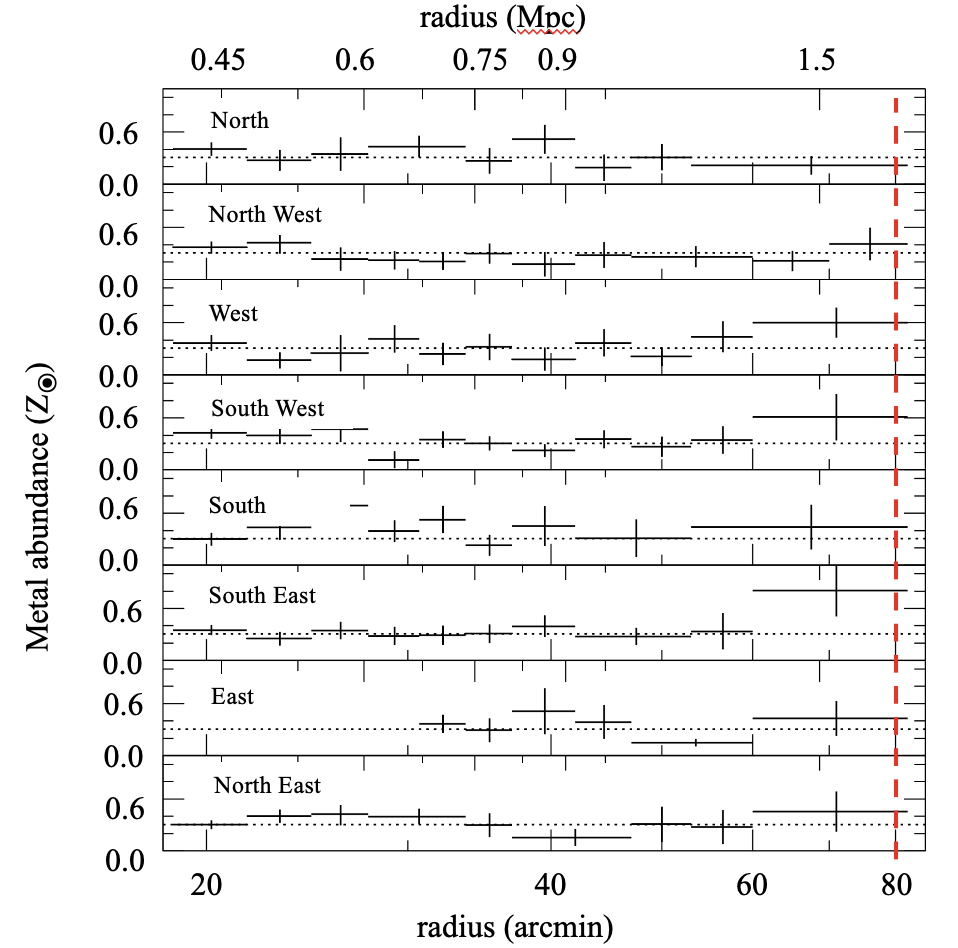}
 }

\caption{Left: Suzaku measurements of the metal abundance in the cluster outskirts (blue) from \cite{Urban2017} compared to numerical simulations from \cite{Biffi2018}. Figure from \cite{Biffi2018}, reprinted with permission. Right: Metal abundance in the outskirts of Perseus along the 8 arms of the Suzaku mosaic adapted from \cite{Werner2013}, reprinted with permission, showing a uniform distribution around an average value of 0.3$Z_{\odot}$. {The location of $R_{200c}$ is shown by the vertical dashed red line. } }
\label{metals}
\end{figure*}

Galaxy clusters, with their deep potential wells, are `closed-box' systems ideal for studying galaxy formation physics. In particular, the metallicity in the outskirts of galaxy clusters is a powerful probe of feedback physics \cite{Werner2013, Biffi2018, Mernier2018}, as the metal distribution in the ICM is strongly dependent on the chemical enrichment histories. {For further details on the metal abundance of galaxy clusters, see the chapter `Chemical enrichment in groups and clusters'.}

Feedback from active galactic nuclei (AGN) at early times ($z \geq 2$) is found to be effective at removing pre-enriched gas from galaxies, spreading metals uniformly throughout the cluster outskirts (see the solid black curve in Fig.\ref{metals}, left). On the other hand, late-time enrichment leads to inhomogeneous distributions of ICM metals that rapidly decline with cluster-centric radius (red curve in Fig.\ref{metals}). The relative composition of various elements of the ICM originating from different types of supernova explosions (core-collapse versus Ia) also place constraints on the star formation histories and the chemical evolution of the universe \citep{Simionescu2015}. 

Measurements of metal abundance in the outskirts are extremely challenging, and are currently limited to within half the virial radius for \textit{Chandra} and \textit{XMM-Newton}. With \textit{Suzaku} only very few clusters have measurements reaching the virial radius (blue crosses in Fig.\ref{metals} from \cite{Urban2017}) which also come with large uncertainties.

In the Perseus cluster, the metal abundance measured by Suzaku in \cite{Werner2013} is found to be very uniform and flat in the outskirts (Fig. \ref{metals}, right panel). The Suzaku metal measurements in the outskirts generally show the metal abundance to level out at around 0.3$Z_{\odot}$ \cite{Simionescu17, Urban2017, MirakhorA2199}, {using the abundance tables from \cite{Asplund2009}}. This is consistent with models of early metal enrichment (black solid line in Fig. \ref{metals}, left panel), in which the most of the intragalactic medium is enriched by metals before galaxy clusters formed, likely more than 10 billion years ago, during the epoch of utmost star formation and black hole activity.

\subsection{Connections to the cosmic web}
\label{sec:cosmicweb}

\begin{figure*}[t]
 \hbox{ 
     \includegraphics[width=0.55\textwidth]{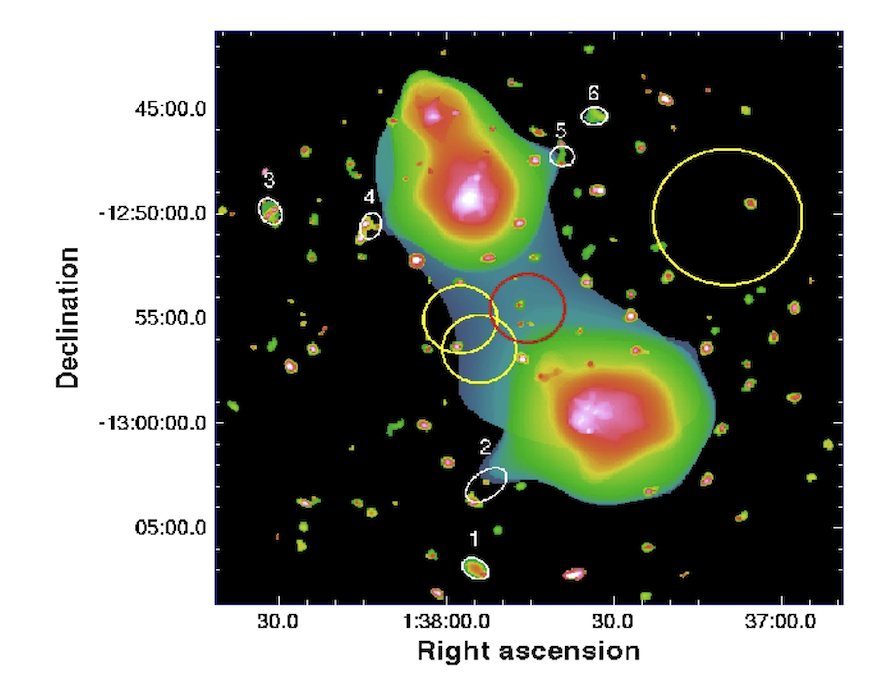}
  \includegraphics[width=0.4\textwidth]{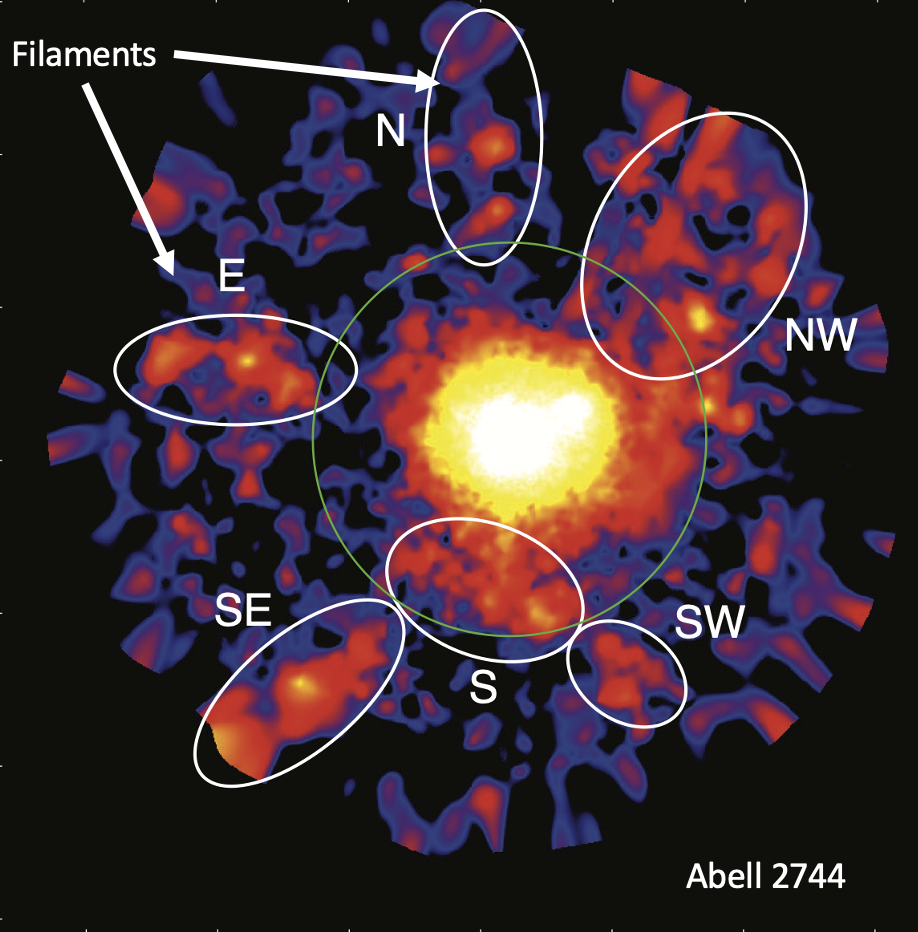}
 }

\caption{Left: Soft band XMM-Newton image of the filament connecting the galaxy cluster system Abell 222/223, from \cite{Werner2008detection}, reprinted with permission. Right: Soft band XMM-Newton image  of the galaxy cluster Abell 2744 and its surrounding cosmic web filaments adapted from \cite{Eckert15_Nature}, reprinted with permission.  }
\label{WHIM}
\end{figure*}

 Absorption studies are the focus of a companion work in this series (see `Absorption studies of the most diffuse gas in the LSS'). Here we will discuss attempts to observe the cosmic web filaments {in the vicinity of massive clusters of galaxies directly from their X-ray emission.}

Based on measurements of the cosmic microwave background, we expect around 5 percent of total energy content of the Universe to be in baryons \citep{ade2016planck}. However, observing all of these baryons has proved difficult, leading to only around half of the expected value being accounted for \citep[e.g.][]{Fukugita1998,Bregman2007,Sinha2010}. Numerical simulations of structure formation \citep[e.g.][]{Cen1999where,Dave2001,Dolag2006Simulating,Cui2019the} predict that these `missing baryons' should reside in the cosmic web filaments which connect galaxy clusters. The gas in these filaments is known as the warm-hot intergalactic medium (WHIM), within which simulations predict the gas to have a very low density (around 5--200 times the mean baryonic density of the Universe) and a warm-hot temperature range of $10^5 < T < 10^7$ K \citep[e.g.][]{Bregman2007}. 

This low density and low temperature make it extremely challenging to observe these WHIM filaments from their X-ray emission. One approach, which attempts to get around this limitation, is to search for the WHIM absorption lines along the line of sight to distant active galactic nuclei and quasars using high-resolution spectroscopy \citep[e.g.][]{Nicastro2005mass,Zappacosta2010studying,Nicastro2018observations}. 

A second approach to detect the WHIM directly through its X-ray emission is to observe large-scale filaments between a close pair of galaxy clusters, which are aligned such that the line of sight looks along the length of the filament (see Fig. \ref{WHIM}, left hand panel), maximising the volume of gas we are looking through and thus increasing the level of visible X-ray emission \citep[e.g.][]{Werner2008detection,Alvarez2018chandra}. Such filaments provide the best opportunity to trace the gas near the outskirts of cluster pairs, and to detect the densest and hottest parts of the WHIM. 

Studying the filaments through their emission has the advantage that it allows us to spatially map all of their structure, whereas absorption studies rely on bright background objects, and are limited to a small number of sight lines through the filament. Detection of filaments through their X-ray emission remain rare and limited to a handful of cases, e.g., Abell 222/223 \citep{Werner2008detection}, Abell 399/401 \citep{Sakelliou2004xmm}, Abell 2804/2811 \citep{Sato2010study}, Abell 3556/3558 \citep{Mitsuishi2012search}, Abell 2029/2033 \citep{MirakhorA2029}, Abell 133 \cite{Connor2018} and Abell 3391/3395 \citep{Alvarez2018chandra,Reiprich2021}.

Filaments have also been seen directly in emission from the galaxy cluster Abell 2744 \cite{Eckert15_Nature}, which is an extremely massive ($1.5\times10^{15} M_{\odot}$) cluster currently undergoing a merger consisting of at least four components (see the right hand panel of Fig. \ref{WHIM}). These filaments coincide with the overdensity of galaxies and dark matter surrounding the cluster, and have been measured to have temperatures around $10^{7}$K.

\section{Theory and Simulations}
\label{sec:theory}

Here we describe our current state of the art theoretical understanding of the cluster outskirts. This section covers the self-similar growth of clusters (section~\ref{sec:self-similar}), the predicted ICM  profiles in cluster outskirts (section~\ref{sec:icm_outskirts}), non thermal gas motions (section~\ref{sec:non-thermal_theory}), gas clumping (section~\ref{sec:clumping_theory}), shocks and electron-ion non-equilibrium (section~\ref{sec:noneq_theory}), and a discussion on future simulations (section~\ref{sec:future_theory}).





\subsection{Self-similarity in Cluster Outskirts}\label{sec:self-similar}

{ Under the prevailing \textLambda CDM picture of structure formation, dark matter halos form and grow from the primordial, nearly scale-free matter density field of the early universe through gravitational collapse. Smaller systems such as galaxies, and galaxy groups form first and subsequently merge to form galaxy clusters. Since there is no preferred physical scale in gravitational collapse, the growth of halos is scale-free, or ``self-similar''. }  As a consequence, galaxy clusters are thought to be ``self-similar'' \citep{kaiser86}. This means that, with some appropriate scaling, galaxy clusters of different masses and sizes display remarkable similarity with one another. Self-similarity of galaxy clusters has been particularly important and useful because it is one of the main assumption for studies on mass-observable scaling relations, which are the keys to infer galaxy cluster masses in cluster cosmology. { Note that the self-similarity will be broken if any other physical processes become relevant during the growth of galaxy clusters. }
Specifically, non-gravitational processes, such as radiative cooling, star formation, and feedback from active galactic nuclei, dominate the cluster cores, usually within $0.2 R_{500c}$. { Indeed, cluster cores show the largest cluster-to-cluster variance in terms of density and temperature profiles.} 
This is also why cluster cores within $0.2 R_{500c}$ are often excised to reduce scatter in scaling relations between cluster masses and the measured X-ray luminosities and temperatures. In the radial range of $0.2 \leq r/R_{500c} \leq 1$, the thermodynamical profiles usually display remarkable self-similarity, once their radii are scaled by $R_{\Delta c}$ (with $\Delta_c = 500$ or $200$, etc.)  and their profiles are renormalized with factors that depend only on their mass $M_{\Delta c}$ and redshift \citep{voit05b}. 

\subsection{Thermodynamical Profiles of ICM in Cluster Outskirts}\label{sec:icm_outskirts}

\begin{figure}
    \centering
    \includegraphics[scale=0.4]{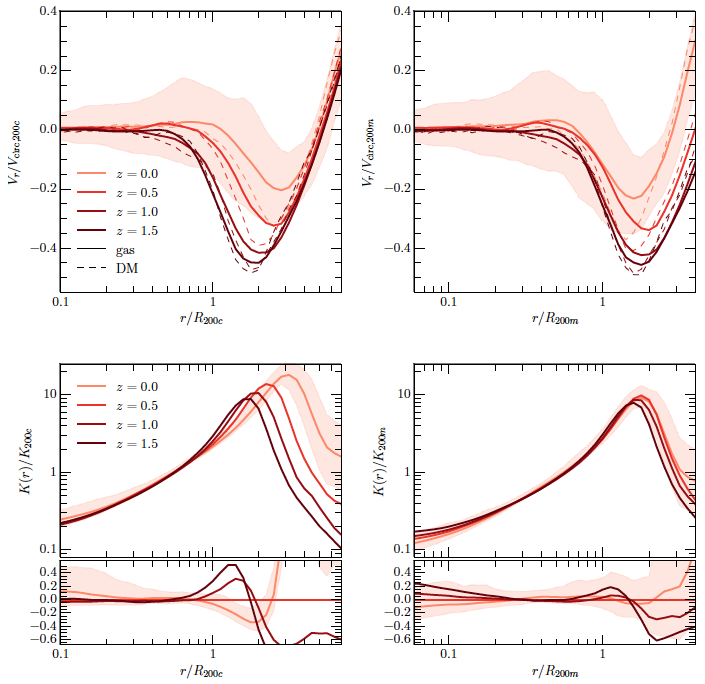}
    \caption{Profiles of radial velocities
(top panels) and gas entropy (bottom panels) of halos at different redshift from the \emph{Omega500} cosmological simulations.  In the radial velocity profiles, the dashed lines indicate the profiles for dark matter. The left and right panels show the profiles scaled with $R_{200c}$ and $R_{200m}$, respectively. The lines in each panel indicate the mean profiles. { We only show the $1\sigma$ scatter for the $z=0$ profiles as shaded regions. The profiles at other redshifts have similar amount of scatter. }  Figure taken from \cite{lau15}, reprinted with permission. }
    \label{fig:entropy_profile_sims}
\end{figure}

{ Under the self-similar, hierarchical picture of cluster growth, the inner regions of the clusters are formed at earlier times, while those in the outskirts are added to the clusters at later times}. Thus, the growth history of the cluster leaves imprints on the shapes of the thermodynamical profiles of the ICM. In particular, the entropy of the ICM, $K$, is the thermodynamical quantity that is most sensitive to the growth history of the cluster.  As accreting gas falls into the gravitational potential well of the cluster, it is shocked as its infalling speed exceeds the local sound speed.  As such, its entropy is increased due to the irreversible shock-heating process. As the cluster grows, its potential also becomes deeper. { Consequently the infalling gas accretes at a higher speed, and thus experiences stronger shocks and larger entropy increases. This results in a radially increasing entropy profile as $K \propto r^{1.1}$: the earlier accreted gas with lower entropy resides in the inner cluster regions, while the later accreted gas with high entropy piles up in the outer regions \citep{voit05}.} In the bottom panels of Fig.~\ref{fig:entropy_profile_sims}, we show the increasing entropy profiles from cosmological simulation \citep{lau15}. The locations of the peaks of the entropy profiles, correspond directly to the most negative values of the gas radial velocities. 
This accretion shock of infalling gas also serves as a natural boundary of the ICM halo \citep[e.g.][]{molnar09, aung20} (see Section~\ref{sec:definition} and Fig. \ref{fig:shock_splashback_map}). 

In the cluster outskirts ($R>R_{200c}$), self-similarity of the ICM profiles behaves differently from the inner regions. The left column in Fig.~\ref{fig:entropy_profile_sims} shows that when the radial velocity and entropy profiles of the gas are normalized by quantities related to $R_{200c}$,  they no longer behave self-similarly, as shown by the large, systematic differences of their profiles at different redshifts. Instead, they become self-similar when they are scaled with respect to $R_{200m}$, as shown in the right columns of the same Figure. An explanation to this is that the thermodynamics of the ICM in the outskirts is predominantly governed by the accretion process, which depends on the mean mass density of the Universe rather than the critical density. This mean-density self-similar scaling also applies to any other accretion-driven phenomena in the cluster outskirts such as non-thermal pressure fraction, and election-ion non-equilibrium, as we discuss below. 

\subsection{Non-thermal Gas Motions in Cluster Outskirts}\label{sec:non-thermal_theory}

The hot gas in galaxy clusters is not in a steady state. Ongoing mergers and accretion processes during cluster growth create considerable amount of gas motions in the ICM, providing extra ``non-thermal'' pressure support for the gas. The term ``non-thermal'' is used here to distinguish from the thermal, microscopic gas motions with a Maxwellian velocity distribution, as one would expect for gas in thermodynamical equilibrium. 
These gas motions arise from the incomplete thermalization of the ICM after mergers or infall as the gas is shocked \citep[e.g.][]{vazza09}. While shocks are efficient engines in converting the kinetic energy of the infalling gas to thermal energy, such conversion is always incomplete. Even in the strong shock limit, only about half of the kinetic energy is converted to thermal energy. The remaining kinetic energy of shocked, subsonic gas will eventually turn into heat via turbulent dissipation, with timescales (i.e. eddy turnover time) on the order of the dynamical time { $t_{\rm dyn}$ (typical time for particle to complete its orbit) of the cluster \citep{shi14}. Here the dynamical time at a radius $r$ is defined as $t_{\rm dyn} \equiv r/v_{\rm circ}$, and $v_{\rm circ} \equiv \sqrt{GM(<r)/r}$ is the circular velocity}. The dissipation time is usually much longer than the { time needed for any gas perturbations to travel across the cluster, i.e., the sound-crossing time, defined as $t_{\rm cross} \equiv r/c_{\rm s}$ where $c_{\rm s}$ is the sound speed of the gas. Therefore, for $r > R_{500c}$,  non-thermal turbulent motions are dynamically important until they dissipate in a few Gyrs, as shown in cosmological simulations \citep[e.g.][]{nelson12}. }

\begin{figure*}[t]
    \centering
    \includegraphics[scale=0.24]{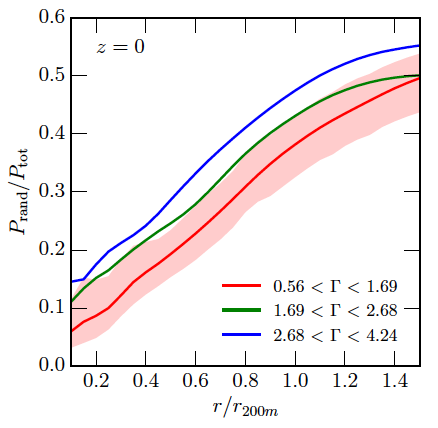}
    \includegraphics[scale=0.24]{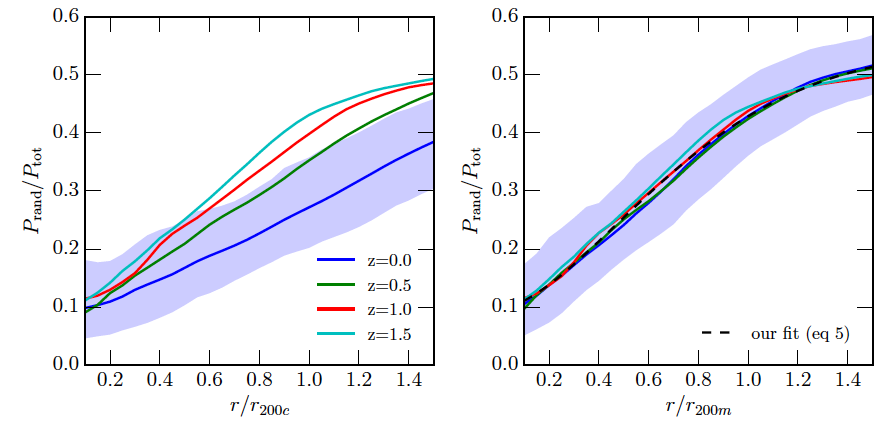}
    \caption{Profiles of the non-thermal pressure fraction for clusters in cosmological simulations. Leftmost panel shows the profiles for $z=0$ halos in different bins of halo mass accretion rate $\Gamma$. Higher $\Gamma$ means higher rate of mass accretion. The middle and right panels show the profiles at different redshifts, with radii scaled with respect to $R_{200c}$ and $R_{200m}$ respectively.  { In each panel, the shaded region represents the $1\sigma$ scatter. } Figure taken from \cite{nelson14b}, reprinted with permission. }
    \label{fig:nelson14}
\end{figure*}

In the cluster outskirts, with longer dynamical time (thus longer turbulent dissipation time) and where accretion is happening, the relative contribution of non-thermal gas motions to that of thermal motion is larger compared to the inner regions of the cluster. We can quantify the contribution of non-thermal gas motions with the non-thermal pressure fraction, defined as
\begin{equation}\label{eq:fnt}
    f_{nt} \equiv \frac{P_{nt}}{P_{nt}+P_{th}} = \frac{\sigma_{gas}^2}{\sigma_{gas}^2 + 3kT_{gas}/(\mu m_p)},
\end{equation}
where $\sigma_{gas} = \sqrt{\sigma_x^2 + \sigma_y^2+\sigma_z^2}$ is the 3-dimensional gas velocity dispersion, and $T_{gas}$ is the gas temperature. { Note that in the second equality in Equation~\ref{eq:fnt}, we assume that other potential sources of non-thermal pressure, such as the magnetic field and cosmic rays, are negligible compared to non-thermal gas motions \citep{vazza09}. } The left panel in Fig.~\ref{fig:nelson14} shows the non-thermal pressure fraction profile predicted from cosmological simulations \citep{nelson14}. It shows that non-thermal pressure fraction profile increases with radius. The non-thermal pressure fraction also depends on the mass accretion rate of the halo. In simulations we can measure the halo mass accretion rate (Equation~\ref{eq:mar}). Higher mass accretion rate leads to higher non-thermal pressure fraction.  

The non-thermal pressure fraction is also redshift dependent, since we expect clusters to be accreting faster at higher redshift when they are embedded in a denser universe which enhances the strength of the gravitational forces, and hence have more non-thermal gas motions. The middle panel shows the redshift-dependence of the non-thermal pressure fraction profile, with radius scaled with respect to $R_{200c}$. 
If we normalize the { non-thermal pressure fraction profiles} with respect with $R_{200m}$, they become self-similar (c.f. Sec.~\ref{sec:self-similar}), as shown in the rightmost panel in Fig.~\ref{fig:nelson14}.   

There is an important implication of the presence of non-thermal pressure in the ICM. Cluster masses measured in X-ray or microwave, are often estimated under the assumption of hydrostatic equilibrium. The total mass of the cluster can be expressed in terms of radial profiles of thermal and non-thermal pressure, as well as gas density \citep{rasia04,lau09}
\begin{equation}\label{eq:hydrostatic_mass}
M (<r) = \frac{-r^2}{G\rho_{\rm gas}(r)}\left(\frac{dP_{th}}{dr} + \frac{dP_{nt}}{dr} \right).
\end{equation}
Under hydrostatic equilibrium, only the thermal pressure gradient $dP_{th}/{dr}$ contributes to the support against the gravity of the cluster,  and the additional support coming from the non-thermal pressure is neglected. Cluster mass computed under this assumption will be underestimated compared to the true mass \citep{rasia06,Nagai07}. 
The difference between the hydrostatic mass and the true mass of the cluster is known as hydrostatic mass bias. Such bias is one of the main systematic uncertainties on cluster cosmology and may explain some of the tension in the cosmological constraints derived from cluster abundances and the cosmic microwave background \citep{Planck15CluterCosmology}. { We refer the reader to the chapter of `X-ray Cluster Cosmology` for more details about the cosmological implications of hydrostatic mass bias. }
Non-thermal pressure support from merger- and accretion-induced gas motions are predicted to be one of the main drivers of the bias. As non-thermal pressure fraction increases with radius, we also expect the hydrostatic mass bias to be increasing as well \citep[e.g.][]{lau09, biffi2016}.

Note that other terms contribute to the hydrostatic mass bias in addition to the gas motions. Under the general non-steady state condition, acceleration of the infalling gas also contributes to the support against gravity \citep{suto13}. This gas acceleration contribution can also become important in cluster outskirts. { Other non-thermal components, such as magnetic field and cosmic rays, may also contribute to the total non-thermal pressure.  }

\subsection{Gas Density Inhomogeneities, or Gas Clumping in Cluster Outskirts}\label{sec:clumping_theory}

The gas density distribution in the ICM is expected to be inhomogeneous. This naturally arises from the hierarchical growth of galaxy clusters from accretion and mergers of smaller clusters, groups and galaxies. The gas residing in the potential wells of these infalling halos is often denser than the surrounding diffuse ICM. Eventually, this gas will be stripped off from these infalling objects through ram-pressure stripping by the ICM. { The ram pressure $P_{\rm ram}$ is roughly equal to the product of ambient ICM gas density $\rho_{\rm gas}$ and the square of the relative velocity of the infalling object with respect to the ambient ICM $v_{\rm rel}^2$,
\begin{equation}
    P_{\rm ram} \approx \rho_{\rm gas} v_{\rm rel}^2,
\end{equation}
up to a factor of order unity due to the geometrical shape of the infalling object. }
{ We refer the reader to the chapter on ``The Merger Dynamics of the X-ray Emitting Plasma in Clusters of Galaxies'' for more details of ram pressure stripping during mergers. }
As the infalling halos fall deeper in the cluster potential well, their gas experiences stronger ram pressure stripping, due to both the increase in the ICM density towards the cluster center, as well as the increase in the speed of the infalling object as it approaches pericenter. This results in a radial dependence in the gas inhomogeneity: towards the halo center, the gas is more homogeneous because the gas of most of the infalled galaxies has already been stripped and mixed with the ICM. 

Gas density inhomogeneities are often quantified in terms of the dimensionless gas clumping factor $C$, which is defined as
\begin{equation}
    C \equiv \frac{\langle \rho_{\rm gas} ^2 \rangle}{\langle \rho_{\rm gas} \rangle^2} \geq 1,
\end{equation}
where $\langle \rho _{\rm gas}^2 \rangle$ is the average of square gas density within a given volume (e.g. a thin radial shell), and $\langle \rho_{\rm gas} \rangle^2$ is the square of the average of gas density. Note that mathematically, $C$ is always greater than or equal to one. In the case where $C = 1$, the gas is perfectly homogeneous in density. 

The level of gas clumping has direct implications on inferring the thermodynamical properties of the ICM from X-ray observations { (see section~\ref{sec:observed_biases})}. 
The X-ray surface brightness of the hot diffuse ICM is proportional to the square of gas density. 
\begin{equation}\label{eq:xsb}
    S_X \propto \int \rho_{\rm gas}^2 \Lambda (T,Z) dl
\end{equation}
where $\rho_{\rm gas}$ is the gas density, and $\Lambda(T,Z)$ is the X-ray cooling function that depends on gas temperature $T$ and metallicity $Z$, and $l$ is the coordinate along the line-of-sight.  As observations always measure the average X-ray surface brightness over an area of the sky $\langle S_X \rangle \propto \langle \rho_{\rm gas}^2 \rangle$, the inferred gas density from the averaged X-ray surface brightness $\langle S_X \rangle$ will always be biased high, specifically by the square root of the clumping factor:
\begin{equation}\label{eq:clumping}
    \hat{\rho}_{\rm gas} = C^{1/2} \langle {\rho}_{\rm gas} \rangle,
\end{equation}
where the hat symbol denotes an estimated quantity from observations. Thus, other X-ray estimated thermodynamic quantities, specifically thermal pressure $P_{th} \propto \rho_{\rm gas} $ and entropy $K \propto \rho_{\rm gas}^{-2/3}$, will be biased by clumping: $
    \hat{K} = C^{-1/3} \langle K \rangle,
    \hat{P}_{th} = C^{1/2} \langle P_{th} \rangle.$ 
Thus, whenever the ICM has a significant clumping factor, { the X-ray measurement of entropy and thermal pressure derived from the gas density will be biased low and high, respectively}. The clumpy ICM will also bias high the X-ray derived gas mass \cite{mathiesen99, vazza13}, since $\hat{M}_{\rm gas} = \int C(r)^{1/2}\langle\rho_{\rm gas}(r)\rangle dV$, and bias low the hydrostatic mass (see Eq.~\ref{eq:hydrostatic_mass}), which is inversely proportional to gas density \cite{roncarelli13}. Finally, another effect of clumping is that it will bias high the gas mass fraction $f_{\rm gas} = M_{\rm gas}/M_{\rm total}$, if either gas mass or total mass, or both, are estimated from X-ray observations.  

\begin{figure*}[t]
    \centering
    \includegraphics[scale=0.20]{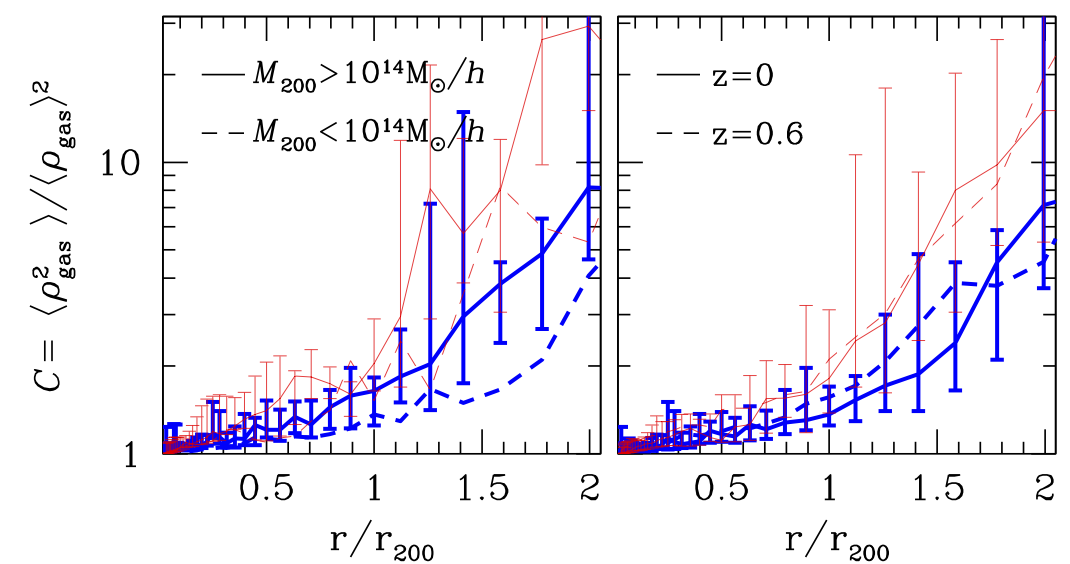}
    \includegraphics[scale=0.17]{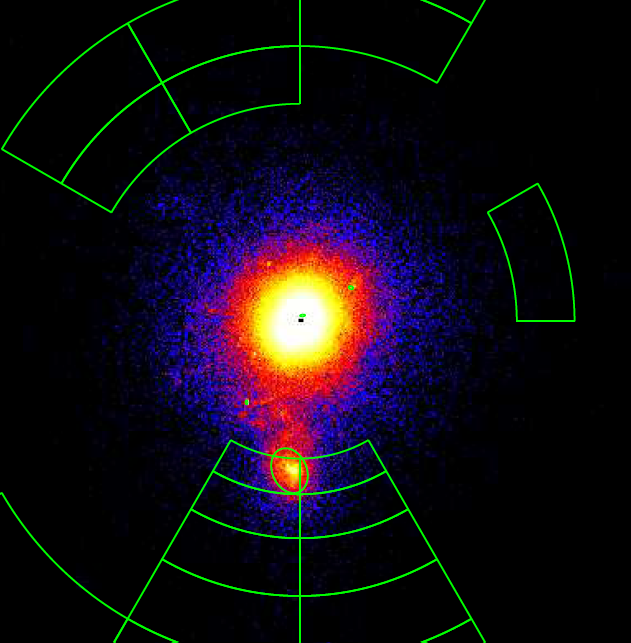}
    \caption{Profiles of the gas density clumping factor for the hot ICM gas ($T > 10^{6}$~K) of simulated clusters. The blue and red lines indicate simulations performed with and without dissipative baryonic physics (cooling, star formation, feedback), respectively. Solid and dashed lines indicate the median profiles of two sub-samples divided by mass (left panel) and by redshift (middle panel). Associated to the solid line, the interquartile range is shown. Both panels are taken from \cite{nagai11}, reprinted with permission. The right panel shows a mock X-ray map from one of the simulated clusters, with the infalling halos circled and filamentary structures marked in annular regions. Figure taken from \cite{avestruz14}, reprinted with permission.  }
    \label{fig:clumping}
\end{figure*}

Cosmological simulations show that { the clumping factor of the ICM outside $R_{200c}$ can be higher than 1.5 and even reaches extreme values ($C>10$) at $2R_{200c}$. On the other hand, gas within $R_{200c}$ typically has $C \sim 1$.} Consequently, the biases in the X-ray derived gas thermodynamics, gas mass, and hydrostatic mass, due to gas density clumping, are only important in cluster outskirts. The left two panels in Fig.~\ref{fig:clumping} show the gas clumping factor from a sample of cluster taken from cosmological simulation \citep{nagai11}, { showing its trend as a function of radius}. The large clumping factors in the outskirts are due to X-ray emitting gas in infalling subhalos as well as the hot diffuse gas in the filaments which these infalling objects travel along. The right panel in Fig.~\ref{fig:clumping} shows mock X-ray maps from cosmological simulated clusters, where infalling gas clumps and filamentary structures are identified. 

{The predictions on the level of clumping factor from cosmological simulations are in general consistent with those estimated from observations in Section~\ref{sec:observed_biases}. }

\subsection{Shocks and Electron-Ion Non-Equilibration}
\label{sec:noneq_theory}

Accretion and merger shocks during cluster growth also lead to biases in our estimation of the thermal energy of the gas in cluster outskirts. In cluster outskirts,  the typical gas density is low with $n_{\rm gas} \sim 10^{-5}\,{\rm cm^{-3}}$. The mean free path between electrons and ions can be quite large ($\sim 100 $~kpc), meaning that these shocks are collisionless. Specifically, these collisionless shocks can create differences between the temperatures of electrons, and the temperature of ions. As the bulk of the kinetic energy of the pre-shock gas lies in the more massive ions than the less massive electrons, the ions possess most of the thermal energy after { the ions and electrons} are shock heated. The temperature of the electrons will increase subsequently as they undergo Coulomb collisions with the hotter ions, until they eventually reach thermal equilibrium with the ions. This equilibration process can be long in cluster outskirts. The equilibration timescale $t_{ei}$ for a fully ionized plasma with electrons, protons and Helium nuclei is given by \cite{spitzer62}
\begin{equation}\label{eq:tei}
    t_{ei} \approx 0.65\,{\rm Gyr}\left(\frac{T_e}{10^{7}{\rm K}}\right)^{3/2}\left(\frac{n_i}{10^{-5}{\rm cm^{-3}}}\right)^{-1}.
\end{equation}
In the cluster outskirts, the ion number density can reach below $n_i \sim 10^{-5}\,{\rm cm^{-3}}$, while the electron temperature can still be on the order of the virial temperature with $T_e \sim 10^7$~K. This means that the equilibration time can be as long as $t_{ei} \sim 1$~Gyr, which can be longer than { the dynamical time in cluster outskirts}. Unless there are other mechanisms to equilibrate the ion and electron temperatures, such as interactions via the magnetic field, the electrons will remain colder than the ions. Since the bulk of the X-ray emissions in the hot diffuse ICM comes from free-free emissions of electrons (thermal bremsstrahlung) which depends on electron temperature, the gas temperature (i.e. average temperature for both ions and electrons) estimated from X-ray observations can be underestimated due to electron-ion non-equilibrium in the cluster outskirts. 

Analytic models \cite{fox97,wong2009} and cosmological simulations \cite{rudd09, avestruz15} can be used to predict the deviation of electron temperatures from the ion temperature, by adopting a two-fluid approach where electrons and ions have their own separate temperatures, with the equilibration process explicitly computed. 

\begin{figure*}[t]
    \centering
    \includegraphics[scale=0.16]{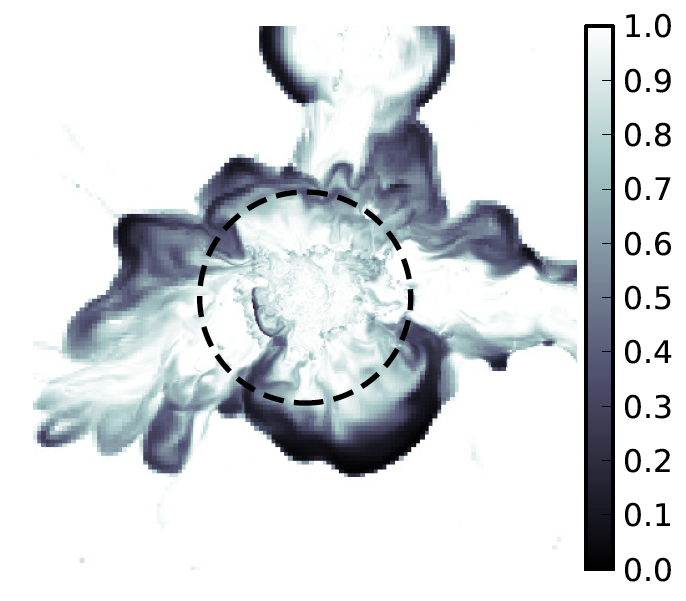}
    \includegraphics[scale=0.18]{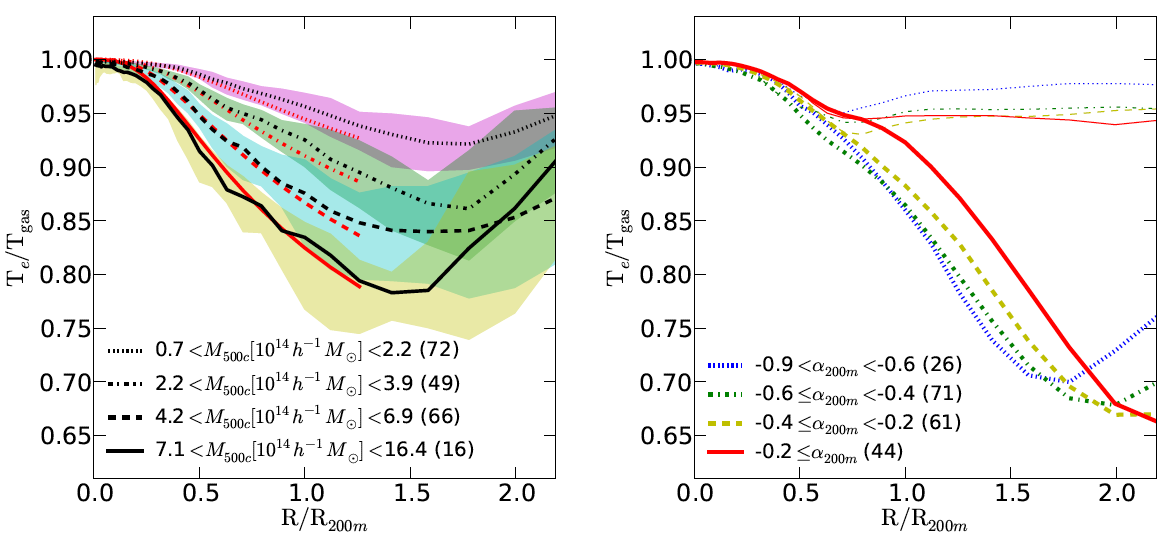}
    \caption{Left panel: Map of showing a thin slice of $T_e/T_{\rm gas}$ for a massive and rapidly accreting cluster taken from cosmological simulation \citep{avestruz15}. The colorbar indicates the $T_e/T_{\rm gas}$ ratio. The width of the map is roughly $16$~Mpc, with thickness of $8$~kpc. The dashed circle indicates $R_{200m}$ of the cluster. The right two panels show the spherically averaged $T_e/T_{\rm gas}$ profile for a sample of cosmologically simulated clusters in different bins of cluster mass (middle panel) and mass accretion rate $\alpha$ (right panel). Here $\alpha$ is defined as the averaged DM infalling velocity normalized by the circular velocity of the cluster. The more negative $\alpha$ is, the more rapid the cluster is accreting its mass. { The lines and shaded regions in the middle panel indicates the mean ratios and their $1\sigma$ scatter; and the thin lines in the right panel indicate the mean ratio measured inside filaments, which show much higher $T_e/T_{\rm gas}$, compared to regions outside filaments. } Figures taken from \cite{avestruz15}, reprinted with permission.  }
    \label{fig:ei_eq}
\end{figure*}

The left panel in Fig.~\ref{fig:ei_eq} shows a map of the ratio of electron-to-total gas temperature $T_e/T_{\rm gas}$ for a galaxy cluster taken from cosmological simulations that tracks electron and ion temperatures separately \cite{avestruz15}. Here the total gas temperature is just the weighted average among electrons and ions: $T_{\rm gas} \equiv (n_e T_e + n_iT_i)/(n_e+n_i)$. The map shows that the smallest $T_e/T_{\rm gas}$ ratio occurs right behind the strong accretion shocks (with Mach number $\sim 10-100$) which are located roughly at around $1.6 R_{200m}$ of the cluster. The distribution of $T_e/T_{\rm gas}$ is also strongly anisotropic. Regions of low $T_e/T_{\rm gas}$ indicate the post-shock regions, effectively tracing the accretion shocks that surround not only the cluster but also the cosmic web filaments that feed into the cluster. 
On the other hand, the infalling gas inside the filaments does not experience any recent strong shocks, thus the electron temperature remains similar to the total gas temperature. However, the filament gas will experience shocks as its infalling velocity exceeds the local sound speed, leading to lower electron temperatures. This typically occurs at radii of about $0.5 R_{200m}$.  The right panels of Fig.~\ref{fig:ei_eq} show the spherically averaged $T_e/T_{\rm gas}$ ratio as a function of radius from the cluster center for the { 65} simulated clusters in the sample. In the inner regions, the equilibration timescale is small because of high gas densities (Eq.~\ref{eq:tei}), so the electrons are in thermal equilibrium with the ions. As density decreases with radius, the $T_e/T_{\rm gas}$ ratio also decreases because of the increasing equilibration timescale. The $T_e/T_{\rm gas}$ ratio reaches the minimum, where the infalling gas is just shocked at the accretion shock radius. 
Note that the overall $T_e/T_{\rm gas}$ ratio also depends on cluster mass and mass accretion rate. The more massive the cluster, the higher the post-shock temperature is, which leads to a longer equilibration time and thus a lower $T_e/T_{\rm gas}$ ratio. Similarly, the gas in clusters with higher mass accretion rates experiences more or stronger merger and accretion shocks, leading to lower $T_e/T_{\rm gas}$. 

Simulations predict that $T_e/T_{\rm gas}$ can be as low as $0.6-0.7$ for most massive and merging clusters around the accretion shock radius $R_{sh} \sim 1.5-2.0 R_{200m}$. At these radii, the gas density is very low, with $n_{\rm gas}\sim 10^{-7}-10^{-6}\,\mathrm{cm}^{-3}$, therefore it will be very difficult to probe gas there even for the next generation of X-ray telescopes, such as ATHENA and the proposed {\em Lynx} mission. For the inner regions at around $0.3 R_{200m}$, the $T_e/T_{\rm gas}$ is predicted to be closer to unity $\sim 0.9-1.0$, meaning that the temperature bias due to electron-ion non-equilibrium is unlikely to be important. Note that the $T_e/T_{\rm gas}$ values predicted in these simulations are also likely to be lower limits, given that other physical effects, which may speed up the equilibration, were absent. If magnetic fields are present, the equilibration time will likely decrease, because the gyroradii of the electrons are smaller than the mean free path, and thus the electrons and ions can interact more rapidly. 

\subsection{Future Simulations and Modeling Efforts}
\label{sec:future_theory}

The simulation results outlined above are mostly from cosmological hydrodynamical simulations. These simulations did not include microphysics such as magnetic fields or cosmic rays, which can become dynamically and energetically important in the outskirts, given the very low densities in these regions. One of the future directions is to examine the physical roles of these microphysical processes in the cluster outskirts. 

An equally challenging aspect in modeling cluster outskirt physics is that the physical processes involved span a wide range of physical scales. Higher mass and spatial resolution will improve the modeling of the largely non-equilibrium physics in cluster outskirts, where the matters is constantly accreting and stirring up the ICM, while dynamical times are long given the low densities in the outskirts. High resolution simulations are needed to resolve the full turbulence cascade in non-thermal gas motions, the small-scales gas clumps in the cluster outskirts and their interactions with the ambient ICM, and the merger and accretion shocks that generate the thermal non-equilibrium between electrons and ions.  

At the same time, to anticipate large-amount of data from upcoming surveys of galaxy clusters, there is a need to perform cosmological simulations with a large enough volume (Gpc scales) to sample enough massive clusters that span the whole spectrum of cluster accretion histories. The dynamical range needed for high-resolution, large-box simulation is very large, with physical scales spanning from sub-kpc scale to Gpc scales, meaning that it will be very computationally expensive to perform these simulations. To tackle this, future efforts will likely focus on multi-scale modeling with (1) dedicated small-box, high resolution simulations to study small-scale accretion physics in detail, (2) large-box, low resolution simulations to sample the range of cluster mass accretion histories, and (3) physically motivated, well-calibrated sub-grid models bridging the simulations of large and small scales.


\begin{figure*}[t]
 \hbox{ 
  \includegraphics[width=0.5\textwidth]{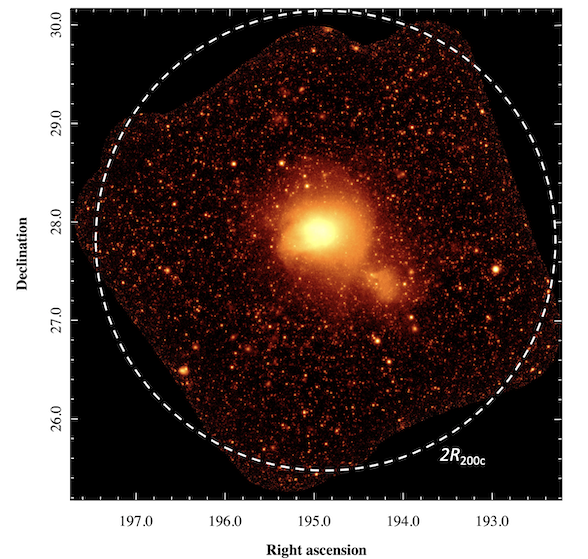}
    \includegraphics[width=0.5\textwidth]{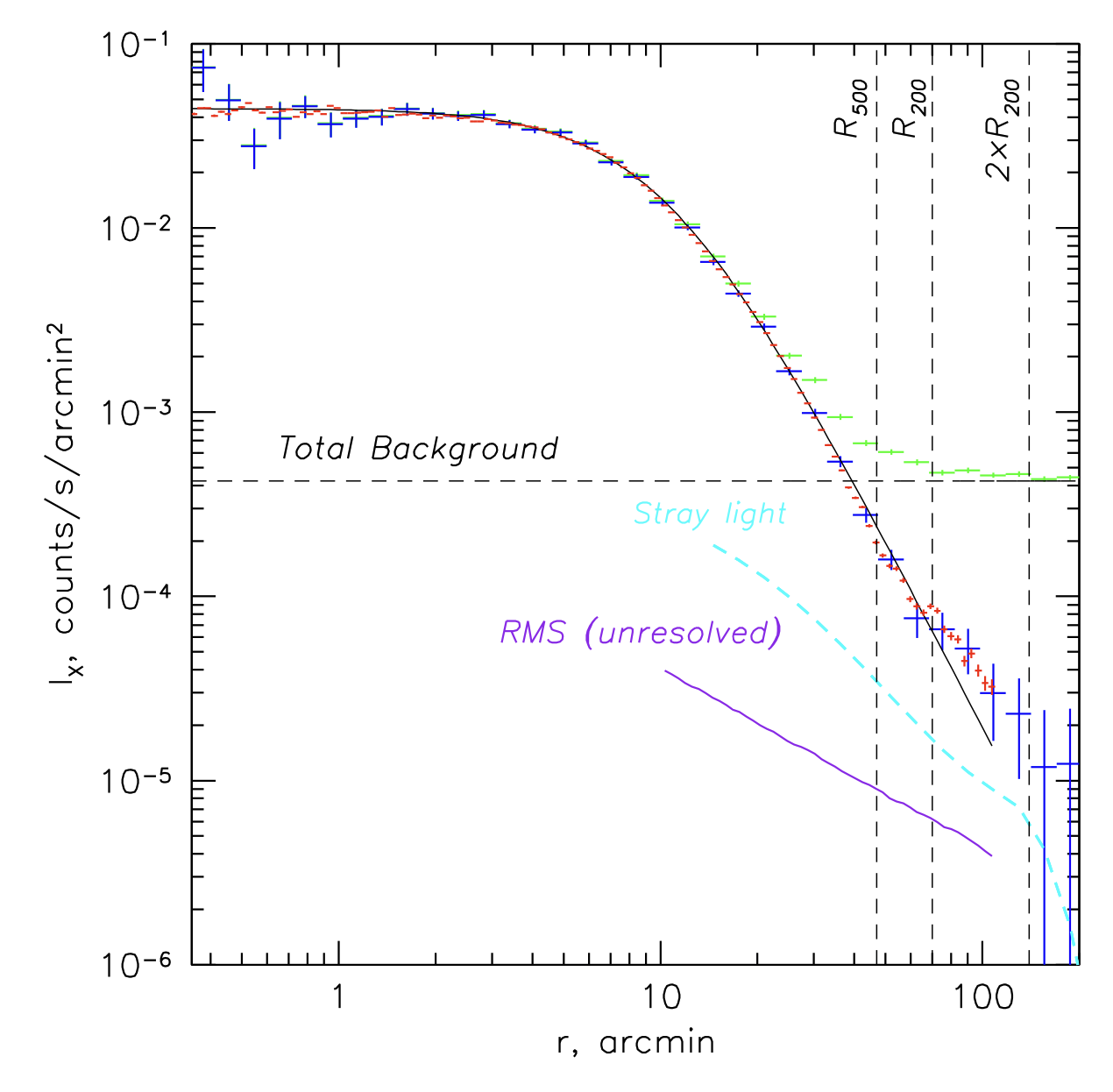}
    } 
    
     \hbox{ 
  \includegraphics[width=0.53\textwidth]{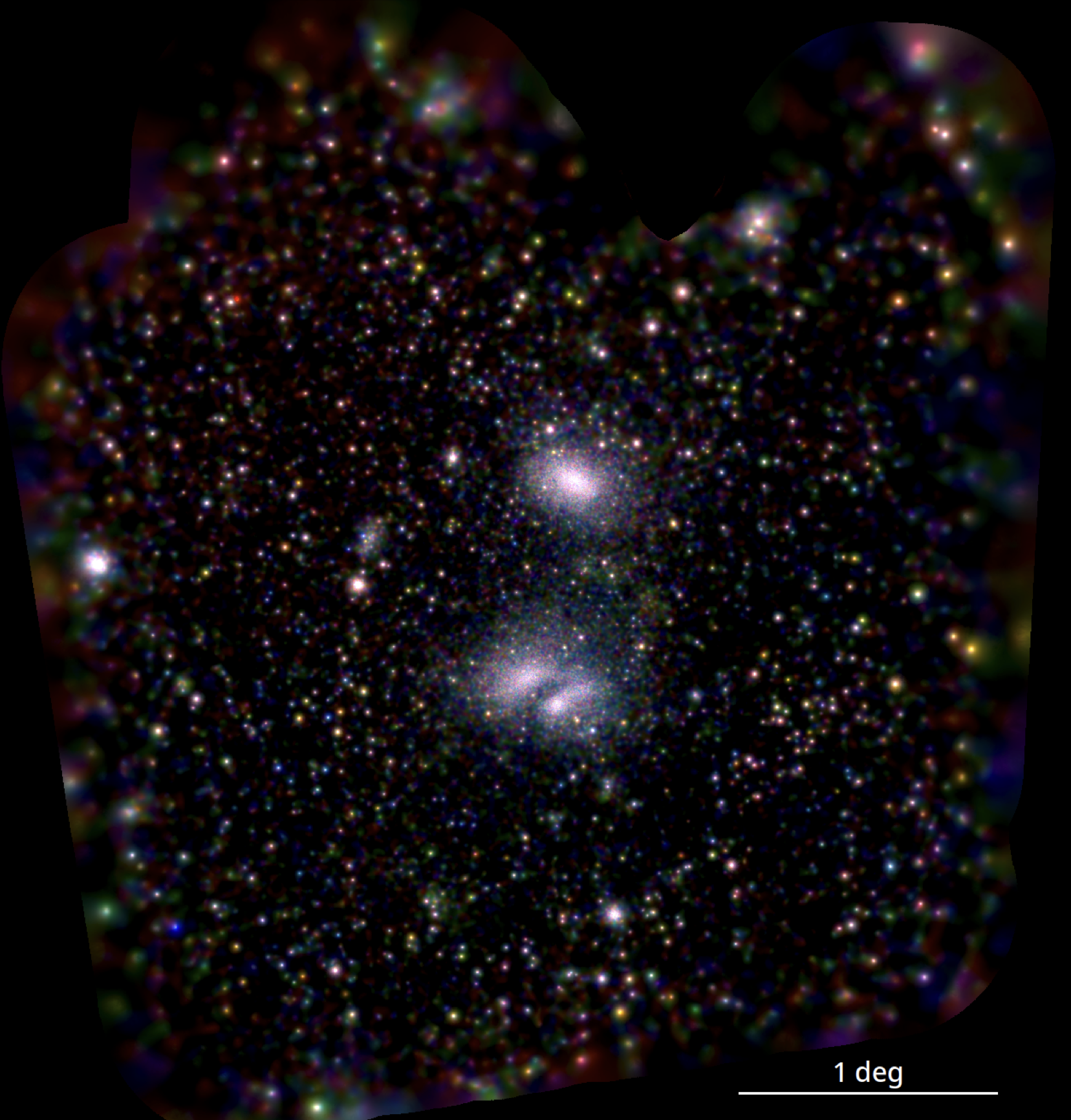}
  \hspace{1.0cm}
  \includegraphics[width=0.42\textwidth]{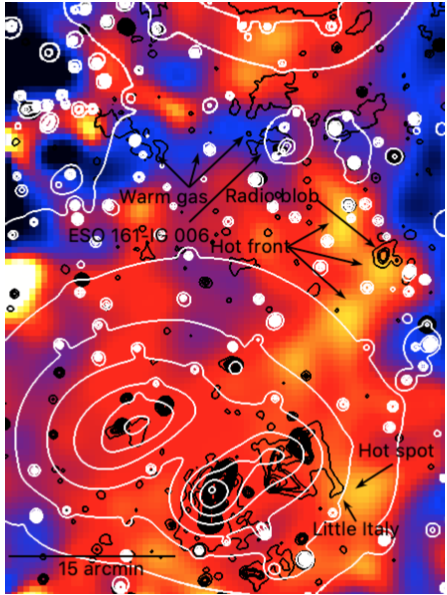}
}
  
  \caption{New eROSITA results. Top panels show results for the Coma cluster from \cite{Churazov2021}, reprinted with permission, showing that X-ray emission is seen out to 2$R_{200c}$. Bottom panels show the Abell 3391/3395 system as studied in \cite{Reiprich2021}, reprinted with permission, with the left panel showing the full X-ray image of the system, and the right hand panel showing an image {of the same system} in a narrow band between 512-658 eV in which emission from the oxygen lines OVII and OVIII is expected from warm gas. Evidence is seen for this warm gas (blue regions).}
\label{eRosita}
\end{figure*}

\section{Upcoming and Future X-ray measurements}
\label{sec:futureobservations}

In 2019 eROSITA \citep{Predehl2021} launched, and it has conducted the first all sky X-ray survey since the ROSAT. eROSITA is the first X-ray telescope to be launched to L2, and this location gives it a stable background. Its short focal length also reduces the instrumental background. eROSITA's large field of view (a 1.03 degree diameter circle) allows it to image clusters out to their outskirts without the need for mosaics, and also allows a local background measurement without the need for dedicated background observations. Early results from eROSITA on the Coma cluster \citep{Churazov2021} are shown in the top of Fig. \ref{eRosita}, showing that X-ray emission is detected out to 2$R_{200c}$. 

eROSITA's high effective area in the soft X-ray band will also allow it to probe the X-ray emission from bridges between neighbouring galaxy clusters. \cite{Reiprich2021} have explored a bridge of X-ray emission in the Abell 3391/3395 galaxy cluster system, shown in the bottom panel of Fig. \ref{eRosita}. The combination of high effective area and good spectral resolution in the soft X-ray band has allowed eROSITA to produce images in the narrow band in which we expect to see oxygen emission lines (in particular OVII at 574 eV and OVIII at 654 eV) from warm ($\leq$1keV) plasma (see the bottom right hand panel of Fig. \ref{eRosita}).

The last decade has seen a tremendous amount of progress in our observational understanding of the cluster outskirts. To make further progress, more powerful X-ray observatories are needed, with greater effective area, a low and stable instrumental background, and high spatial resolution. In the near future, eROSITA's large soft band collecting area, and large field of view combined with good spectral resolution promise a dramatic improvement in our understanding of cluster outskirts, by providing us with routine measurements out to 2$R_{200c}$, and probing the way cosmic web filaments connect to clusters. {With its all-sky coverage, eROSITA will allow us to investigate the outskirts of clusters for a large number of objects, increasing the sample size and allowing population studies possibly down to the group regime.} 

XRISM, scheduled to launch in 2023, will also have the same low earth orbit as Suzaku, but will benefit from a field of view 4 times as large. 

ATHENA is scheduled to launch in the early 2030s. Its large collecting area, combined with {an array of microcalorimeter detectors}, will allow line {shifts and} broadening due to gas motions to be directly probed in the outskirts of clusters, providing a direct measurement of {turbulence and bulk motions}, {in the vein of the pioneering measurements by the Hitomi satellite in the core of the Perseus cluster.} {Prospective possible future missions such as Lynx, AXIS, HUBS or the Cosmic Web Explorer would further combine a large collecting area, a low and stable background and a very high spatial and spectral resolution (see \cite{Walker2019} for a review of these future missions).}

\begin{figure}
\begin{center}
    \includegraphics[width=0.6\textwidth]{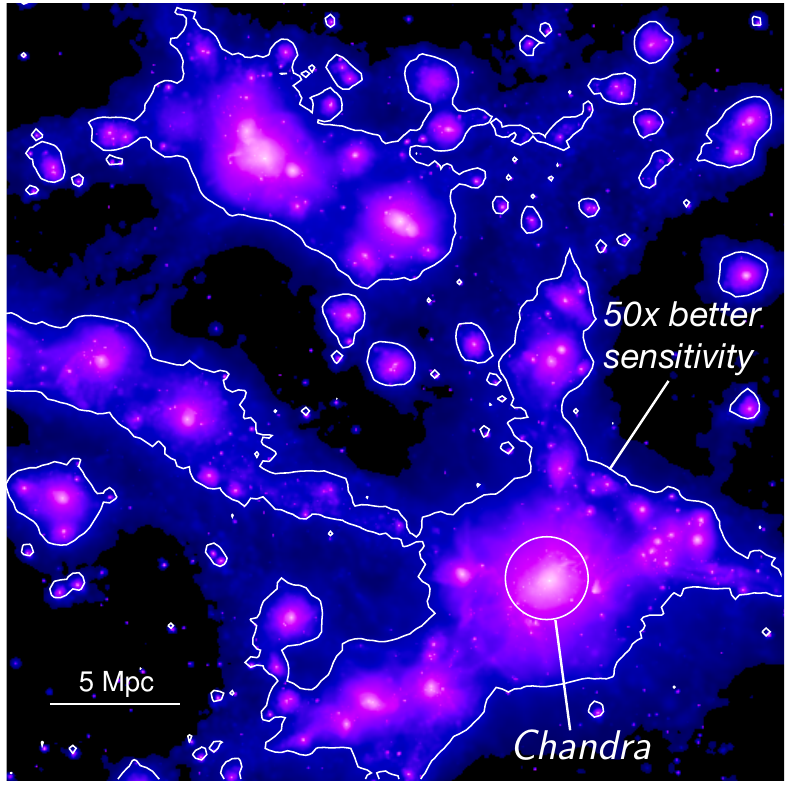}

\end{center}

\caption{Hydrodynamical cosmological simulation of large-scale structure formation, showing that galaxy clusters reside at the nodes of cosmic web filaments. With \textit{Chandra} and \textit{XMM-Newton}, only the central regions of clusters can be explored in detail (inner white circle). To begin to explore large-scale structure in its entirety (outer white contours) requires at least a factor of 50 improvement in sensitivity to low surface brightness extended emission. Figure from \cite{Walker_whitepaper}, reprinted with permission. } 

\label{future}
\end{figure}

With these future missions, it will be possible to probe down to very low levels
of X-ray surface brightness. Fig. \ref{future} shows that an increase in sensitivity by a factor of 50 will allow us to reveal the full picture of the way the cluster outskirts connect with the cosmic web.

\vspace{0.3cm}


\vspace{0.5cm}


\vspace{0.1cm}


\vspace{0.5cm}


\vspace{0.1cm}















\section{Cross-References}

Chapter: Absorption studies of the most diffuse gas in the LSS \\
Chapter: Chemical enrichment in groups and clusters
\\
Chapter: Thermodynamical profiles of clusters and groups, and their evolution \\
Chapter: The Merger Dynamics of the X-ray Emitting Plasma in Clusters of Galaxies \\
Chapter: X-ray cluster cosmology \\

\bibliographystyle{ieeetr} 
\bibliography{all_references}

\begin{thebibliography}{100}

\bibitem{planck18_cosmo}
{Planck Collaboration}, N.~{Aghanim}, Y.~{Akrami}, M.~{Ashdown}, J.~{Aumont},
  C.~{Baccigalupi}, M.~{Ballardini}, A.~J. {Banday}, R.~B. {Barreiro},
  N.~{Bartolo}, S.~{Basak}, R.~{Battye}, K.~{Benabed}, J.~P. {Bernard},
  M.~{Bersanelli}, P.~{Bielewicz}, J.~J. {Bock}, J.~R. {Bond}, J.~{Borrill},
  F.~R. {Bouchet}, F.~{Boulanger}, M.~{Bucher}, C.~{Burigana}, R.~C. {Butler},
  E.~{Calabrese}, J.~F. {Cardoso}, J.~{Carron}, A.~{Challinor}, H.~C. {Chiang},
  J.~{Chluba}, L.~P.~L. {Colombo}, C.~{Combet}, D.~{Contreras}, B.~P. {Crill},
  F.~{Cuttaia}, P.~{de Bernardis}, G.~{de Zotti}, J.~{Delabrouille}, J.~M.
  {Delouis}, E.~{Di Valentino}, J.~M. {Diego}, O.~{Dor{\'e}}, M.~{Douspis},
  A.~{Ducout}, X.~{Dupac}, S.~{Dusini}, G.~{Efstathiou}, F.~{Elsner}, T.~A.
  {En{\ss}lin}, H.~K. {Eriksen}, Y.~{Fantaye}, M.~{Farhang}, J.~{Fergusson},
  R.~{Fernandez-Cobos}, F.~{Finelli}, F.~{Forastieri}, M.~{Frailis}, A.~A.
  {Fraisse}, E.~{Franceschi}, A.~{Frolov}, S.~{Galeotta}, S.~{Galli},
  K.~{Ganga}, R.~T. {G{\'e}nova-Santos}, M.~{Gerbino}, T.~{Ghosh},
  J.~{Gonz{\'a}lez-Nuevo}, K.~M. {G{\'o}rski}, S.~{Gratton}, A.~{Gruppuso},
  J.~E. {Gudmundsson}, J.~{Hamann}, W.~{Handley}, F.~K. {Hansen}, D.~{Herranz},
  S.~R. {Hildebrandt}, E.~{Hivon}, Z.~{Huang}, A.~H. {Jaffe}, W.~C. {Jones},
  A.~{Karakci}, E.~{Keih{\"a}nen}, R.~{Keskitalo}, K.~{Kiiveri}, J.~{Kim},
  T.~S. {Kisner}, L.~{Knox}, N.~{Krachmalnicoff}, M.~{Kunz}, H.~{Kurki-Suonio},
  G.~{Lagache}, J.~M. {Lamarre}, A.~{Lasenby}, M.~{Lattanzi}, C.~R. {Lawrence},
  M.~{Le Jeune}, P.~{Lemos}, J.~{Lesgourgues}, F.~{Levrier}, A.~{Lewis},
  M.~{Liguori}, P.~B. {Lilje}, M.~{Lilley}, V.~{Lindholm},
  M.~{L{\'o}pez-Caniego}, P.~M. {Lubin}, Y.~Z. {Ma}, J.~F.
  {Mac{\'\i}as-P{\'e}rez}, G.~{Maggio}, D.~{Maino}, N.~{Mandolesi},
  A.~{Mangilli}, A.~{Marcos-Caballero}, M.~{Maris}, P.~G. {Martin},
  M.~{Martinelli}, E.~{Mart{\'\i}nez-Gonz{\'a}lez}, S.~{Matarrese}, N.~{Mauri},
  J.~D. {McEwen}, P.~R. {Meinhold}, A.~{Melchiorri}, A.~{Mennella},
  M.~{Migliaccio}, M.~{Millea}, S.~{Mitra}, M.~A. {Miville-Desch{\^e}nes},
  D.~{Molinari}, L.~{Montier}, G.~{Morgante}, A.~{Moss}, P.~{Natoli}, H.~U.
  {N{\o}rgaard-Nielsen}, L.~{Pagano}, D.~{Paoletti}, B.~{Partridge},
  G.~{Patanchon}, H.~V. {Peiris}, F.~{Perrotta}, V.~{Pettorino},
  F.~{Piacentini}, L.~{Polastri}, G.~{Polenta}, J.~L. {Puget}, J.~P. {Rachen},
  M.~{Reinecke}, M.~{Remazeilles}, A.~{Renzi}, G.~{Rocha}, C.~{Rosset},
  G.~{Roudier}, J.~A. {Rubi{\~n}o-Mart{\'\i}n}, B.~{Ruiz-Granados},
  L.~{Salvati}, M.~{Sandri}, M.~{Savelainen}, D.~{Scott}, E.~P.~S. {Shellard},
  C.~{Sirignano}, G.~{Sirri}, L.~D. {Spencer}, R.~{Sunyaev}, A.~S. {Suur-Uski},
  J.~A. {Tauber}, D.~{Tavagnacco}, M.~{Tenti}, L.~{Toffolatti}, M.~{Tomasi},
  T.~{Trombetti}, L.~{Valenziano}, J.~{Valiviita}, B.~{Van Tent}, L.~{Vibert},
  P.~{Vielva}, F.~{Villa}, N.~{Vittorio}, B.~D. {Wandelt}, I.~K. {Wehus},
  M.~{White}, S.~D.~M. {White}, A.~{Zacchei}, and A.~{Zonca}, ``{Planck 2018
  results. VI. Cosmological parameters},'' {\em \aap}, vol.~641, p.~A6, Sept.
  2020.

\bibitem{aung20}
H.~{Aung}, D.~{Nagai}, and E.~T. {Lau}, ``{Shock and splash: gas and dark
  matter halo boundaries around {\ensuremath{\Lambda}}CDM galaxy clusters},''
  {\em \mnras}, vol.~508, pp.~2071--2078, Dec. 2021.

\bibitem{bryan_norman98}
G.~L. {Bryan} and M.~L. {Norman}, ``{Statistical Properties of X-Ray Clusters:
  Analytic and Numerical Comparisons},'' {\em \apj}, vol.~495, pp.~80--+, Mar.
  1998.

\bibitem{vikhlinin06}
A.~{Vikhlinin}, A.~{Kravtsov}, W.~{Forman}, C.~{Jones}, M.~{Markevitch}, S.~S.
  {Murray}, and L.~{Van Speybroeck}, ``{Chandra Sample of Nearby Relaxed Galaxy
  Clusters: Mass, Gas Fraction, and Mass-Temperature Relation},'' {\em \apj},
  vol.~640, pp.~691--709, Apr. 2006.

\bibitem{diemer14}
B.~{Diemer} and A.~V. {Kravtsov}, ``{Dependence of the Outer Density Profiles
  of Halos on Their Mass Accretion Rate},'' {\em \apj}, vol.~789, p.~1, July
  2014.

\bibitem{adhikari14}
S.~{Adhikari}, N.~{Dalal}, and R.~T. {Chamberlain}, ``{Splashback in accreting
  dark matter halos},'' {\em \jcap}, vol.~11, p.~019, Nov. 2014.

\bibitem{more15}
S.~{More}, B.~{Diemer}, and A.~V. {Kravtsov}, ``{The Splashback Radius as a
  Physical Halo Boundary and the Growth of Halo Mass},'' {\em \apj}, vol.~810,
  p.~36, Sept. 2015.

\bibitem{urban14}
O.~{Urban}, A.~{Simionescu}, N.~{Werner}, S.~W. {Allen}, S.~{Ehlert},
  I.~{Zhuravleva}, R.~G. {Morris}, A.~C. {Fabian}, A.~{Mantz}, P.~E.~J.
  {Nulsen}, J.~S. {Sanders}, and Y.~{Takei}, ``{Azimuthally resolved X-ray
  spectroscopy to the edge of the Perseus Cluster},'' {\em \mnras}, vol.~437,
  pp.~3939--3961, Feb. 2014.

\bibitem{Simionescu17}
A.~{Simionescu}, N.~{Werner}, A.~{Mantz}, S.~W. {Allen}, and O.~{Urban},
  ``{Witnessing the growth of the nearest galaxy cluster: thermodynamics of the
  Virgo Cluster outskirts},'' {\em \mnras}, vol.~469, pp.~1476--1495, Aug.
  2017.

\bibitem{MirakhorA2199}
M.~S. {Mirakhor} and S.~A. {Walker}, ``{A high coverage view of the
  thermodynamics and metal abundance in the outskirts of the nearby galaxy
  cluster Abell 2199},'' {\em \mnras}, vol.~497, pp.~3943--3952, Sept. 2020.

\bibitem{Vikhlinin2005}
A.~{Vikhlinin}, M.~{Markevitch}, S.~S. {Murray}, C.~{Jones}, W.~{Forman}, and
  L.~{Van Speybroeck}, ``{Chandra Temperature Profiles for a Sample of Nearby
  Relaxed Galaxy Clusters},'' {\em \apj}, vol.~628, pp.~655--672, Aug. 2005.

\bibitem{Leccardi2008}
A.~{Leccardi} and S.~{Molendi}, ``{Radial temperature profiles for a large
  sample of galaxy clusters observed with XMM-Newton},'' {\em \aap}, vol.~486,
  pp.~359--373, Aug. 2008.

\bibitem{simionescu11}
A.~{Simionescu}, S.~W. {Allen}, A.~{Mantz}, N.~{Werner}, Y.~{Takei}, R.~G.
  {Morris}, A.~C. {Fabian}, J.~S. {Sanders}, P.~E.~J. {Nulsen}, M.~R. {George},
  and G.~B. {Taylor}, ``{Baryons at the Edge of the X-ray-Brightest Galaxy
  Cluster},'' {\em Science}, vol.~331, p.~1576, Mar. 2011.

\bibitem{walker13}
S.~A. {Walker}, A.~C. {Fabian}, J.~S. {Sanders}, A.~{Simionescu}, and
  Y.~{Tawara}, ``{X-ray exploration of the outskirts of the nearby Centaurus
  cluster using Suzaku and Chandra},'' {\em \mnras}, vol.~432, pp.~554--569,
  June 2013.

\bibitem{kawaharada10}
M.~{Kawaharada}, N.~{Okabe}, K.~{Umetsu}, M.~{Takizawa}, K.~{Matsushita},
  Y.~{Fukazawa}, T.~{Hamana}, S.~{Miyazaki}, K.~{Nakazawa}, and T.~{Ohashi},
  ``{Suzaku Observation of A1689: Anisotropic Temperature and Entropy
  Distributions Associated with the Large-scale Structure},'' {\em \apj},
  vol.~714, pp.~423--441, May 2010.

\bibitem{walker12}
S.~A. {Walker}, A.~C. {Fabian}, J.~S. {Sanders}, and M.~R. {George}, ``{Further
  X-ray observations of the galaxy cluster PKS 0745-191 to the virial radius
  and beyond},'' {\em \mnras}, vol.~424, pp.~1826--1840, Aug. 2012.

\bibitem{Ghirardini2019}
V.~{Ghirardini}, D.~{Eckert}, S.~{Ettori}, E.~{Pointecouteau}, S.~{Molendi},
  M.~{Gaspari}, M.~{Rossetti}, S.~{De Grandi}, M.~{Roncarelli}, H.~{Bourdin},
  P.~{Mazzotta}, E.~{Rasia}, and F.~{Vazza}, ``{Universal thermodynamic
  properties of the intracluster medium over two decades in radius in the X-COP
  sample},'' {\em \aap}, vol.~621, p.~A41, Jan. 2019.

\bibitem{Walker2019}
S.~{Walker}, A.~{Simionescu}, D.~{Nagai}, N.~{Okabe}, D.~{Eckert},
  T.~{Mroczkowski}, H.~{Akamatsu}, S.~{Ettori}, and V.~{Ghirardini}, ``{The
  Physics of Galaxy Cluster Outskirts},'' {\em \ssr}, vol.~215, p.~7, Jan.
  2019.

\bibitem{Arnaud10}
M.~{Arnaud}, G.~W. {Pratt}, R.~{Piffaretti}, H.~{B{\"o}hringer}, J.~H.
  {Croston}, and E.~{Pointecouteau}, ``{The universal galaxy cluster pressure
  profile from a representative sample of nearby systems (REXCESS) and the
  Y$_{SZ}$ - M$_{500}$ relation},'' {\em \aap}, vol.~517, p.~A92, July 2010.

\bibitem{voit05}
G.~M. {Voit}, S.~T. {Kay}, and G.~L. {Bryan}, ``{The baseline intracluster
  entropy profile from gravitational structure formation},'' {\em \mnras},
  vol.~364, pp.~909--916, Dec. 2005.

\bibitem{Ghirardini2018}
V.~{Ghirardini}, S.~{Ettori}, D.~{Eckert}, S.~{Molendi}, F.~{Gastaldello},
  E.~{Pointecouteau}, G.~{Hurier}, and H.~{Bourdin}, ``{The XMM Cluster
  Outskirts Project (X-COP): Thermodynamic properties of the intracluster
  medium out to R$_{200}$ in Abell 2319},'' {\em \aap}, vol.~614, p.~A7, June
  2018.

\bibitem{MirakhorVirgo2021}
M.~S. {Mirakhor} and S.~A. {Walker}, ``{Gas clumping in the outskirts of the
  Virgo cluster},'' {\em \mnras}, vol.~506, pp.~139--148, Sept. 2021.

\bibitem{zhuravleva13}
I.~{Zhuravleva}, E.~{Churazov}, A.~{Kravtsov}, E.~T. {Lau}, D.~{Nagai}, and
  R.~{Sunyaev}, ``{Quantifying properties of ICM inhomogeneities},'' {\em
  \mnras}, vol.~428, pp.~3274--3287, Feb. 2013.

\bibitem{Eckert15}
D.~{Eckert}, M.~{Roncarelli}, S.~{Ettori}, S.~{Molendi}, F.~{Vazza},
  F.~{Gastaldello}, and M.~{Rossetti}, ``{Gas clumping in galaxy clusters},''
  {\em \mnras}, vol.~447, pp.~2198--2208, Mar. 2015.

\bibitem{vazza13}
F.~{Vazza}, D.~{Eckert}, A.~{Simionescu}, M.~{Br{\"u}ggen}, and S.~{Ettori},
  ``{Properties of gas clumps and gas clumping factor in the intra-cluster
  medium},'' {\em \mnras}, vol.~429, pp.~799--814, Feb. 2013.

\bibitem{roncarelli13}
M.~{Roncarelli}, S.~{Ettori}, S.~{Borgani}, K.~{Dolag}, D.~{Fabjan}, and
  L.~{Moscardini}, ``{Large-scale inhomogeneities of the intracluster medium:
  improving mass estimates using the observed azimuthal scatter},'' {\em
  \mnras}, vol.~432, pp.~3030--3046, July 2013.

\bibitem{Eckert18}
D.~{Eckert}, V.~{Ghirardini}, S.~{Ettori}, E.~{Rasia}, V.~{Biffi},
  E.~{Pointecouteau}, M.~{Rossetti}, S.~{Molendi}, F.~{Vazza},
  F.~{Gastaldello}, M.~{Gaspari}, S.~{De Grandi}, S.~{Ghizzardi}, H.~{Bourdin},
  C.~{Tchernin}, and M.~{Roncarelli}, ``{Non-thermal pressure support in X-COP
  galaxy clusters},'' {\em \aap}, vol.~621, p.~A40, Jan. 2019.

\bibitem{Hoshino10}
A.~{Hoshino}, J.~P. {Henry}, K.~{Sato}, H.~{Akamatsu}, W.~{Yokota},
  S.~{Sasaki}, Y.~{Ishisaki}, T.~{Ohashi}, M.~{Bautz}, Y.~{Fukazawa},
  N.~{Kawano}, A.~{Furuzawa}, K.~{Hayashida}, N.~{Tawa}, J.~P. {Hughes},
  M.~{Kokubun}, and T.~{Tamura}, ``{X-Ray Temperature and Mass Measurements to
  the Virial Radius of Abell 1413 with Suzaku},'' {\em \pasj}, vol.~62,
  pp.~371--389, Apr. 2010.

\bibitem{markevitch07}
M.~{Markevitch} and A.~{Vikhlinin}, ``{Shocks and cold fronts in galaxy
  clusters},'' {\em \physrep}, vol.~443, pp.~1--53, May 2007.

\bibitem{Ascasibar2006}
Y.~{Ascasibar} and M.~{Markevitch}, ``{The Origin of Cold Fronts in the Cores
  of Relaxed Galaxy Clusters},'' {\em \apj}, vol.~650, pp.~102--127, Oct. 2006.

\bibitem{Roediger2011}
E.~{Roediger}, M.~{Br{\"u}ggen}, A.~{Simionescu}, H.~{B{\"o}hringer},
  E.~{Churazov}, and W.~R. {Forman}, ``{Gas sloshing, cold front formation and
  metal redistribution: the Virgo cluster as a quantitative test case},'' {\em
  \mnras}, vol.~413, pp.~2057--2077, May 2011.

\bibitem{ZuHone2011}
J.~A. {ZuHone}, M.~{Markevitch}, and D.~{Lee}, ``{Sloshing of the Magnetized
  Cool Gas in the Cores of Galaxy Clusters},'' {\em \apj}, vol.~743, p.~16,
  Dec. 2011.

\bibitem{simionescu12}
A.~{Simionescu}, N.~{Werner}, O.~{Urban}, S.~W. {Allen}, A.~C. {Fabian}, J.~S.
  {Sanders}, A.~{Mantz}, P.~E.~J. {Nulsen}, and Y.~{Takei}, ``{Large-scale
  Motions in the Perseus Galaxy Cluster},'' {\em \apj}, vol.~757, p.~182, Oct.
  2012.

\bibitem{Walker2017}
S.~A. {Walker}, J.~{Hlavacek-Larrondo}, M.~{Gendron-Marsolais}, A.~C. {Fabian},
  H.~{Intema}, J.~S. {Sanders}, J.~T. {Bamford}, and R.~{van Weeren}, ``{Is
  there a giant Kelvin-Helmholtz instability in the sloshing cold front of the
  Perseus cluster?},'' {\em \mnras}, vol.~468, pp.~2506--2516, June 2017.

\bibitem{Rossetti13}
M.~{Rossetti}, D.~{Eckert}, S.~{De Grandi}, F.~{Gastaldello}, S.~{Ghizzardi},
  E.~{Roediger}, and S.~{Molendi}, ``{Abell 2142 at large scales: An extreme
  case for sloshing?},'' {\em \aap}, vol.~556, p.~A44, Aug. 2013.

\bibitem{Walker14}
S.~A. {Walker}, A.~C. {Fabian}, and J.~S. {Sanders}, ``{Large-scale gas
  sloshing out to half the virial radius in the strongest cool core REXCESS
  galaxy cluster, RXJ2014.8-2430},'' {\em \mnras}, vol.~441, pp.~L31--L35, June
  2014.

\bibitem{Douglass2018}
E.~M. {Douglass}, E.~L. {Blanton}, S.~W. {Randall}, T.~E. {Clarke}, L.~O.~V.
  {Edwards}, Z.~{Sabry}, and J.~A. {ZuHone}, ``{The Megaparsec-scale
  Gas-sloshing Spiral in the Remnant Cool Core Cluster Abell 1763},'' {\em
  \apj}, vol.~868, p.~121, Dec. 2018.

\bibitem{zuhone11}
J.~A. {ZuHone}, M.~{Markevitch}, and D.~{Lee}, ``{Sloshing of the Magnetized
  Cool Gas in the Cores of Galaxy Clusters},'' {\em \apj}, vol.~743, p.~16,
  Dec. 2011.

\bibitem{Walker2018}
S.~A. {Walker}, J.~{ZuHone}, A.~{Fabian}, and J.~{Sanders}, ``{The split in the
  ancient cold front in the Perseus cluster},'' {\em Nature Astronomy}, vol.~2,
  pp.~292--296, Feb. 2018.

\bibitem{Akamatsu2013a}
H.~{Akamatsu} and H.~{Kawahara}, ``{Systematic X-Ray Analysis of Radio Relic
  Clusters with Suzaku},'' {\em \pasj}, vol.~65, p.~16, Feb. 2013.

\bibitem{Ogrean2014}
G.~A. {Ogrean}, M.~{Br{\"u}ggen}, R.~{van Weeren}, H.~{R{\"o}ttgering},
  A.~{Simionescu}, M.~{Hoeft}, and J.~H. {Croston}, ``{Multiple density
  discontinuities in the merging galaxy cluster CIZA J2242.8+5301},'' {\em
  \mnras}, vol.~440, pp.~3416--3425, June 2014.

\bibitem{Akamatsu2015}
H.~{Akamatsu}, R.~J. {van Weeren}, G.~A. {Ogrean}, H.~{Kawahara}, A.~{Stroe},
  D.~{Sobral}, M.~{Hoeft}, H.~{R{\"o}ttgering}, M.~{Br{\"u}ggen}, and J.~S.
  {Kaastra}, ``{Suzaku X-ray study of the double radio relic galaxy cluster
  CIZA J2242.8+5301},'' {\em \aap}, vol.~582, p.~A87, Oct. 2015.

\bibitem{vanWeeren2010}
R.~J. {van Weeren}, H.~J.~A. {R{\"o}ttgering}, M.~{Br{\"u}ggen}, and
  M.~{Hoeft}, ``{Particle Acceleration on Megaparsec Scales in a Merging Galaxy
  Cluster},'' {\em Science}, vol.~330, p.~347, Oct. 2010.

\bibitem{vanWeeren2019}
R.~J. {van Weeren}, F.~{de Gasperin}, H.~{Akamatsu}, M.~{Br{\"u}ggen},
  L.~{Feretti}, H.~{Kang}, A.~{Stroe}, and F.~{Zandanel}, ``{Diffuse Radio
  Emission from Galaxy Clusters},'' {\em \ssr}, vol.~215, p.~16, Feb. 2019.

\bibitem{Akamatsu13}
H.~{Akamatsu}, S.~{Inoue}, T.~{Sato}, K.~{Matsusita}, Y.~{Ishisaki}, and C.~L.
  {Sarazin}, ``{Suzaku X-Ray Observations of the Accreting NGC 4839 Group of
  Galaxies and a Radio Relic in the Coma Cluster},'' {\em \pasj}, vol.~65,
  p.~89, Aug. 2013.

\bibitem{Akamatsu2017}
H.~{Akamatsu}, M.~{Mizuno}, N.~{Ota}, Y.~Y. {Zhang}, R.~J. {van Weeren},
  H.~{Kawahara}, Y.~{Fukazawa}, J.~S. {Kaastra}, M.~{Kawaharada},
  K.~{Nakazawa}, T.~{Ohashi}, H.~J.~A. {R{\"o}ttgering}, M.~{Takizawa},
  J.~{Vink}, and F.~{Zandanel}, ``{Suzaku observations of the merging galaxy
  cluster Abell 2255: The northeast radio relic},'' {\em \aap}, vol.~600,
  p.~A100, Apr. 2017.

\bibitem{Stroe2014}
A.~{Stroe}, J.~J. {Harwood}, M.~J. {Hardcastle}, and H.~J.~A. {R{\"o}ttgering},
  ``{Spectral age modelling of the `Sausage' cluster radio relic},'' {\em
  \mnras}, vol.~445, pp.~1213--1222, Dec. 2014.

\bibitem{Urban2017}
O.~{Urban}, N.~{Werner}, S.~W. {Allen}, A.~{Simionescu}, and A.~{Mantz}, ``{A
  uniform metallicity in the outskirts of massive, nearby galaxy clusters},''
  {\em \mnras}, vol.~470, pp.~4583--4599, Oct 2017.

\bibitem{Biffi2018}
V.~{Biffi}, S.~{Planelles}, S.~{Borgani}, E.~{Rasia}, G.~{Murante},
  D.~{Fabjan}, and M.~{Gaspari}, ``{The origin of ICM enrichment in the
  outskirts of present-day galaxy clusters from cosmological hydrodynamical
  simulations},'' {\em \mnras}, vol.~476, pp.~2689--2703, May 2018.

\bibitem{Werner2013}
N.~{Werner}, O.~{Urban}, A.~{Simionescu}, and S.~W. {Allen}, ``{A uniform metal
  distribution in the intergalactic medium of the Perseus cluster of
  galaxies},'' {\em \nat}, vol.~502, pp.~656--658, Oct 2013.

\bibitem{Mernier2018}
F.~{Mernier}, V.~{Biffi}, H.~{Yamaguchi}, P.~{Medvedev}, A.~{Simionescu},
  S.~{Ettori}, N.~{Werner}, J.~S. {Kaastra}, J.~{de Plaa}, and L.~{Gu},
  ``{Enrichment of the Hot Intracluster Medium: Observations},'' {\em \ssr},
  vol.~214, p.~129, Dec. 2018.

\bibitem{Simionescu2015}
A.~{Simionescu}, N.~{Werner}, O.~{Urban}, S.~W. {Allen}, Y.~{Ichinohe}, and
  I.~{Zhuravleva}, ``{A Uniform Contribution of Core-collapse and Type Ia
  Supernovae to the Chemical Enrichment Pattern in the Outskirts of the Virgo
  Cluster},'' {\em \apj}, vol.~811, p.~L25, Oct 2015.

\bibitem{Asplund2009}
M.~{Asplund}, N.~{Grevesse}, A.~J. {Sauval}, and P.~{Scott}, ``{The Chemical
  Composition of the Sun},'' {\em \araa}, vol.~47, pp.~481--522, Sept. 2009.

\bibitem{Werner2008detection}
N.~{Werner}, A.~{Finoguenov}, J.~S. {Kaastra}, A.~{Simionescu}, J.~P.
  {Dietrich}, J.~{Vink}, and H.~{B{\"o}hringer}, ``{Detection of hot gas in the
  filament connecting the clusters of galaxies Abell 222 and Abell 223},'' {\em
  \aap}, vol.~482, pp.~L29--L33, May 2008.

\bibitem{Eckert15_Nature}
D.~{Eckert}, M.~{Jauzac}, H.~{Shan}, J.-P. {Kneib}, T.~{Erben}, H.~{Israel},
  E.~{Jullo}, M.~{Klein}, R.~{Massey}, J.~{Richard}, and C.~{Tchernin},
  ``{Warm-hot baryons comprise 5-10 per cent of filaments in the cosmic web},''
  {\em \nat}, vol.~528, pp.~105--107, Dec. 2015.

\bibitem{ade2016planck}
{Planck Collaboration XIII}, ``Planck 2015 results-xiii. cosmological
  parameters,'' {\em \aap}, vol.~594, p.~A13, 2016.

\bibitem{Fukugita1998}
M.~{Fukugita}, C.~J. {Hogan}, and P.~J.~E. {Peebles}, ``{The Cosmic Baryon
  Budget},'' {\em \apj}, vol.~503, pp.~518--530, Aug. 1998.

\bibitem{Bregman2007}
J.~N. {Bregman}, ``{The Search for the Missing Baryons at Low Redshift},'' {\em
  \araa}, vol.~45, pp.~221--259, Sept. 2007.

\bibitem{Sinha2010}
M.~{Sinha} and K.~{Holley-Bockelmann}, ``{Balancing the baryon budget: the
  fraction of the IGM due to galaxy mergers},'' {\em \mnras}, vol.~405,
  pp.~L31--L35, June 2010.

\bibitem{Cen1999where}
R.~{Cen} and J.~P. {Ostriker}, ``{Where Are the Baryons?},'' {\em \apj},
  vol.~514, pp.~1--6, Mar. 1999.

\bibitem{Dave2001}
R.~{Dav{\'e}}, R.~{Cen}, J.~P. {Ostriker}, G.~L. {Bryan}, L.~{Hernquist},
  N.~{Katz}, D.~H. {Weinberg}, M.~L. {Norman}, and B.~{O'Shea}, ``{Baryons in
  the Warm-Hot Intergalactic Medium},'' {\em \apj}, vol.~552, pp.~473--483, May
  2001.

\bibitem{Dolag2006Simulating}
K.~{Dolag}, M.~{Meneghetti}, L.~{Moscardini}, E.~{Rasia}, and A.~{Bonaldi},
  ``{Simulating the physical properties of dark matter and gas inside the
  cosmic web},'' {\em \mnras}, vol.~370, pp.~656--672, Aug. 2006.

\bibitem{Cui2019the}
W.~{Cui}, A.~{Knebe}, N.~I. {Libeskind}, S.~{Planelles}, X.~{Yang}, W.~{Cui},
  R.~{Dav{\'e}}, X.~{Kang}, R.~{Mostoghiu}, L.~{Staveley-Smith}, H.~{Wang},
  P.~{Wang}, and G.~{Yepes}, ``{The large-scale environment from cosmological
  simulations II: The redshift evolution and distributions of baryons},'' {\em
  \mnras}, vol.~485, pp.~2367--2379, May 2019.

\bibitem{Nicastro2005mass}
F.~{Nicastro}, S.~{Mathur}, M.~{Elvis}, J.~{Drake}, T.~{Fang}, A.~{Fruscione},
  Y.~{Krongold}, H.~{Marshall}, R.~{Williams}, and A.~{Zezas}, ``{The mass of
  the missing baryons in the X-ray forest of the warm-hot intergalactic
  medium},'' {\em \nat}, vol.~433, pp.~495--498, Feb. 2005.

\bibitem{Zappacosta2010studying}
L.~{Zappacosta}, F.~{Nicastro}, R.~{Maiolino}, G.~{Tagliaferri}, D.~A. {Buote},
  T.~{Fang}, P.~J. {Humphrey}, and F.~{Gastaldello}, ``{Studying the WHIM
  Content of Large-scale Structures Along the Line of Sight to H 2356-309},''
  {\em \apj}, vol.~717, pp.~74--84, July 2010.

\bibitem{Nicastro2018observations}
F.~{Nicastro}, J.~{Kaastra}, Y.~{Krongold}, S.~{Borgani}, E.~{Branchini},
  R.~{Cen}, M.~{Dadina}, C.~W. {Danforth}, M.~{Elvis}, F.~{Fiore}, A.~{Gupta},
  S.~{Mathur}, D.~{Mayya}, F.~{Paerels}, L.~{Piro}, D.~{Rosa-Gonzalez},
  J.~{Schaye}, J.~M. {Shull}, J.~{Torres-Zafra}, N.~{Wijers}, and
  L.~{Zappacosta}, ``{Observations of the missing baryons in the warm-hot
  intergalactic medium},'' {\em \nat}, vol.~558, pp.~406--409, June 2018.

\bibitem{Alvarez2018chandra}
G.~E. {Alvarez}, S.~W. {Randall}, H.~{Bourdin}, C.~{Jones}, and
  K.~{Holley-Bockelmann}, ``{Chandra and XMM-Newton Observations of the Abell
  3395/Abell 3391 Intercluster Filament},'' {\em \apj}, vol.~858, p.~44, May
  2018.

\bibitem{Sakelliou2004xmm}
I.~{Sakelliou} and T.~J. {Ponman}, ``{XMM-Newton observations of the binary
  cluster system Abell 399/401},'' {\em \mnras}, vol.~351, pp.~1439--1456, July
  2004.

\bibitem{Sato2010study}
K.~{Sato}, R.~L. {Kelley}, Y.~{Takei}, T.~{Tamura}, N.~Y. {Yamasaki},
  T.~{Ohashi}, A.~{Gupta}, and M.~{Galeazzi}, ``{Study of the Intracluster and
  Intergalactic Medium in the Sculptor Supercluster with Suzaku},'' {\em
  \pasj}, vol.~62, p.~1423, Dec. 2010.

\bibitem{Mitsuishi2012search}
I.~{Mitsuishi}, A.~{Gupta}, N.~Y. {Yamasaki}, Y.~{Takei}, T.~{Ohashi},
  K.~{Sato}, M.~{Galeazzi}, J.~P. {Henry}, and R.~L. {Kelley}, ``{Search for
  X-Ray Emission Associated with the Shapley Supercluster with Suzaku},'' {\em
  \pasj}, vol.~64, p.~18, Feb. 2012.

\bibitem{MirakhorA2029}
M.~S. {Mirakhor}, S.~A. {Walker}, and J.~{Runge}, ``{A detailed study of the
  bridge of excess X-ray emission between the galaxy clusters Abell 2029 and
  Abell 2033},'' {\em \mnras}, vol.~509, pp.~1109--1118, Jan. 2022.

\bibitem{Connor2018}
T.~{Connor}, D.~D. {Kelson}, J.~{Mulchaey}, A.~{Vikhlinin}, S.~G. {Patel},
  M.~L. {Balogh}, G.~{Joshi}, R.~{Kraft}, D.~{Nagai}, and S.~{Starikova},
  ``{Wide-field Optical Spectroscopy of Abell 133: A Search for Filaments
  Reported in X-Ray Observations},'' {\em \apj}, vol.~867, p.~25, Nov. 2018.

\bibitem{Reiprich2021}
T.~H. {Reiprich}, A.~{Veronica}, F.~{Pacaud}, M.~E. {Ramos-Ceja}, N.~{Ota},
  J.~{Sanders}, M.~{Kara}, T.~{Erben}, M.~{Klein}, J.~{Erler}, J.~{Kerp}, D.~N.
  {Hoang}, M.~{Br{\"u}ggen}, J.~{Marvil}, L.~{Rudnick}, V.~{Biffi}, K.~{Dolag},
  J.~{Aschersleben}, K.~{Basu}, H.~{Brunner}, E.~{Bulbul}, K.~{Dennerl},
  D.~{Eckert}, M.~{Freyberg}, E.~{Gatuzz}, V.~{Ghirardini}, F.~{K{\"a}fer},
  A.~{Merloni}, K.~{Migkas}, K.~{Nandra}, P.~{Predehl}, J.~{Robrade},
  M.~{Salvato}, B.~{Whelan}, A.~{Diaz-Ocampo}, D.~{Hernandez-Lang},
  A.~{Zenteno}, M.~J.~I. {Brown}, J.~D. {Collier}, J.~M. {Diego}, A.~M.
  {Hopkins}, A.~{Kapinska}, B.~{Koribalski}, T.~{Mroczkowski}, R.~P. {Norris},
  A.~{O'Brien}, and E.~{Vardoulaki}, ``{The Abell 3391/95 galaxy cluster
  system. A 15 Mpc intergalactic medium emission filament, a warm gas bridge,
  infalling matter clumps, and (re-) accelerated plasma discovered by combining
  SRG/eROSITA data with ASKAP/EMU and DECam data},'' {\em \aap}, vol.~647,
  p.~A2, Mar. 2021.

\bibitem{kaiser86}
N.~{Kaiser}, ``{Evolution and clustering of rich clusters},'' {\em \mnras},
  vol.~222, pp.~323--345, Sept. 1986.

\bibitem{voit05b}
G.~M. {Voit}, ``{Tracing cosmic evolution with clusters of galaxies},'' {\em
  Rev.Mod.Phys.}, vol.~77, p.~207, 2005.

\bibitem{lau15}
E.~T. {Lau}, D.~{Nagai}, C.~{Avestruz}, K.~{Nelson}, and A.~{Vikhlinin},
  ``{Mass Accretion and its Effects on the Self-similarity of Gas Profiles in
  the Outskirts of Galaxy Clusters},'' {\em \apj}, vol.~806, p.~68, June 2015.

\bibitem{molnar09}
S.~M. {Molnar}, N.~{Hearn}, Z.~{Haiman}, G.~{Bryan}, A.~E. {Evrard}, and
  G.~{Lake}, ``{Accretion Shocks in Clusters of Galaxies and Their SZ Signature
  from Cosmological Simulations},'' {\em \apj}, vol.~696, pp.~1640--1656, May
  2009.

\bibitem{vazza09}
F.~{Vazza}, G.~{Brunetti}, A.~{Kritsuk}, R.~{Wagner}, C.~{Gheller}, and
  M.~{Norman}, ``{Turbulent motions and shocks waves in galaxy clusters
  simulated with adaptive mesh refinement},'' {\em \aap}, vol.~504, pp.~33--43,
  Sept. 2009.

\bibitem{shi14}
X.~{Shi} and E.~{Komatsu}, ``{Analytical model for non-thermal pressure in
  galaxy clusters},'' {\em \mnras}, vol.~442, pp.~521--532, July 2014.

\bibitem{nelson12}
K.~{Nelson}, D.~H. {Rudd}, L.~{Shaw}, and D.~{Nagai}, ``{Evolution of the
  Merger-induced Hydrostatic Mass Bias in Galaxy Clusters},'' {\em \apj},
  vol.~751, p.~121, June 2012.

\bibitem{nelson14}
K.~{Nelson}, E.~T. {Lau}, D.~{Nagai}, D.~H. {Rudd}, and L.~{Yu}, ``{Weighing
  Galaxy Clusters with Gas. II. On the Origin of Hydrostatic Mass Bias in
  {$\Lambda$}CDM Galaxy Clusters},'' {\em \apj}, vol.~782, p.~107, Feb. 2014.

\bibitem{nelson14b}
K.~{Nelson}, E.~T. {Lau}, and D.~{Nagai}, ``{Hydrodynamic Simulation of
  Non-thermal Pressure Profiles of Galaxy Clusters},'' {\em \apj}, vol.~792,
  p.~25, Sept. 2014.

\bibitem{rasia04}
E.~{Rasia}, G.~{Tormen}, and L.~{Moscardini}, ``{A dynamical model for the
  distribution of dark matter and gas in galaxy clusters},'' {\em \mnras},
  vol.~351, pp.~237--252, June 2004.

\bibitem{lau09}
E.~T. {Lau}, A.~V. {Kravtsov}, and D.~{Nagai}, ``{Residual Gas Motions in the
  Intracluster Medium and Bias in Hydrostatic Measurements of Mass Profiles of
  Clusters},'' {\em \apj}, vol.~705, pp.~1129--1138, Nov. 2009.

\bibitem{rasia06}
E.~{Rasia}, S.~{Ettori}, L.~{Moscardini}, P.~{Mazzotta}, S.~{Borgani},
  K.~{Dolag}, G.~{Tormen}, L.~M. {Cheng}, and A.~{Diaferio}, ``{Systematics in
  the X-ray cluster mass estimators},'' {\em \mnras}, vol.~369, pp.~2013--2024,
  July 2006.

\bibitem{Nagai07}
D.~{Nagai}, A.~{Vikhlinin}, and A.~V. {Kravtsov}, ``{Testing X-Ray Measurements
  of Galaxy Clusters with Cosmological Simulations},'' {\em \apj}, vol.~655,
  pp.~98--108, Jan. 2007.

\bibitem{Planck15CluterCosmology}
{Planck Collaboration}, P.~A.~R. {Ade}, N.~{Aghanim}, M.~{Arnaud},
  M.~{Ashdown}, J.~{Aumont}, C.~{Baccigalupi}, A.~J. {Banday}, R.~B.
  {Barreiro}, J.~G. {Bartlett}, and et~al., ``{Planck 2015 results. XXIV.
  Cosmology from Sunyaev-Zeldovich cluster counts},'' {\em \aap}, vol.~594,
  p.~A24, Sept. 2016.

\bibitem{biffi2016}
V.~{Biffi}, S.~{Borgani}, G.~{Murante}, E.~{Rasia}, S.~{Planelles}, G.~L.
  {Granato}, C.~{Ragone-Figueroa}, A.~M. {Beck}, M.~{Gaspari}, and K.~{Dolag},
  ``{On the Nature of Hydrostatic Equilibrium in Galaxy Clusters},'' {\em
  \apj}, vol.~827, p.~112, Aug. 2016.

\bibitem{suto13}
D.~{Suto}, H.~{Kawahara}, T.~{Kitayama}, S.~{Sasaki}, Y.~{Suto}, and R.~{Cen},
  ``{Validity of Hydrostatic Equilibrium in Galaxy Clusters from Cosmological
  Hydrodynamical Simulations},'' {\em \apj}, vol.~767, p.~79, Apr. 2013.

\bibitem{mathiesen99}
B.~{Mathiesen}, A.~E. {Evrard}, and J.~J. {Mohr}, ``{The Effects of Clumping
  and Substructure on Intracluster Medium Mass Measurements},'' {\em \apjl},
  vol.~520, pp.~L21--L24, July 1999.

\bibitem{nagai11}
D.~{Nagai} and E.~T. {Lau}, ``{Gas Clumping in the Outskirts of {$\Lambda$}CDM
  Clusters},'' {\em \apjl}, vol.~731, p.~L10, Apr. 2011.

\bibitem{avestruz14}
C.~{Avestruz}, E.~T. {Lau}, D.~{Nagai}, and A.~{Vikhlinin}, ``{Testing X-Ray
  Measurements of Galaxy Cluster Outskirts with Cosmological Simulations},''
  {\em \apj}, vol.~791, p.~117, Aug. 2014.

\bibitem{spitzer62}
L.~{Spitzer}, {\em {Physics of Fully Ionized Gases}}.
\newblock Physics of Fully Ionized Gases, New York: Interscience (2nd edition),
  1962, 1962.

\bibitem{fox97}
D.~C. {Fox} and A.~{Loeb}, ``{Do the Electrons and Ions in X-Ray Clusters Share
  the Same Temperature?},'' {\em \apj}, vol.~491, pp.~459--+, Dec. 1997.

\bibitem{wong2009}
K.-W. {Wong} and C.~L. {Sarazin}, ``{Effects of the Non-Equipartition of
  Electrons and Ions in the Outskirts of Relaxed Galaxy Clusters},'' {\em
  \apj}, vol.~707, pp.~1141--1159, Dec. 2009.

\bibitem{rudd09}
D.~H. {Rudd} and D.~{Nagai}, ``{Non-equilibrium Electrons and the
  Sunyaev-Zel'dovich Effect of Galaxy Clusters},'' {\em accepted to the ApJL},
  June 2009.

\bibitem{avestruz15}
C.~{Avestruz}, D.~{Nagai}, E.~T. {Lau}, and K.~{Nelson}, ``{Non-equilibrium
  Electrons in the Outskirts of Galaxy Clusters},'' {\em \apj}, vol.~808,
  p.~176, Aug. 2015.

\bibitem{Churazov2021}
E.~{Churazov}, I.~{Khabibullin}, N.~{Lyskova}, R.~{Sunyaev}, and A.~M. {Bykov},
  ``{Tempestuous life beyond R$_{500}$: X-ray view on the Coma cluster with
  SRG/eROSITA. I. X-ray morphology, recent merger, and radio halo
  connection},'' {\em \aap}, vol.~651, p.~A41, July 2021.

\bibitem{Predehl2021}
P.~{Predehl}, R.~{Andritschke}, V.~{Arefiev}, V.~{Babyshkin}, O.~{Batanov},
  W.~{Becker}, H.~{B{\"o}hringer}, A.~{Bogomolov}, T.~{Boller}, K.~{Borm},
  W.~{Bornemann}, H.~{Br{\"a}uninger}, M.~{Br{\"u}ggen}, H.~{Brunner},
  M.~{Brusa}, E.~{Bulbul}, M.~{Buntov}, V.~{Burwitz}, W.~{Burkert}, N.~{Clerc},
  E.~{Churazov}, D.~{Coutinho}, T.~{Dauser}, K.~{Dennerl}, V.~{Doroshenko},
  J.~{Eder}, V.~{Emberger}, T.~{Eraerds}, A.~{Finoguenov}, M.~{Freyberg},
  P.~{Friedrich}, S.~{Friedrich}, M.~{F{\"u}rmetz}, A.~{Georgakakis},
  M.~{Gilfanov}, S.~{Granato}, C.~{Grossberger}, A.~{Gueguen}, P.~{Gureev},
  F.~{Haberl}, O.~{H{\"a}lker}, G.~{Hartner}, G.~{Hasinger}, H.~{Huber},
  L.~{Ji}, A.~v. {Kienlin}, W.~{Kink}, F.~{Korotkov}, I.~{Kreykenbohm},
  G.~{Lamer}, I.~{Lomakin}, I.~{Lapshov}, T.~{Liu}, C.~{Maitra},
  N.~{Meidinger}, B.~{Menz}, A.~{Merloni}, T.~{Mernik}, B.~{Mican}, J.~{Mohr},
  S.~{M{\"u}ller}, K.~{Nandra}, V.~{Nazarov}, F.~{Pacaud}, M.~{Pavlinsky},
  E.~{Perinati}, E.~{Pfeffermann}, D.~{Pietschner}, M.~E. {Ramos-Ceja},
  A.~{Rau}, J.~{Reiffers}, T.~H. {Reiprich}, J.~{Robrade}, M.~{Salvato},
  J.~{Sanders}, A.~{Santangelo}, M.~{Sasaki}, H.~{Scheuerle}, C.~{Schmid},
  J.~{Schmitt}, A.~{Schwope}, A.~{Shirshakov}, M.~{Steinmetz}, I.~{Stewart},
  L.~{Str{\"u}der}, R.~{Sunyaev}, C.~{Tenzer}, L.~{Tiedemann},
  J.~{Tr{\"u}mper}, V.~{Voron}, P.~{Weber}, J.~{Wilms}, and V.~{Yaroshenko},
  ``{The eROSITA X-ray telescope on SRG},'' {\em \aap}, vol.~647, p.~A1, Mar.
  2021.

\bibitem{Walker_whitepaper}
S.~{Walker}, D.~{Nagai}, A.~{Simionescu}, M.~{Markevitch}, H.~{Akamatsu},
  M.~{Arnaud}, C.~{Avestruz}, M.~{Bautz}, V.~{Biffi}, S.~{Borgani},
  E.~{Bulbul}, E.~{Churazov}, K.~{Dolag}, D.~{Eckert}, S.~{Ettori},
  Y.~{Fujita}, M.~{Gaspari}, V.~{Ghirardini}, R.~{Kraft}, E.~T. {Lau},
  A.~{Mantz}, K.~{Matsushita}, M.~{McDonald}, E.~{Miller}, T.~{Mroczkowski},
  P.~{Nulsen}, N.~{Okabe}, N.~{Ota}, E.~{Pointecouteau}, G.~{Pratt}, K.~{Sato},
  X.~{Shi}, G.~{Tremblay}, M.~{Tremmel}, F.~{Vazza}, I.~{Zhuravleva},
  E.~{Zinger}, and J.~{ZuHone}, ``{Unveiling the Galaxy Cluster - Cosmic Web
  Connection with X-ray observations in the Next Decade},'' {\em \baas},
  vol.~51, p.~218, May 2019.

\end{thebibliography}


\end{document}